\documentclass[11pt]{article}
\usepackage{amsmath}
\usepackage{graphicx}
\usepackage{amsthm}
\usepackage{amsfonts}
\usepackage{color}
\usepackage{orcidlink}
\usepackage{hyperref}
\hypersetup{colorlinks=true, linkcolor=blue, citecolor=blue, urlcolor=blue}
\usepackage{caption}
\usepackage{subcaption}
\usepackage{setspace}
\usepackage{multirow}
\usepackage{multicol}
\usepackage{amssymb}
\RequirePackage[numbers,sort&compress]{natbib}
\begin{document}
\baselineskip=20pt

\begin{center}
\LARGE{\bf Effects of spontaneous Lorentz Symmetry breaking on Letelier-AdS charged black holes within Kalb-Ramond gravity }
\par\end{center}


\begin{center}
{\bf Faizuddin Ahmed\orcidlink{0000-0003-2196-9622}}\footnote{\bf faizuddinahmed15@gmail.com}\\

{\it Department of Physics, The Assam Royal Global University, Guwahati, 781035, Assam, India}\\

{\bf Ahmad Al-Badawi\orcidlink{0000-0002-3127-3453}}\footnote{\bf ahmadbadawi@ahu.edu.jo}\\
{\it Department of Physics, Al-Hussein Bin Talal University, 71111,
Ma'an, Jordan} \\

{\bf \.{I}zzet Sakall{\i}\orcidlink{0000-0001-7827-9476}}\footnote{\bf izzet.sakalli@emu.edu.tr (Corresponding author)}\\
{\it Physics Department, Eastern Mediterranean University, Famagusta 99628, North Cyprus via Mersin 10, Turkey}

\end{center}

\vspace{0.2cm}

\begin{abstract}
In this study, we investigate the geodesic motion of massless particles-specifically photons-in the spacetime of a charged anti-de Sitter (AdS) black hole (BH) surrounded by a cloud of strings (CoS) within the framework of Kalb-Ramond (KR) gravity. We analyze the effective potential that governs photon trajectories, explore the properties and location of the photon sphere (PS), and examine the effective radial force acting on photons. The resulting BH shadow is also studied, highlighting the roles of both the CoS parameter $\alpha$ and the KR field parameter $\ell$ in shaping its geometry. We constrain these parameters using observational data from M87* and Sgr A* obtained by the Event Horizon Telescope (EHT). Furthermore, we extend our investigation to the motion of neutral test particles in the same gravitational background. By examining the impact of the CoS and KR field, we show how these additional fields modify the dynamics relative to standard charged BH scenarios. Finally, we study the fundamental frequencies associated with quasiperiodic oscillations (QPOs) of test particles, demonstrating how these frequencies are affected by the presence of the CoS and KR field. 
\end{abstract}

\section{Introduction}\label{sec:1}

{\color{black} Einstein’s general relativity (GR) is widely regarded as one of the greatest scientific achievements of the twentieth century. It provides an accurate description of gravitational phenomena across scales-from solar system dynamics to the large-scale structure of the universe and is supported by extensive observational evidence. GR has also led to profound insights into modern cosmology and astrophysics, with BHs standing out as one of its most striking predictions \cite{Einstein1916}. The groundbreaking images of the supermassive BHs M87* and Sgr A*, released by the Event Horizon Telescope (EHT), have opened a new era in both theoretical and observational BH physics \cite{EHTL1,EHTL4,EHTL6,EHTL12,EHTL25,EHTL26}. These observations not only confirm GR’s predictions but also offer valuable insights into the nature of strong gravity and extreme cosmic environments, paving the way for a new branch of BH astronomy. The BH shadow-an inner dark region-arises from the dynamics of photon orbits and the underlying spacetime geometry, while the surrounding bright, asymmetric photon ring is formed by strongly lensed light. Recent EHT observations have revealed both features with unprecedented detail.}

From a theoretical perspective, BHs represent some of the simplest yet most profound solutions to Einstein’s field equations. Under assumptions such as stationarity, asymptotic flatness, and the absence of external fields, an isolated BH is fully characterized by only three observable parameters-mass, angular momentum, and electric charge \cite{BB1,BB2}. This result is formalized in the no-hair theorem, which states that all higher multipole moments of the gravitational and electromagnetic fields are uniquely determined by these quantities \cite{BB3,BB4,BB5}. Despite this apparent simplicity, BHs play a central role in modern astrophysics and cosmology. Observations confirm the existence of stellar-mass BHs formed from collapsed massive stars and supermassive BHs (SMBHs) residing in galactic centers \cite{BB6}. The first direct detection of gravitational waves from binary BH mergers by LIGO-Virgo provided striking confirmation of GR in the strong-field regime \cite{BB7}. BHs also exhibit rich dynamical and thermodynamic behavior: accretion processes in active galactic nuclei (AGN) produce extreme luminosities and relativistic jets \cite{BB8}, while SMBH feedback shapes galaxy evolution and underlies empirical correlations such as the $M$–$\sigma$ relation \cite{BB9,BB10}. Thus, although BHs are defined by only a few parameters, their physical manifestations are remarkably diverse, placing them at the intersection of GR, quantum theory, thermodynamics, and high-energy astrophysics.

{\color{black} General relativity (GR) extends the principles of special relativity by incorporating gravity while preserving local Lorentz symmetry at every point on the spacetime manifold. Although extensive experimental and observational evidence supports Lorentz symmetry as a fundamental feature of nature, several theoretical frameworks suggest that it may be violated at sufficiently high energy scales. Such possibilities arise in string theory \cite{VAK1989}, loop quantum gravity \cite{Alfaro2002}, Hořava-Lifshitz gravity \cite{Horava2009}, and noncommutative field theory \cite{Carroll2001}. These approaches offer important insights into the microscopic structure of spacetime and the foundations of fundamental physics. Lorentz symmetry breaking (LSB) can occur either explicitly or spontaneously: explicit LSB arises when the Lagrangian density itself is not Lorentz invariant, implying that the physical laws differ between reference frames, whereas spontaneous LSB occurs when the ground state of a system violates Lorentz symmetry despite a Lorentz-invariant Lagrangian. The Standard-Model Extension (SME) \cite{VAK2004} provides a comprehensive framework for investigating spontaneous LSB and its phenomenological implications.}

Finding a unified theory-commonly expected to be a theory of quantum gravity-remains one of the most fundamental challenges in theoretical physics. Several candidates have been proposed, with string theory standing out as one of the most promising approaches. Modified gravity theories often emerge as low-energy effective descriptions of such underlying quantum gravity frameworks. A compelling feature that may serve as a signal of Planck-scale physics is Lorentz violation (LV). Among the various formulations of LV, explicit Lorentz violation is particularly significant. It is characterized by vacuum expectation values (VEVs) that remain invariant under active (or particle) Lorentz transformations. To consistently describe such VEVs that violate local Lorentz symmetry while preserving general coordinate invariance, a suitable geometric framework is required. In this regard, Riemann-Cartan geometry-which allows for torsion-is especially appropriate \cite{CC1}. The SME provides a comprehensive phenomenological framework for describing explicit Lorentz violation within an effective field theory approach \cite{CC2}. Interestingly, in flat spacetime, Lorentz violation is closely tied to CPT (Charge, Parity, and Time) violation, as has been demonstrated in various studies \cite{CC3,CC4}. In the Minkowski spacetime limit of the SME, Lorentz-violating terms can be systematically classified based on their behavior under CPT transformations. One widely studied and attractive mechanism for explicit Lorentz violation is spontaneous Lorentz violation \cite{CC5}. Originally introduced within the context of string theory, this mechanism was initially believed to be a unique feature of string models and unlikely to appear in four-dimensional renormalizable gauge theories. Another approach that gives rise to LV involves noncommutative field theory, which can be seen as a specific realization of the SME \cite{CC6}. In such theories, the noncommutative nature of spacetime-embodied by the non-vanishing commutators of coordinate operators-can be interpreted as arising from VEVs of certain fields in the string spectrum. This structure emerges naturally within string theory \cite{CC7}. 

A particularly well-studied framework for spontaneous Lorentz violation is the Bumblebee model, originally proposed in Ref.~\cite{CC8}. In this scenario, the VEV of a vector field selects a preferred spacetime direction, thereby breaking local Lorentz symmetry. The model further imposes constraints on the compactification of extra dimensions \cite{CC8}. Despite its minimal structure, the Bumblebee model effectively captures the essential phenomenology associated with spontaneous LSB. The Bumblebee field itself may be interpreted as a vector field emerging from the first excited state of the open bosonic string spectrum. More recently, significant attention has been devoted to BH solutions in the presence of the KR field \cite{Kalb1974}, a self-interacting, second-rank antisymmetric tensor field that modifies the Einstein–Hilbert action. This modification is closely connected with heterotic string theory \cite{Gross1985}. A modified BH solution incorporating the KR field was introduced in \cite{CC13}, followed by investigations of test-particle dynamics and gravitational lensing phenomena in \cite{CC14}. {\color{black} In addition, traversable wormhole solutions supported by a background KR field have been constructed in Refs.~\cite{Lessa2021,Maluf2022}, while the consequences of LSB on Bianchi type I cosmology were analyzed in \cite{Neves2022}.} A static and spherically symmetric BH solution in the presence of a non-dynamical KR field-where the field assumes a fixed VEV configuration was obtained in Ref.~\cite{CC9}. This work was later extended to electrically charged configurations in \cite{CC10}. A more general treatment of static, neutral, spherically symmetric BHs with a KR field was presented in Ref.~\cite{CC11}. Numerous subsequent studies have broadened the exploration of BH solutions in KR gravity, examining diverse theoretical and phenomenological aspects. For example, quasinormal modes and greybody factor bounds with generalized uncertainty principle corrections were studied in \cite{CC16}. The interplay between geodesic motion, scattering processes, and quasinormal modes in charged BH spacetimes was discussed in \cite{CC17}. The geodesic structure and scalar-field dynamics in BH geometries with a surrounding quintessence field were investigated in \cite{CC19}. The role of perfect fluid dark matter in KR-modified BH environments was analyzed in \cite{CC20}. Moreover, circular motion and QPOs were examined in \cite{CC21}, and the radiative properties and QPO behavior of charged BHs in a KR background were explored in \cite{CC22}. {\color{black} Further investigations of BH solutions in KR-gravity in several directions have been reported in Refs.~\cite{ref1,ref2,Panting2025,Ahmad2024,Sucu2025,Mangut2025,AAAF5,Hassan}. Recently, a BH space-time in KR gravity incorporating non-commutative corrections derived from a gauge-theoretic approach was introduced in \cite{AAAF1}. Subsequently, various physical aspects of this new BH space-time have been examined, including optical phenomena \cite{AAAF2}, particle motion and thermal effects \cite{AAAF3}, as well as neutrino dynamics \cite{AAAF4}. Collectively, these studies underscore the growing interest in the phenomenology of KR gravity, highlighting its potential to provide new insights into BH physics, gravitational lensing, and observable astrophysical signatures. It is noteworthy, however, that BH space-times in KR gravity with a global monopole have been investigated only in a limited number of works \cite{ref3,Baruah2025,Fathi2025}, and, to the best of our knowledge, such systems have not yet been studied in the presence of a CoS \cite{PSL}.}

QPOs represent a significant phenomenon observed in the X-ray emissions emanating from neutron stars and BHs, offering invaluable insights into the extreme gravitational environments surrounding these compact objects. These oscillations, characterized by their nearly periodic variations in brightness, are believed to arise from fundamental physical mechanisms, including the dynamic behavior of accretion discs and the complex gravitational interactions within these systems. The detection of QPOs provides a unique window into probing the nature of spacetime in strong gravity regimes, where relativistic effects become prominent \cite{JR2023a}. Among the various types of QPOs, twin-peak QPOs have garnered particular attention due to their distinctive frequency patterns, which are thought to be linked to oscillatory modes or resonances within the accretion disc structure. These discoveries have motivated the development of sophisticated theoretical models aimed at explaining the origin and properties of QPOs, alongside efforts to enhance observational capabilities for more accurate measurements. The initial identification of QPOs through spectral analysis and studies of X-ray binary fluxes \cite{LA1989} laid the foundation for extensive research into their physical nature. Notably, models that relate QPO phenomena to the dynamics of charged particle motion and the resultant modulations in their orbital trajectories have shown great promise. These models often emphasize the synchronized or resonant motions of charged test particles in the strong gravitational and magnetic fields near compact objects, leading to observable features such as accretion disc oscillations \cite{DHW2015, PCF2016, MOR2020, ZS2020, AM2020, JR2023b, Vrba2025,Mitra2025,Vrba2023, Reggie2025, Zdenek2022, Grigoris2021, Zdenek2015,Stuchlik2021}. Continued exploration of QPOs not only advances our understanding of accretion physics but also serves as a crucial tool for testing the predictions of general relativity and uncovering the fundamental properties of matter under extreme conditions.

{\color{black}
From an astrophysical perspective, gaining a deeper understanding of the underlying spacetime geometry and the fields surrounding compact objects has become increasingly important. Such fields can substantially modify the geodesic structure of particles and photons, thereby influencing key observable quantities, including the photon-sphere radius, the resulting shadow size, and the innermost stable circular orbit (ISCO). These observables play a crucial role in probing not only the geometry of the background spacetime but also the physical fields present around astrophysical BHs. Consequently, exploring the properties of BH solutions-and identifying features that distinguish them from their potential mimickers in alternative theories of gravity-provides valuable insight, particularly in regions close to the event horizon. Previous studies have examined the influence of the KR field on the geodesics of null and time-like particles, as well as on gravitational lensing in the presence of a global monopole charge \cite{ref3,Baruah2025,Fathi2025}. Motivated by these considerations, in this work, we investigate} the optical properties of a static, spherically symmetric, charged BH in an AdS background within the framework of KR gravity. Moreover, the BH is surrounded by a CoS, where only the electric sector contributes to the background geometry. This configuration was originally proposed by Letelier \cite{PSL}, who introduced a phenomenological model describing a CoS as a matter source in Einstein’s field equations. {\color{black}At this point, it is also worth noting that BHs in various configurations in $(1+2)$- and $(1+3)$-dimensional space-times have been extensively studied in the presence of a CoS (see, for example, \cite{Girgoris2018,Angel2018,Ahmed2025,Saeed2025,Marcos2025,Edilberto2025,Bouzenada2025} and references therein).} In this paper, we derive several key optical features of the solution, including the photon sphere, BH shadow, and the effective radial force experienced by photons. Furthermore, we analyze the topological properties of the photon sphere associated with this geometry, providing deeper insight into the behavior of null orbits under the influence of the background fields. Special emphasis is placed on understanding how the KR field parameter and the string cloud density affect these optical characteristics. We show that both fields induce non-trivial modifications to photon dynamics and shadow morphology, thereby offering a potential observational window into physics beyond general relativity. Finally, we investigate the dynamics of neutral test particles in this background, with particular attention to the innermost stable circular orbits and the fundamental frequencies characterizing particle motion, demonstrating that the KR field and the string cloud significantly affect these properties as well.  

{\color{black}
\section{Letelier-AdS Charged BH with LSB: Background Space-time Geometry}\label{sec:2}

In Ref.~\cite{ref3}, the authors investigated BH solutions in Einstein-KR bumblebee gravity sourced by a global monopole both in the absence and presence of the cosmological constant. The static and spherically symmetric metric describing this BH solution is given by
\begin{equation}
    ds^2=-\left(\frac{1-k\,\eta^2}{1-\ell}-\frac{2\,M}{r}-\frac{\Lambda_\text{eff}}{3}\,r^2\right)\,dt^2+\left(\frac{1-k\,\eta^2}{1-\ell}-\frac{2\,M}{r}-\frac{\Lambda_\text{eff}}{3}\,r^2\right)^{-1}\,dr^2+r^2\,(d\theta^2+\sin \theta^2\,d\phi^2).\label{aa2}
\end{equation}
Here $\eta$ is the global monopole charge.

Inspired by these works, we are interested in a static and spherically symmetric AdS BH solution in KR field gravity coupled with a CoS instead of global monopole as does in Ref.~\cite{ref3}. Moreover, we further extend the same solutions for electrically charged analogue to the solution in Ref.~\cite{CC10}. The complete action for this theory, including the nonminimal coupling between the KR field and the Ricci tensor, is presented in Appendix~\ref{app:A} [see Eq.~\eqref{action}]. The corresponding modified Einstein equations and energy-momentum tensors are derived in Eqs.~\eqref{einstein_eq}--\eqref{T_epsilon}.

\subsection{Cloud of strings (C\lowercase{o}S): Matter Content}

The action that describes the CoS is given by \cite{PSL}:
\begin{equation}
S_{\text{CoS}} = \int d^4x \, \sqrt{-g} \, \mathcal{L}_{\text{CoS}} 
= \int (-\gamma)^{1/2} \, \mathcal{M} \, d\lambda^0 d\lambda^1 
= \int \mathcal{M} \left(-\frac{1}{2} \Sigma_{\mu\nu} \Sigma^{\mu\nu} \right)^{1/2} d\lambda^0 d\lambda^1,
\end{equation}
where $\mathcal{M}$ is a dimensionless constant associated with each string, and $\gamma$ is the determinant of the induced metric $\gamma_{AB}$, defined as
\begin{equation}
\gamma_{AB} = g_{\mu\nu} \frac{\partial x^{\mu}}{\partial \lambda^A} \frac{\partial x^{\nu}}{\partial \lambda^B},
\end{equation}
with $x^\mu = x^\mu(\lambda^A)$ describing the string world sheet. The parameters $\lambda^0$ and $\lambda^1$ are, respectively, time-like and space-like coordinates on the world sheet. The bivector $\Sigma^{\mu\nu}$, associated with the world sheet of the string, is defined as:
\begin{equation}
\Sigma^{\mu\nu} = \epsilon^{AB} \frac{\partial x^\mu}{\partial \lambda^A} \frac{\partial x^\nu}{\partial \lambda^B},
\end{equation}
where $\epsilon^{AB}$ is the Levi–Civita symbol with $\epsilon^{01} = -\epsilon^{10} = 1$.

Due to the symmetries of the spacetime, the only non-zero component of the bivector $\Sigma^{\mu\nu}$ is:
\begin{equation}
\Sigma^{01} = \frac{\alpha}{\rho\, r^2},
\end{equation}
which depends only on the radial coordinate $r$. Here, $\alpha$ is an integration constant related to the density of the CoS, constrained by $0 < \alpha < 1$.

A spherically symmetric CoS is described by the stress-energy tensor \cite{PSL}:
\begin{equation}
    T^{\mu\nu}(\mbox{CoS})=\rho_c\,\frac{\Sigma^{\mu\beta}\,\Sigma^{\nu}_{\beta}}{\sqrt{-\gamma}}.\label{aa3}
\end{equation}
The bivector $\Sigma^{\mu\nu}$ satisfies the following equations:
\begin{equation}
\nabla_{\mu} \left( \rho_p \Sigma^{\mu\nu} \right) = 0, \qquad 
\Sigma^{\mu\beta} \nabla_{\mu} \left( \frac{\Sigma^{\nu}_{\ \beta}}{\sqrt{-\gamma}} \right) = 0.
\label{eq:SigmaEqs}
\end{equation}

For spherically symmetric solutions, Letelier showed that the only component satisfying $\gamma < 0$ is the ``electric-like'' component $\Sigma^{01}$. Therefore, the non-null components of the stress-energy tensor of the CoS are given by \cite{PSL}:
\begin{align}
T^{t\,\rm CoS}_{t}= T^{r\,\rm CoS}_{r} =\rho_c= \frac{\alpha}{r^2},\quad
T^{\theta\,\rm CoS}_{\theta}= T^{\phi\,\rm CoS}_{\phi}=0,\label{aa5}
\end{align}
where $\rho_c$ is the energy density of the cloud and $\alpha$ is an integration constant related to the string. Following the development of the Letelier BH solution---a spherically symmetric generalization of the Schwarzschild BH incorporating a CoS---numerous researchers have extended this framework by constructing BH solutions in various configurations, both within the context of general relativity and in modified gravity theories (see Refs.~\cite{Motohashi:2018wdq,Samanta:2018hbw,Ahmed:2024qeu}), and references therein).

Following the procedure outlined in previous studies, particularly in \cite{ref3}, one can construct a static, spherically symmetric, charged AdS BH solution within the framework of KR gravity coupled to a CoS. For the KR field, we consider a pseudo-electric configuration in which the field $B_{\mu\nu}$ is frozen to its VEV $b_{\mu\nu}$, with explicit form given by Eq.~\eqref{KR_config} in Appendix~\ref{app:A}. This configuration leads to the constant norm $b_{\mu\nu} b^{\mu\nu} = -|b|^2$, and the field strength $H_{\lambda\mu\nu}$ identically vanishes. The strength of the Lorentz violation is encoded in the dimensionless parameter [cf. Eq.~\eqref{ell_def}]:
\begin{equation}
\ell = \frac{\varepsilon\,|b|^2}{2}.
\label{ell_parameter}
\end{equation}

Considering a non-vanishing cosmological constant, we find that assuming the KR field to be frozen at its vacuum expectation value, namely $V=0$, does not lead to physically acceptable solutions consistent with the equations of motion. Consequently, an alternative formulation must be adopted. This can be achieved by introducing a linear potential of the form $V(X)=\tilde{\lambda} X$, where $\tilde{\lambda}$ now plays the role of a Lagrange multiplier field (see, e.g., \cite{Maluf:2020kgf,Yang:2023wtu}). In this case, the derivative of the potential with respect to $X$ is simply $V_X(X)=\tilde{\lambda}$. 

Therefore, for $\Lambda\neq 0$, the gravitational equations~\eqref{field} (see Appendix~\ref{app:A}) become explicitly:
\begin{align}\label{ee1}
  (\ell -1) r f''+2 (\ell -1) f'-2 \Lambda  r=0,   
\end{align}
and
\begin{align}\label{ee2}
\ell\,  r^2 f''+(\ell +1) r f'-(\ell -1) f+r^2 \left(b^2 \tilde{\lambda} +\Lambda \right)+\alpha -1=0.    
\end{align}
After some manipulations with Eqs.~\eqref{ee1} and \eqref{ee2}, we obtain 
\begin{align}\label{ee3}
r(1-\ell) f'-(\ell -1) f+r^2 \left(b^2 \tilde{\lambda} +\frac{(1-3 \ell) \Lambda }{1-\ell}\right)+\alpha -1=0.   
\end{align}
The above equation can be solved analytically. Its solution is given by 
\begin{align}\label{fwc}
f(r)=\frac{1-\alpha }{1-\ell}-\frac{(1-3\ell)\Lambda+(1-\ell)\tilde{\lambda} b^2}{3(1-\ell)^2}r^2-\frac{2M}{r}.    
\end{align}
By substituting Eq.~\eqref{fwc} into Eq.~\eqref{ee1}, we can show that the on-shell value of $\tilde{\lambda}$ is determined by
\begin{align}\label{lambda_val}
 \tilde{\lambda}=\frac{3\ell\Lambda}{(1-\ell)b^2}.   
\end{align}
Thereby, the metric function of the uncharged BH solution reads
\begin{align}\label{sol2}
f(r)=\frac{1-\alpha}{1-\ell}-\frac{\Lambda}{3(1-\ell)}r^2-\frac{2M}{r}.
\end{align}

For charged solutions, one must include the electromagnetic field contribution. The matter Lagrangian for the electromagnetic field coupled to the KR field is given by Eq.~\eqref{EM_lagrangian} in Appendix~\ref{app:A}. When the KR field acquires a nonzero VEV, the Lagrangian induces LSB of the electromagnetic field and allows for the existence of electrically charged BH solutions. The resulting modified field equation is given by Eq.~\eqref{charged_field_eq}.

The spacetime is described in Schwarzschild coordinates \((x^0 = t,\, x^1 = r,\, x^2 = \theta,\, x^3 = \phi)\) by the line element: 
\begin{equation}
    ds^2=g_{\mu\nu}\,dx^{\mu}\,dx^{\nu}=-f(r)\,dt^2+\frac{1}{f(r)}\,dr^2+r^2\,(d\theta^2+\sin^2 \theta\,d\phi^2),\label{metric}
\end{equation}
where the metric function $f(r)$ is given by \cite{arxiv1,arxiv2}
\begin{equation}
f(r) = 
\begin{cases}
\displaystyle \frac{1 - \alpha}{1 - \ell} - \frac{2\,M}{r} - \frac{\Lambda_\text{eff}}{3} \, r^2, & \text{for } Q = 0, \\[12pt]
\displaystyle \frac{1 - \alpha}{1 - \ell} - \frac{2\,M}{r} + \frac{Q^2_\text{eff}}{r^2} - \frac{\Lambda_\text{eff}}{3} \, r^2, & \text{for } Q \neq 0.
\end{cases}
\label{function}
\end{equation}
where $Q_\text{eff} = Q/(1-\ell)$ is the effective charge and $\Lambda_\text{eff} = \Lambda/(1-\ell)$ is the effective cosmological constant. In the limiting case $\alpha = \ell > 0$, corresponding to equal contributions from the KR field and string clouds, the lapse function reduces to that of the un(-charged) AdS-like BHs in KR field.

\begin{figure}[ht!]
    \centering
    \includegraphics[width=0.45\linewidth]{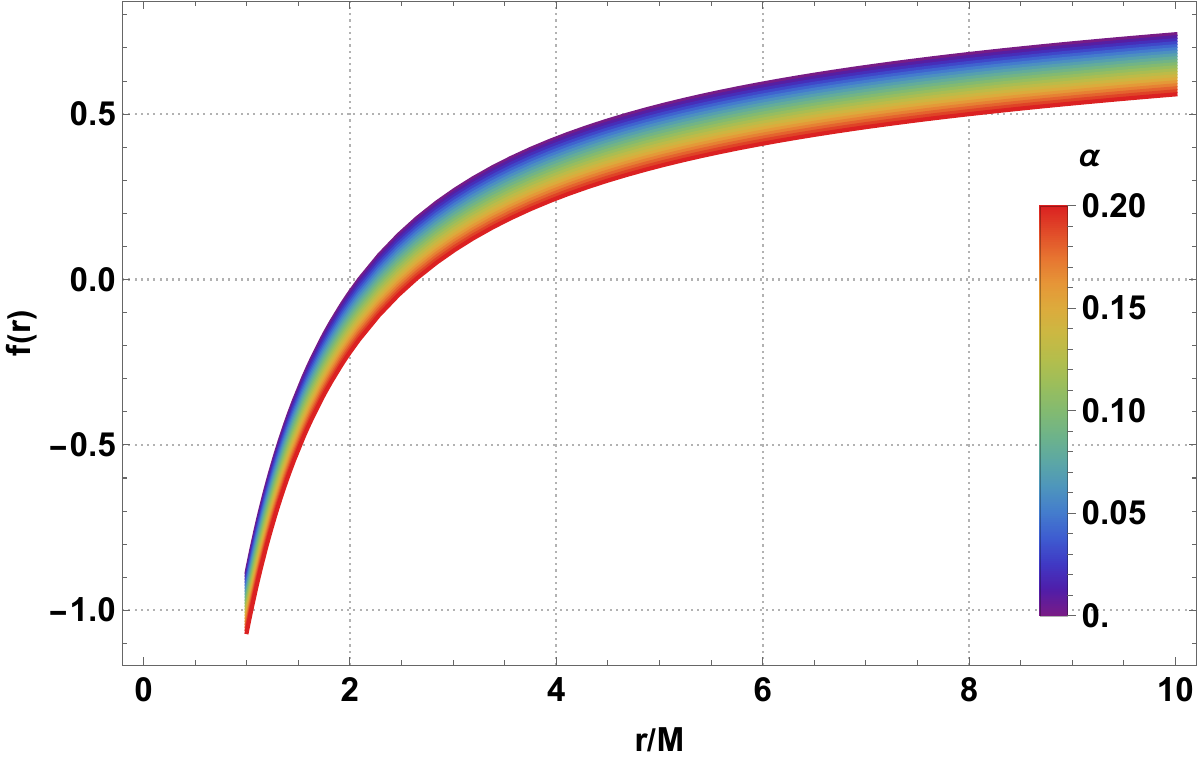}\qquad
    \includegraphics[width=0.45\linewidth]{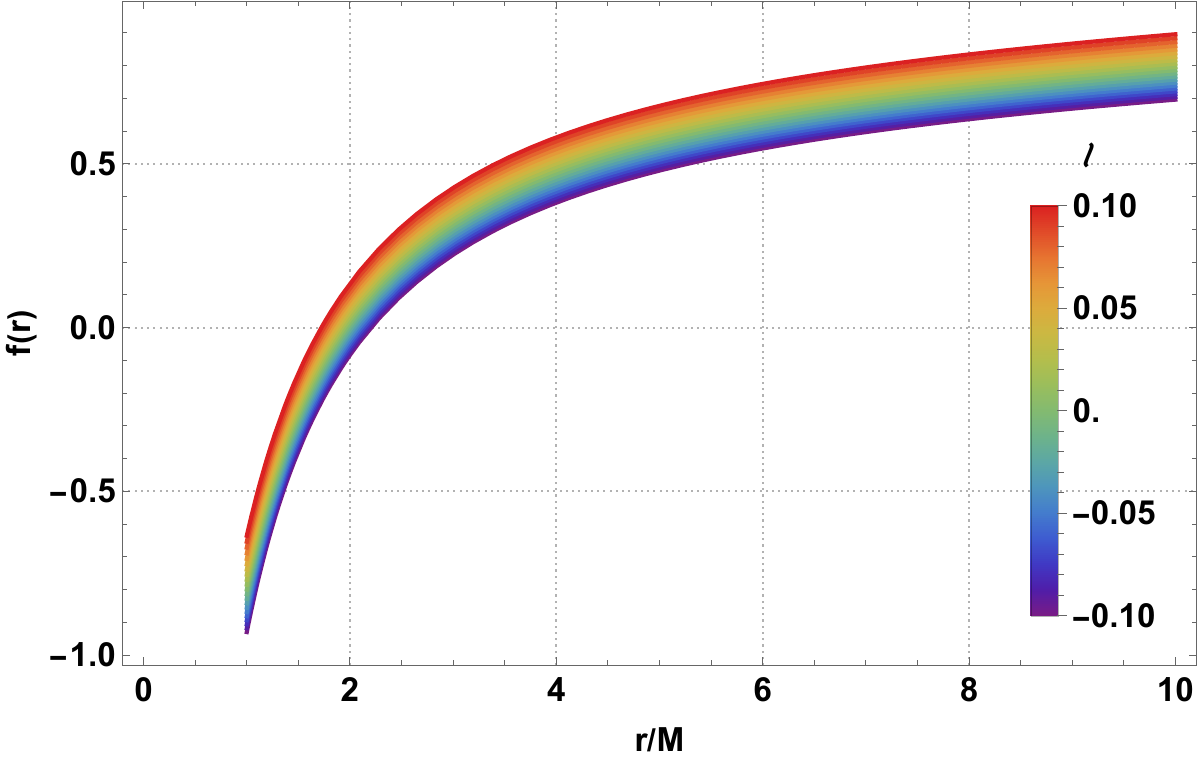}\\
    (i) $\ell=-0.1$ \hspace{7cm} (ii) $\alpha=0.05$
    \caption{ Behavior of the metric function $f(r)$ for a charged BH. Here $Q/M=0.5,\,\,\Lambda=-0.001/M^2$.}
    \label{fig:metric}
\end{figure}

\begin{figure}[ht!]
    \centering
    \includegraphics[width=0.45\linewidth]{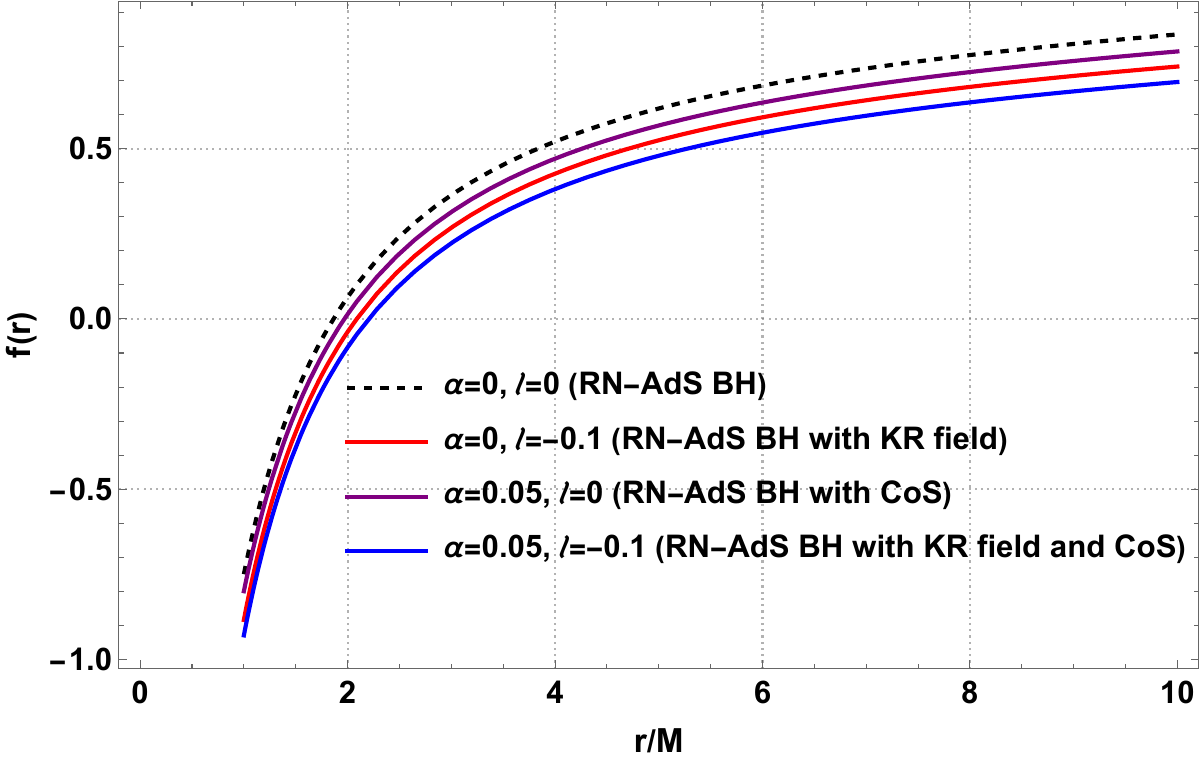}
    \caption{ A comparison of the metric function $f(r)$ for charged BHs in different configurations. Here $Q/M=0.5,\,\,\Lambda=-0.001/M^2$.}
    \label{fig:metric-comp}
\end{figure}

From Figure~\ref{fig:metric}, we observe that as the value of $\alpha$ increases, the horizon of BH increases. In contrast, increasing $\ell$ reduces the horizon radius. Figure~\ref{fig:metric-comp} shows a comparison of the lapse function with and without KR field and string cloud.

The study of curvature invariants, including the Ricci scalar, the quadratic Ricci tensor, and the Kretschmann scalar, provides important information about the properties of space-time geometric and physical properties such as stability and overall structure \cite{Chandrasekhar}. Therefore, in this analysis, we focus on evaluating the curvature invariants to explore characteristics of the space-time geometry of an electrically charged BH with a CoS in KR-gravity in detail.

\begin{figure}[ht!]
    \centering
    \includegraphics[width=0.45\linewidth]{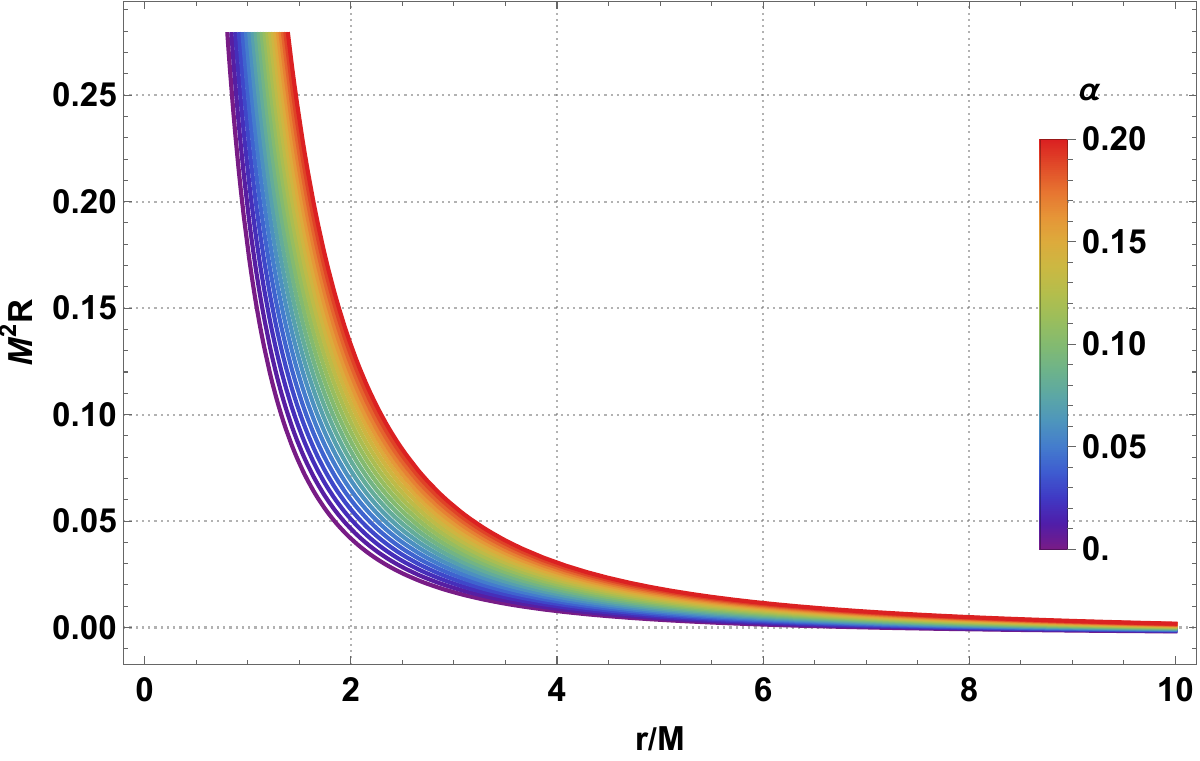}\qquad
    \includegraphics[width=0.45\linewidth]{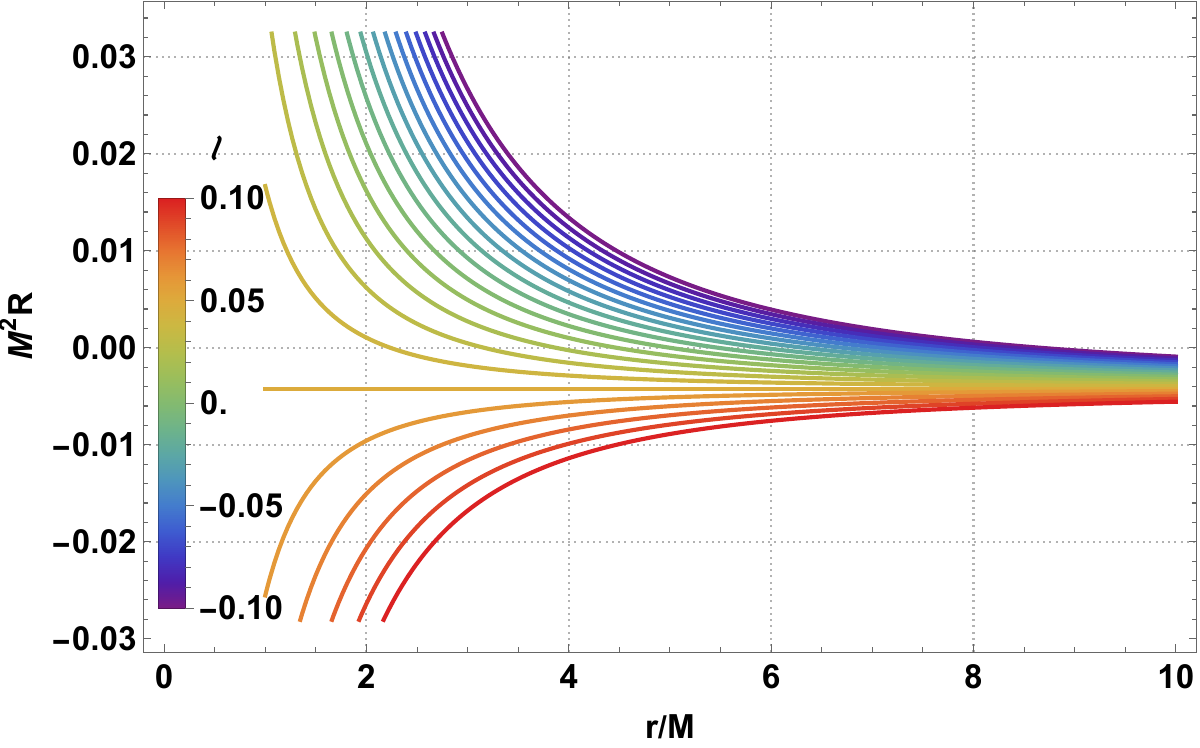}\\
    (i) $\ell=-0.1$ \hspace{7cm} (ii) $\alpha=0.05$
    \caption{The radial profiles of the Ricci scalar for various values of KR field parameter $\ell$ and CoS parameter $\alpha$ with $\Lambda=-0.001/M^2$.}
    \label{fig:Ricci-scalar}
\end{figure}

\begin{itemize}
    \item \textbf{The Ricci scalar:} The mathematical expression for the Ricci scalar $R$ using the space-time~\eqref{metric} is given by
    \begin{equation}
        R=g_{\mu\nu}\,R^{\mu\nu}=\frac{2\,(\alpha-\ell+2\,r^2\,\Lambda)}{r^2\,(1-\ell)}.\label{aaa1}
    \end{equation}
    In the limiting case $\alpha = \ell > 0$, corresponding to equal contributions from the KR field and string clouds, the Ricci scalar simplifies to $R=\frac{4\Lambda}{1-\ell}$.
    
\begin{figure}[ht!]
    \centering
    \includegraphics[width=0.45\linewidth]{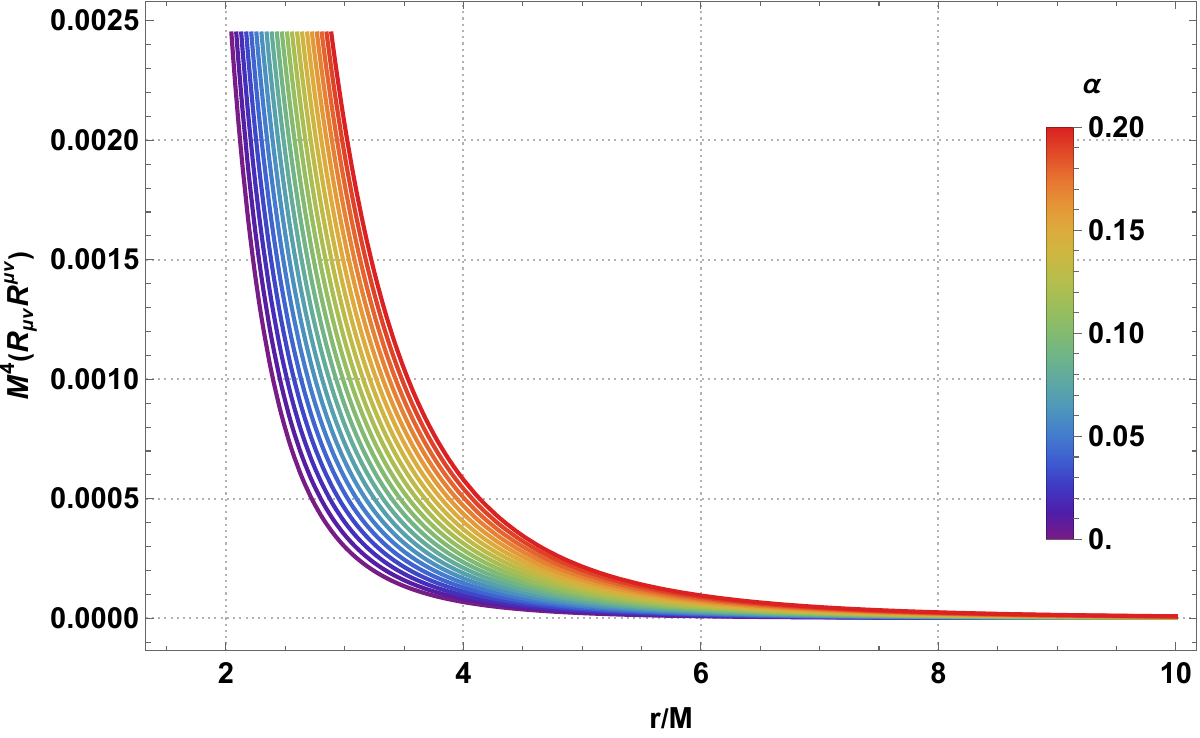}\qquad
    \includegraphics[width=0.45\linewidth]{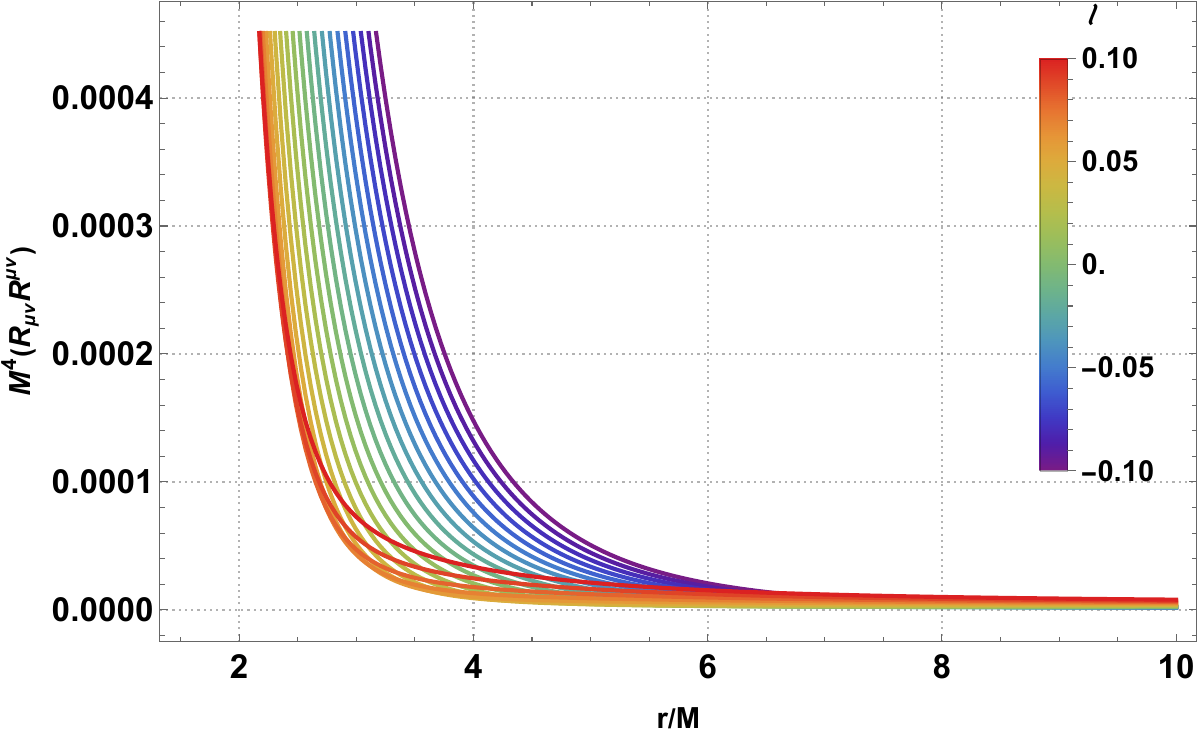}\\
    (i) $\ell=-0.1$ \hspace{6cm} (ii) $\alpha=0.05$
    \caption{The radial profiles of the squared Ricci tensor for various values of KR field parameter $\ell$ and CoS parameter $\alpha$. Here, $Q/M=0.5,\,\,\Lambda=-0.001/M^2$.}
    \label{fig:Quadratic-Ricci-tensor}
\end{figure}

\begin{figure}[ht!]
    \centering
    \includegraphics[width=0.45\linewidth]{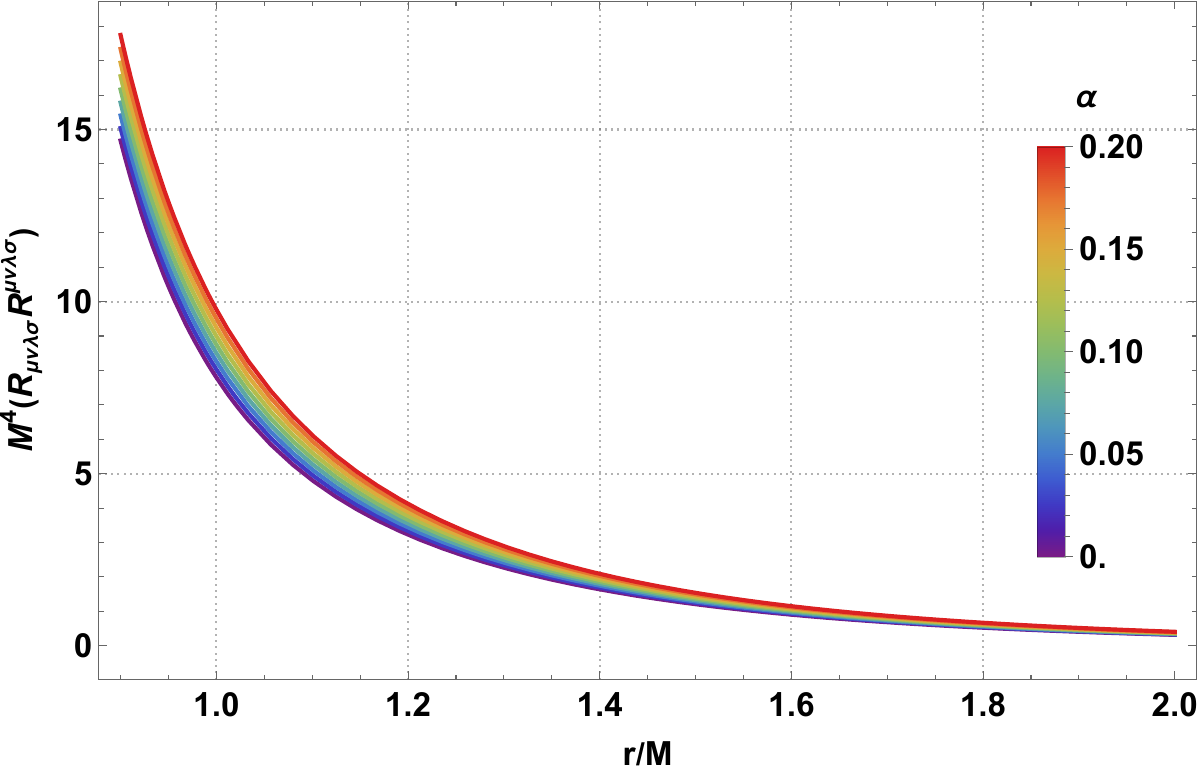}\quad\quad
    \includegraphics[width=0.45\linewidth]{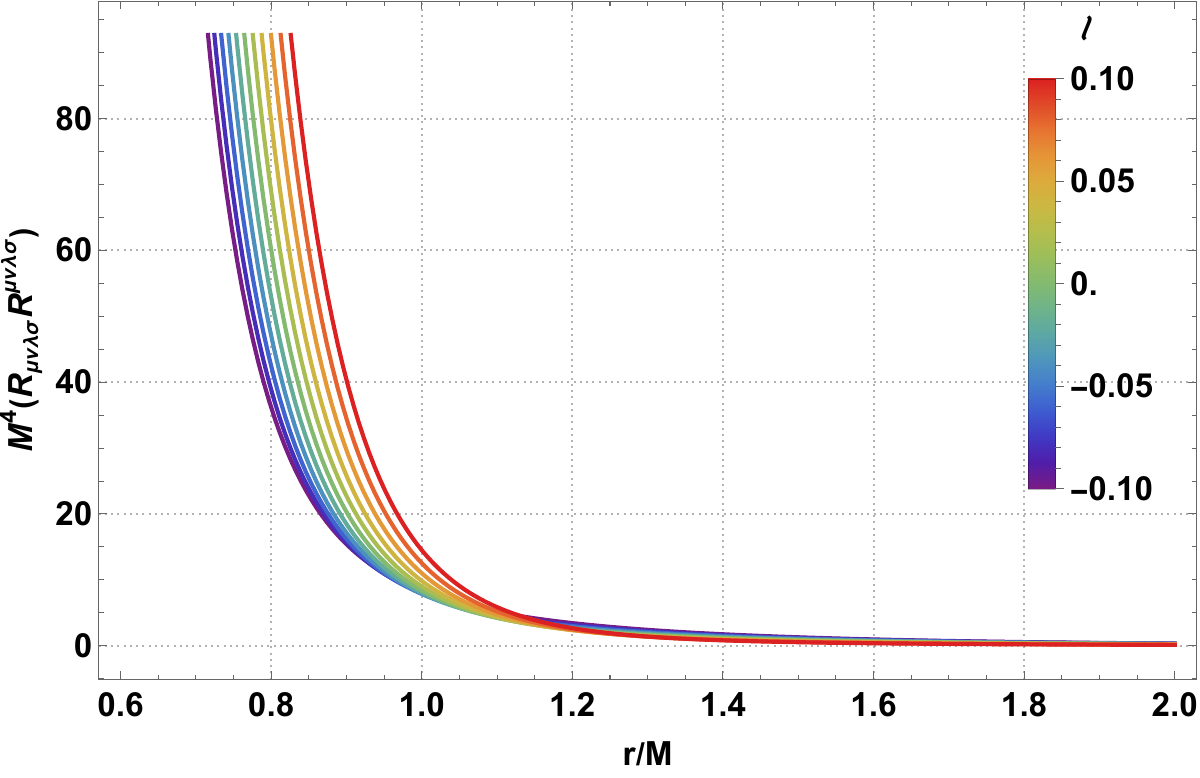}\\
    (i) $\ell=-0.1$ \hspace{6cm} (ii) $\alpha=0.05$
    \caption{The radial profiles of the Kretschmann scalar for various values of KR field parameter $\ell$ and CoS parameter $\alpha$. Here, $Q/M=1,\,\,\Lambda=-0.001/M^2$.}
    \label{fig:kretschmann-scalar}
\end{figure}
    
    \item \textbf{The quadratic Ricci tensor:} The square of the Ricci tensor $R_{\mu\nu}$ can be expressed as
    \begin{equation}
        R_{\mu\nu}\,R^{\mu\nu}=
        \begin{cases}
        \displaystyle 2\,\frac{r^8\, \Lambda^2 + r^4\, (\alpha - \ell + r^2\, \Lambda)^2}{r^8\, (-1 + \ell)^2}, & \text{for } Q=0\\[12pt]
        \displaystyle 2\,\frac{[Q^2 + r^4\, (-1 + \ell)\,\Lambda]^2 +[Q^2 - 
     r^2\, (-1 + \ell)\,(\alpha - \ell+ r^2\, \Lambda)]^2}{r^8\,(-1 + \ell)^4}, & \text{for } Q \neq 0.
        \end{cases}
        \label{aaa2}
    \end{equation}
    In the limiting case $\alpha = \ell > 0$, corresponding to equal contributions from the KR field and string clouds, the squared Ricci tensor simplifies to
\begin{equation}
R_{\mu\nu} R^{\mu\nu} =
\begin{cases}
\dfrac{4 \Lambda^2}{(1-\ell)^2}, & Q = 0,\\[1.0em]
\dfrac{4 \Lambda^2}{(1-\ell)^2} + \dfrac{4 Q^4}{r^8 (1-\ell)^4}, & Q \neq 0.
\end{cases}
\label{eq:aaa2a}
\end{equation}

    \item \textbf{The Kretschmann scalar:} The exact analytical expression for the Kretschmann scalar is given by {\small
\begin{align}
R_{\mu\nu\lambda\sigma}\,R^{\mu\nu\lambda\sigma} &=
\begin{cases}
\displaystyle
\frac{4}{3\,r^8\,(-1 + \ell)^4} \left[
  r^2\,(-1+\ell)^2\,\left\{
    \begin{aligned}
      &3\,r^2(\alpha-\ell)^2 -12\,M\,r\,(\alpha-\ell)\,(-1+\ell)\\
      &+36\,M^2\,(-1+\ell)^2 + 2\,r^4\,(\alpha-\ell)\,\Lambda + 2\,r^6\,\Lambda^2
    \end{aligned}
  \right\}
\right], & \text{for } Q=0, \\[15pt]
\displaystyle
\frac{4}{3\,r^8\,(-1 + \ell)^4} \Bigg[
  42\,Q^2 - 6\,Q^2\,r\,(-1+\ell)\left\{
    12\,M\,(-1+\ell) + r\,(-\alpha+\ell)
  \right\}\\
  + r^2\,(-1+\ell)^2\,\left\{
    \begin{aligned}
      &3\,r^2(\alpha-\ell)^2 -12\,M\,r\,(\alpha-\ell)\,(-1+\ell)\\
      &+36\,M^2\,(-1+\ell)^2 + 2\,r^4\,(\alpha-\ell)\,\Lambda + 2\,r^6\,\Lambda^2
    \end{aligned}
  \right\}
\Bigg], & \text{for } Q \neq 0.
\end{cases}
\label{aaa3}
\end{align}}

In the limiting case $\alpha = \ell > 0$, corresponding to equal contributions from the KR field and string clouds, the Kretschmann scalar simplifies as
\begin{align}
R_{\mu\nu\lambda\sigma} R^{\mu\nu\lambda\sigma} =
\begin{cases}
\dfrac{48 M^2}{r^6} + \dfrac{8 \Lambda^2}{3 (1-\ell)^2}, & Q=0,\\[1.5em]
\dfrac{48 M^2}{r^6} + \dfrac{8 \Lambda^2}{3 (1-\ell)^2} + \dfrac{56 Q^2}{r^8 (1-\ell)^4} - \dfrac{96 Q^2 M}{r^7 (1-\ell)^2}, & Q \neq 0.
\end{cases}
\label{eq:aaa3a}
\end{align}

\end{itemize}

One can study the effects of the KR field ($\ell$), electric charge ($Q$), cosmological constant ($\Lambda$), and the CoS parameter ($\alpha$) on space-time curvature by analyzing key invariants such as the Ricci scalar $R$, the quadratic Ricci tensor $R_{\mu\nu} R^{\mu\nu}$, and the Kretschmann scalar $R_{\mu\nu\rho\sigma} R^{\mu\nu\rho\sigma}$. Each parameter contributes uniquely: $\ell$ modifies curvature via KR field, $Q$ enhances curvature near the origin, $\Lambda$ determines the asymptotic structure, and $\alpha$ affects the local curvature profile. Their interplay shows how these physical quantities collectively shape the geometric behavior of space-time.

In Figure~\ref{fig:Ricci-scalar}, we present the Ricci scalar $R$ for various values of the KR field parameter and the CoS parameter, considering two cases: $Q = 0.5$ and $Q = 1$, while keeping the BH mass fixed at $M = 1$ and the cosmological constant at $\Lambda = -0.01$.

Figure~\ref{fig:Quadratic-Ricci-tensor} shows the behavior of the quadratic Ricci tensor invariant $R_{\mu\nu} R^{\mu\nu}$ under the same parameter settings.

Similarly, Figure~\ref{fig:kretschmann-scalar} illustrates the Kretschmann scalar $R_{\mu\nu\lambda\sigma} R^{\mu\nu\lambda\sigma}$ for varying KR field and CoS parameters with $Q = 0.5$ and $Q = 1$, again with $M = 1$ and $\Lambda = -0.01$.

We now examine the asymptotic behaviour of the lapse function at large radial distances. In the limit $r \to \infty$, the metric function reduces to
\begin{equation}
\lim_{r \to \infty} f(r) = \frac{1-\alpha}{1-\ell} - \frac{\Lambda_{\rm eff}}{3}\, r^2 .
\end{equation}
The presence of the constant prefactor $(1-\alpha)/(1-\ell) \neq 1$ implies that the spacetime does not asymptotically approach the standard anti-de Sitter background. Instead, the asymptotic structure is modified by a deficit (or excess) factor induced by the combined effects of the string cloud and the Kalb-Ramond (KR) field. Interestingly, when the parameters satisfy $\alpha = \ell>0$, corresponding to equal contributions from the string cloud and the KR field, the constant prefactor becomes unity. In this special case, the lapse function asymptotically behaves as
\begin{equation}
\lim_{r \to \infty} f(r) \simeq 1 - \frac{\Lambda_{\rm eff}}{3}\, r^2 ,
\end{equation}
which coincides with the standard AdS spacetime characterized by the effective cosmological constant $\Lambda_{\rm eff}$. Therefore, the condition $\alpha = \ell$ restores the asymptotically AdS behavior, whereas deviations from this equality lead to a modified non-AdS asymptotic geometry. It is worth noting that the space-time discussed in \cite{ref3} also behaves the same characteristic.

}

\section{Geometric Properties }

In general relativity, null geodesics describe the paths followed by massless particles such as photons. These trajectories are essential for understanding the optical appearance of BHs, particularly features such as the photon sphere and the BH shadow. The photon sphere is a region where photons can travel in unstable circular orbits due to the extreme curvature of spacetime \cite{Chandrasekhar,MTW}. Although these orbits are not stable, they play a central role in the formation of the BH shadow, which is observed as a dark region against a bright background when photons are absorbed by the event horizon. The boundary of the shadow corresponds to the apparent position of the photon sphere as seen by a distant observer. The study of null geodesics is not only important in theory but also in practice, as it underpins the interpretation of high-resolution observations of BH shadows.

This theory has found compelling observational relevance with the EHT, which successfully imaged the shadow of the supermassive BH in the galaxy M87* \cite{EHTL1,EHTL4,EHTL6,EHTL12}. Modeling these shadows requires precise understanding of photon trajectories and the influence of BH parameters such as mass, spin, and observer inclination \cite{FalckeL13}. The detailed study of null geodesics, especially those asymptotically approaching unstable circular orbits, is crucial in modeling these observations and inferring BH parameters such as spin and inclination angle.

\begin{center}
    {\bf I.\,Effective Potential of Null Geodesic Motions}
\end{center}

As the selected space-time is static and spherically symmetric, we can, without loss of generality, restrict our analysis of geodesic motion to the equatorial plane, defined by \(\theta = \frac{\pi}{2}\) and \(\dot{\theta} = 0\). Utilizing the Lagrangian formalism, where the Lagrangian density is given by $\mathcal{L}=\frac{1}{2}\,g_{\mu\nu}\,\dot{x}^{\mu}\,\dot{x}^{\nu}$, with the dot denoting differentiation with respect to an affine parameter, we can express the Lagrangian for the metric given in Eq.~(\ref{metric}) as,
\begin{equation}
    \mathcal{L}=\frac{1}{2}\,\left[-f(r)\,\dot{t}^2+\frac{\dot{r}^2}{f(r)}+r^2\,\dot{\phi}^2  \right].\label{b1}
\end{equation}
There are two conserved quantities associated with the temporal coordinate $t$ and the azimuthal coordinate $\phi$. These are given by
\begin{align}
    \mathrm{E}&=f(r)\,\dot{t},\label{b3}\\
    \mathrm{L}&=r^2\,\dot{\phi}.\label{b4}
\end{align}

\begin{figure}[ht!]
    \centering
    \includegraphics[width=0.4\linewidth]{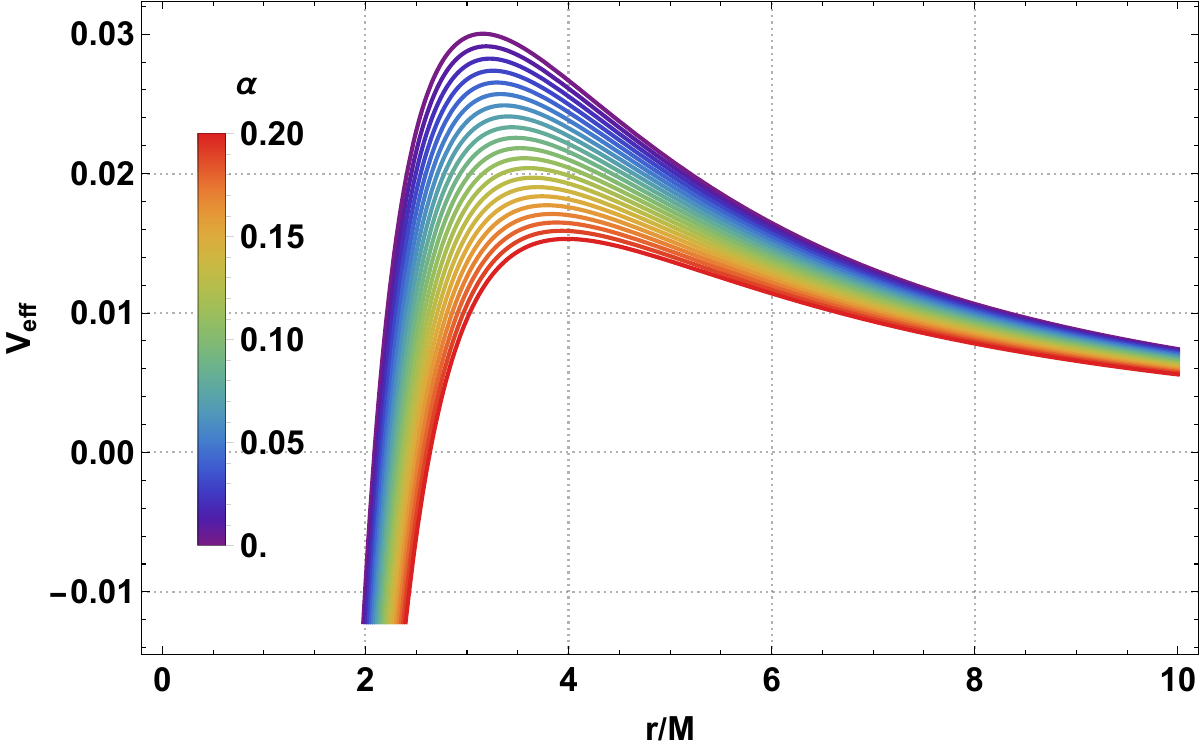}\qquad
    \includegraphics[width=0.4\linewidth]{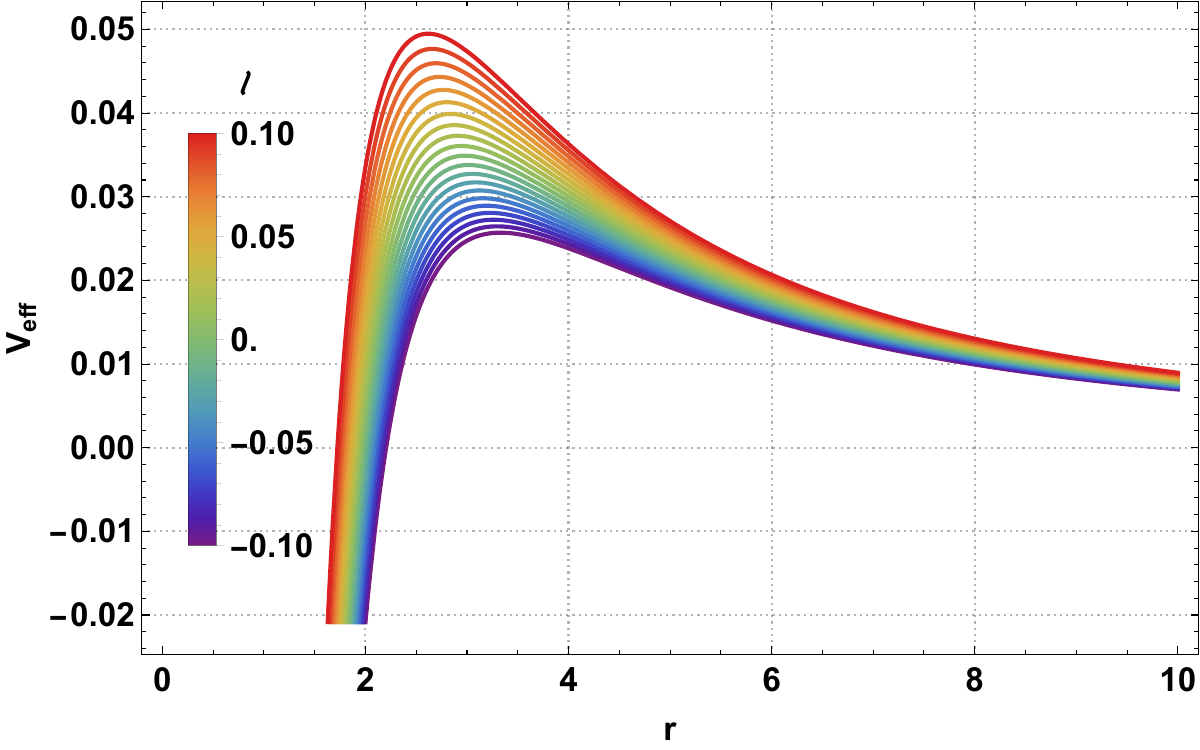}\\
    (i) $\ell=-0.1$ \hspace{6cm} (ii) $\alpha=0.05$
    \caption{Behavior of the effective potential $V_\text{eff}(r)$ as a function of $r/M$ for different values of KR field parameter $\ell$ and CoS parameter $\alpha$. Here $Q/M=0.5,\,\,\mathrm{L}/M=1,\,\,\Lambda=-0.001/M^2$.}
    \label{fig:potential-null}
\end{figure}

\begin{figure}[ht!]
    \centering
    \includegraphics[width=0.4\linewidth]{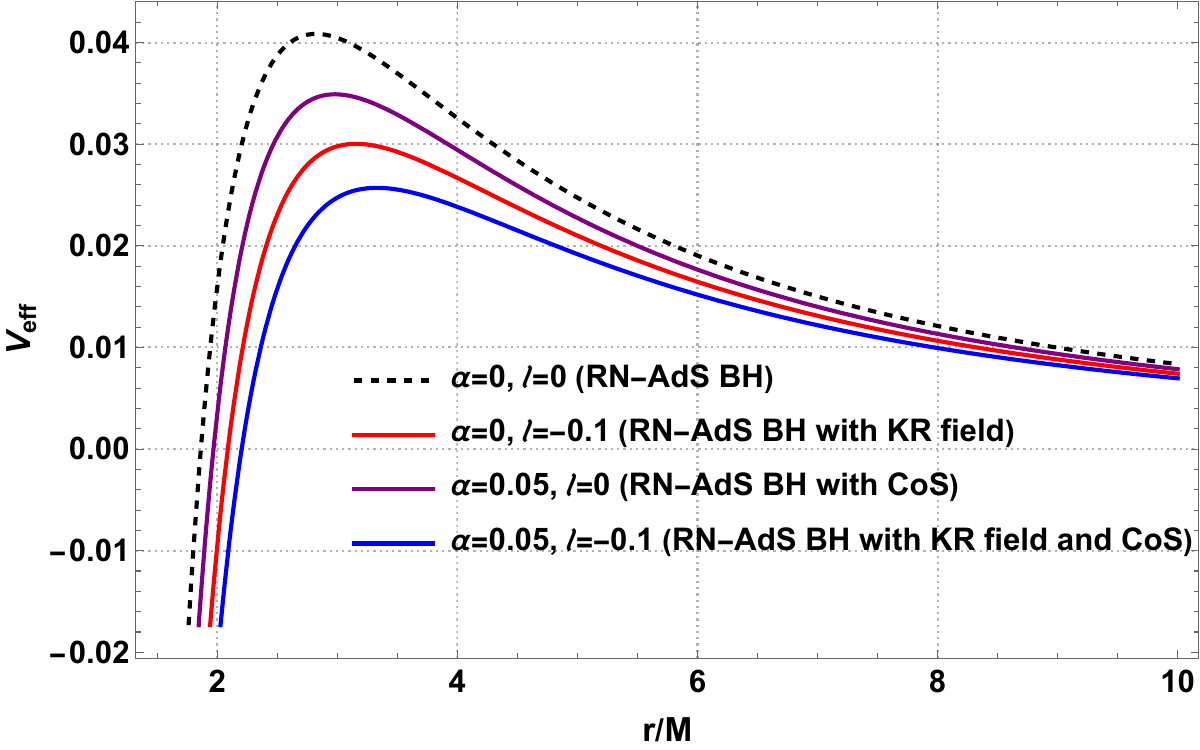}
    \caption{A comparison of the effective potential $V_\text{eff}(r)$ of the null geodesics as a function of $r/M$ for charged BHs in different configurations. Here $Q/M=0.5,\,\,\mathrm{L}/M=1,\,\,\Lambda=-0.001/M^2$.}
    \label{fig:potential-null-comp}
\end{figure}

Eliminating $\dot{t}$ and $\dot{\phi}$ using Eqs. (\ref{b3})--(\ref{b4}) in Eq. (\ref{b1}) for null geodesics, we find 
\begin{equation}
    \dot{r}^2+V_\text{eff}(r)=\mathrm{E}^2,\label{b5}
\end{equation}
where $V_\text{eff}(r)$ is the effective potential of the system and is given by
\begin{align}
V_\text{eff}(r)=\frac{\mathrm{L}^2}{r^2}\,f(r)=
\begin{cases}
\displaystyle \frac{\mathrm{L}^2}{r^2}\,\left(\frac{1 - \alpha}{1 - \ell} - \frac{2\,M}{r} -\frac{\Lambda}{3\,(1-\ell)} \, r^2\right), & \text{for } Q = 0, \\
\displaystyle \frac{\mathrm{L}^2}{r^2}\,\left(\frac{1 - \alpha}{1 - \ell} - \frac{2\,M}{r} + \frac{Q^2}{r^2\,(1-\ell)^2} -\frac{\Lambda}{3\,(1-\ell)} \, r^2\right), & \text{for } Q \neq 0.
\end{cases}.\label{b6}
\end{align}

From expression (\ref{b6}), it becomes evident that the effective potential for null geodesics is influenced by key parameters. These include LSB parameter \(\ell\), string parameter \(\alpha\), the electric charge \(Q\), and the BH mass \(M\). These parameters impact the effective potential governing the dynamics of photon particles in the gravitational field of the chosen BH solution. 

In Figure \ref{fig:potential-null}, we depict the effective potential governs the dynamics of photon particles by varying the string parameter $\alpha$ and the KR field parameter $\ell$ for two different values of the electric charge $Q=0.5$ and $Q=1$. In panels (i)-(ii), we observed that as the value of \(\alpha\) rises, the effective potential decreases. In contrast, in panels (iii)-(iv) this potential rises with increasing the value of $\ell$. {\color{black} Figure \ref{fig:potential-null-comp} shows a comparison of the effective potential governs the photon dynamics for charged BH in different configurations: with and without KR field and string clouds.}

\begin{center}
    {\bf II.\,Effective Radial Force and Photon Trajectories}
\end{center}

Photons, being massless particles of light, are affected by the curvature of spacetime. This effect can be described as an effective radial force that alters their path. For instance, when photons pass near a BH, their trajectory bends-a phenomenon known as gravitational lensing. This bending can be interpreted as the result of an effective inward/outward radial force, though it's actually due to the geometry of spacetime rather than a force in the classical sense. In circular photon orbits around BHs (like the photon sphere), this force balances the photon’s momentum, keeping it on a curved path. Thus, the concept of effective radial force for photons is a relativistic effect arising from spacetime curvature rather than an actual physical force.

The effective radial force experience by photons is defined in terms of the effective potential of null geodesics as, $\mathrm{F}_\text{eff}(r)=-\frac{1}{2}\,\frac{dV_\text{eff}}{dr}$. Using the effective potential given in Eq. (\ref{b6}), we find this effective radial force as,
\begin{equation}
    \mathrm{F}_\text{eff}(r)=
    \begin{cases}
    \displaystyle \frac{\mathrm{L}^2}{r^3}\,\left(\frac{1-\alpha}{1-\ell}-\frac{3\,M}{r}\right), & \text{for } Q=0\\
    \displaystyle \frac{\mathrm{L}^2}{r^3}\,\left(\frac{1-\alpha}{1-\ell}-\frac{3\,M}{r}+\frac{2\,Q^2}{r^2\,(1-\ell)^2}\right), & \text{for } Q \neq 0.
    \end{cases}
    \label{b7}
\end{equation}

\begin{figure}[ht!]
    \centering
    \includegraphics[width=0.45\linewidth]{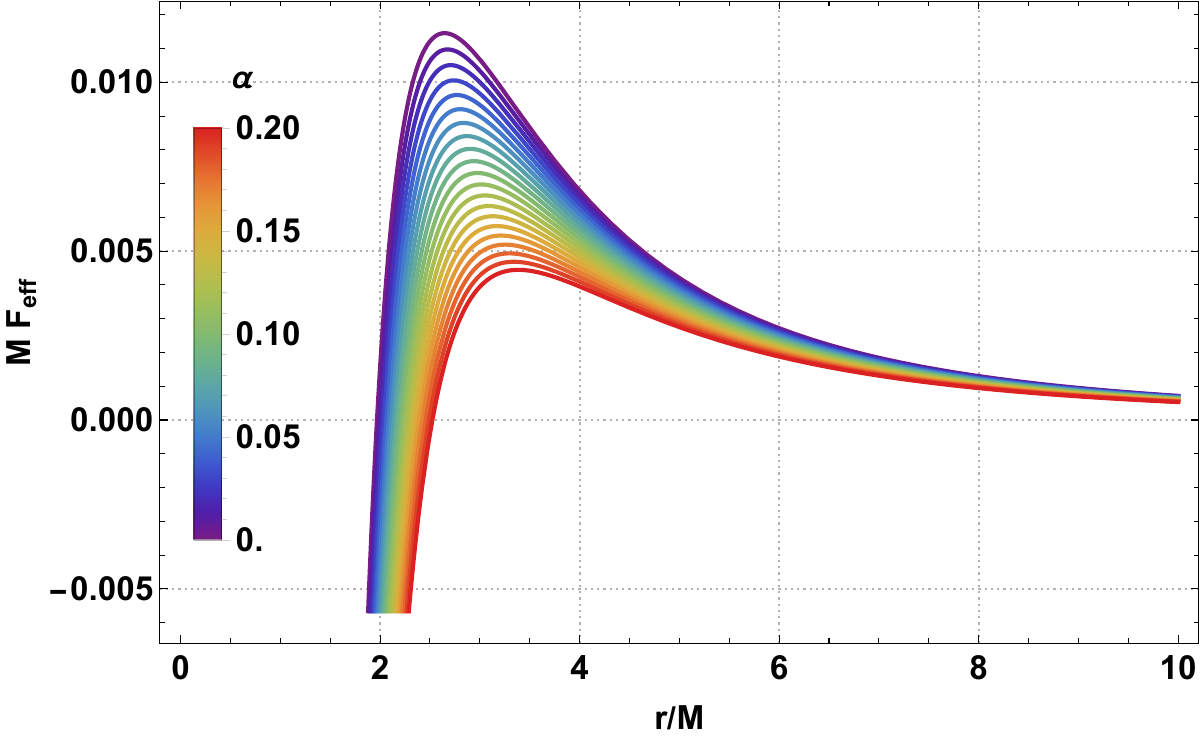}\qquad
    \includegraphics[width=0.45\linewidth]{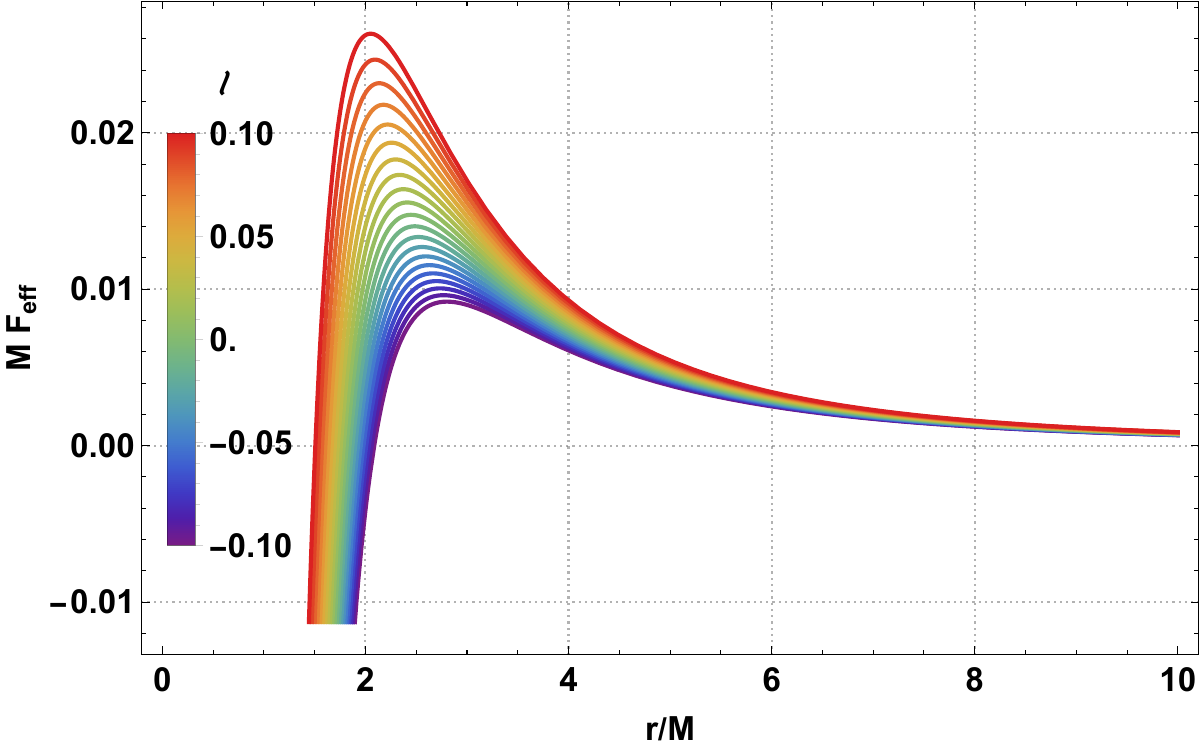}\\
    (i) $\ell=-0.1$ \hspace{6cm} (ii) $\alpha=0.05$
    \caption{Behavior of the effective radial force $F_\text{eff}(r)$ experienced by the photon particles as a function of $r/M$ for different values of KR field parameter $\ell$ and CoS parameter $\alpha$. Here $\mathrm{L}/M=1$ and $\Lambda=-0.001/M^2,\,Q/M=0.5$.}
    \label{fig:force}
\end{figure}

\begin{figure}[ht!]
    \centering
    \includegraphics[width=0.5\linewidth]{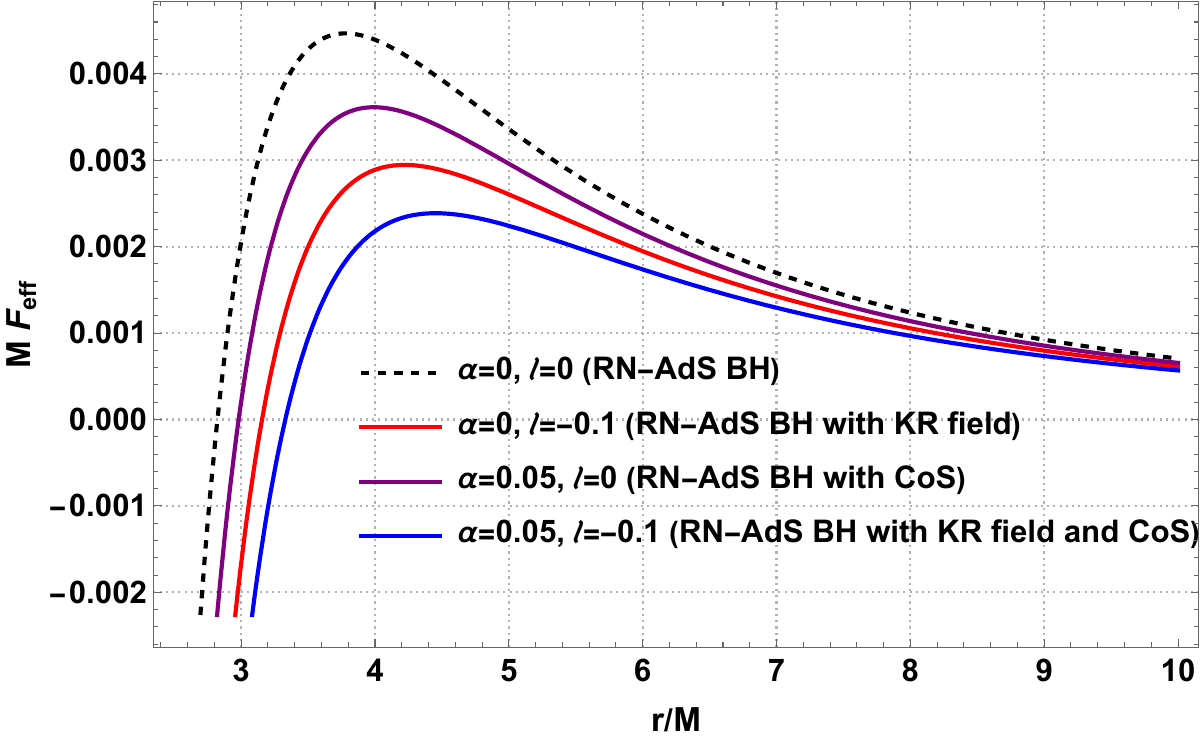}
    \caption{A comparison of the effective radial force $F_\text{eff}(r)$ experienced by the photon particles as a function of $r/M$ for charged BHs in different configurations. Here $\mathrm{L}/M=1$ and $\Lambda=-0.001/M^2,\,Q/M=0.5$.}
    \label{fig:force-comp}
\end{figure}

From expression (\ref{b7}), it becomes evident that the effective radial force on photon particles is influenced by key parameters. These include LSB parameter \(\ell\), string parameter \(\alpha\), the electric charge \(Q\), and the BH mass \(M\). These parameters impact the force which results in modifications from the standard result obtained for the standard charged BH solution.

In Figure \ref{fig:force}, we depict the effective radial force experiences by photons by varying the string parameter $\alpha$ and the KR field parameter $\ell$ for two different values of the electric charge $Q=0.5$ and $Q=1$. In panels (i)-(ii), we observed that as the value of \(\alpha\) rises, the radial force reduces. In contrast, in panels (iii)-(iv), this force increases with increasing the value of $\ell$. {\color{black} Figure \ref{fig:force-comp} shows a comparison of the effective radial force experienced by the photon particles for charged BHs in different configurations: with and without KR field and string clouds.}

Now, we focus into the photon trajectories and show how various parameters involved in the space-time geometry deviate these compared to the Schwarzschild BH solution. The equation of orbit using Eqs. (\ref{b4})--(\ref{b5}) and then (\ref{b6}) is given by 
\begin{equation}
    \left(\frac{1}{r^2}\,\frac{dr}{d\phi}\right)^2+\left(\frac{1-\alpha}{1-\ell}\right)\,\frac{1}{r^2}=
    \begin{cases}
    \displaystyle \frac{1}{\beta^2}+\frac{\Lambda_\text{eff}}{3}+\frac{2\,M}{r^3}, & \text{ for } Q=0\\
    \displaystyle \frac{1}{\beta^2}+\frac{\Lambda_\text{eff}}{3}+\frac{2\,M}{r^3}-\frac{Q^2_\text{eff}}{r^4}. & \text{ for } Q \neq 0\label{b8}
    \end{cases}
\end{equation}
Transforming to an new variable via $r=\frac{1}{u}$ and then differentiating both sides w. r. t. $\phi$ yields the following equation:
\begin{equation}
    \frac{d^2u}{d\phi^2}+\left(\frac{1-\alpha}{1-\ell}\right)\,u=
    \begin{cases}
        \displaystyle 3\,M\,u^2, & \text{ for } Q=0\\
        \displaystyle 3\,M\,u^2-\frac{2\,Q^2}{(1-\ell)^2}\,u^3. & \text{ for } Q \neq 0 \label{b9}
    \end{cases}
\end{equation}

Equation (\ref{b9}) is a non-linear, second-order differential equation representing the photon trajectories in the gravitational field. The photon trajectory is influenced by LV parameter \(\ell\), the string parameter \(\alpha\), the electric charge \(Q\), and the BH mass $M$. {\color{black} In the limiting case $\alpha=\ell> 0$, that is equal contribution for the KR field and string clouds, the photon trajectory equation (\ref{b9}) for uncharged BH reduces to the Schwrazschild case.}

\begin{center}
    {\bf III.\,Photon Sphere and BH Shadow }
\end{center}

Photon sphere is a spherical region around a BH where gravity is strong enough to force photons (light) into unstable circular orbits. Although photons are massless and do not experience force in the Newtonian sense, their motion can be described using an effective potential that arises from the curved spacetime geometry. At the photon sphere, the effective radial force-interpreted from the gradient of this potential-vanishes, indicating a balance between the gravitational pull and the centrifugal effect on light. However, this balance is unstable, so even small disturbances cause photons to either fall into the BH or escape to infinity. On the other hand, the BH shadow, observed for instance by the Event Horizon Telescope, is the dark region in the observer’s sky corresponding to photons that fall into the BH. Its boundary is determined by the critical light rays that asymptotically approach the photon sphere. Thus, the photon sphere defines the apparent size and shape of the BH shadow. Recent research has extended these concepts to more general and dynamical spacetimes. For example, Vertogradov {\it et al.} \cite{Vertogradov2024} showed how additional matter fields can shift the photon sphere radius depending on metric deformations; Paithankar {\it et al.} \cite{Paithankar2023} linked shadow size to acceleration bounds in spherical geometries; and Koga {\it et al.} \cite{Koga2023} analyzed how both the photon sphere and the shadow evolve over time in accreting BH models. These studies emphasize that while the photon sphere and shadow are classical features of static BHs, they are also sensitive probes of more complex, dynamic, or modified gravity environments.

For circular orbits of radius $r=r_c$, the conditions $\dot{r}=0$ and $\ddot{r}=0$ must satisfied. These conditions implies
\begin{equation}
    \mathrm{E}^2=\frac{\mathrm{L}^2}{r^2}\,f(r)\label{b10}
\end{equation}
And
\begin{equation}
    V'_\text{eff}(r)=0.\label{b11}
\end{equation}

Using the effective potential given in Eq.~(\ref{b6}) into the above relation (\ref{b11}), one can determine the photon sphere radius $r=r_s$. We find the following equation:
\begin{equation}
    \frac{1-\alpha}{1-\ell}\,r^2_s-3\,M\,r_s+2\,Q^2_\text{eff}=0.\label{b12}
\end{equation}
Due to the inherent spherical symmetry, the photons can occupy all circular orbits, resulting in the formation of the photon sphere.  Solving the above equation results the photon sphere radius 
\begin{equation}
    r_s=
    \begin{cases}
    \displaystyle 3\,M\,\left(\frac{1-\ell}{1-\alpha}\right), & \text{for } Q=0\\
    \displaystyle \frac{3M}{2}\frac{1-\ell}{1-\alpha}\left[1+\sqrt{1-\frac{8\,Q^2\,(1-\alpha)}{9\,M^2\,(1-\ell)^3}}\right], & \text{for } Q \neq 0.
    \end{cases}
    \label{b13}
\end{equation}

{\color{black} In case of charged BH, the existence of a real photon sphere solution depends on the discriminant $1 -\tfrac{8 Q^2 (1-\alpha)}{9M^2(1-\ell)^3}$ remaining non-negative. The term $\frac{8 Q^2(1-\alpha)}{9M^2(1-\ell)^3}$ quantifies the balance between electromagnetic and gravitational fields effect on null geodesics. The numerator contains the electromagnetic field and string cloud contribution $Q^2 (1-\alpha)$, while the denominator $M^2(1-\ell)^3$ captures gravitational strength modified by the LV parameter. When this ratio approaches unity, the photon sphere transitions to a critical configuration where electromagnetic repulsion nearly counterbalances gravitational attraction. For the photon sphere radius to be real and finite, we require the constraint on the parameters $M > \frac{2Q}{3(1-\ell)}\sqrt{\frac{2 (1-\alpha)}{1-\ell}}$. This generalizes the extremality condition from RN BHs: configurations violating this bound have excessive charge-to-mass ratios that prevent circular photon orbit formation. The LV parameter modifies this threshold through $(1-\ell)^{-3/2}$ scaling and the string cloud through $(1-\alpha)^{1/2}$ scaling, demonstrating how spontaneous Lorentz violation and string cloud affects the photon sphere existence domain.

Few limiting cases of the photon sphere radius show the parameter dependencies:

\noindent When $\alpha \to 0$, it recovers the result of the (un)-charged AdS BH in KR-gravity {\it i. e.,}
\begin{equation}
r_s =
\begin{cases}
\displaystyle 3M(1-\ell), & \text{for } Q = 0, \\[0.8em]
\displaystyle \frac{3M}{2}(1-\ell)\!\left(1 + \sqrt{1 - \frac{8Q^{2}}{9M^{2}(1-\ell)^{3}}}\right), & \text{for } Q \neq 0,
\end{cases}
\label{b13a}
\end{equation}

\noindent For $\ell \to 0$, one can recover the results of Letelier-AdS BH and/or RN-AdS BH with a string cloud in Einstein gravity, {\it i.e.,}
\begin{equation}
    r_s=
    \begin{cases}
    \displaystyle \frac{3M}{1-\alpha}, & \text{for } Q=0\\
    \displaystyle \frac{3M}{2 (1-\alpha)}\left[1+\sqrt{1-\frac{8 Q^2 (1-\alpha)}{9 M^2}}\right], & \text{for } Q \neq 0.
    \end{cases}
    \label{b13b}
\end{equation}
which further reduces to the result of the RN-(A)dS BH \cite{Eiroa2002} provided $\alpha \to 0$.

\noindent Finally, for $\alpha=\ell>0$, that is equal contribution from the KR-field and string cloud effects, we find the photon sphere radius as,
\begin{equation}
    r_s=
    \begin{cases}
    \displaystyle 3M, & \text{for } Q=0\\
    \displaystyle \frac{3M}{2}\left[1+\sqrt{1-\frac{8 Q^2}{9 M^2 (1-\ell)^2}}\right], & \text{for } Q \neq 0.
    \end{cases}
    \label{b13c}
\end{equation}
From the above equation (\ref{b13c}), we observe that when the contributions from the KR field and the string cloud effects are equal, the photon sphere, and consequently, the shadow radius of the uncharged BH coincides with the corresponding result for the Schwarzschild case.
}

Next, we focus into the BH shadows cast by the BH solution, and examine how the geometric and physical parameters influence the size of the shadow. The shadow size $ R_s $ of a spherically symmetric BH (as seen by a distant observer) is determined by the photon sphere radius $ r_s$ and is typically expressed using the critical impact parameter $ \beta_c $. {\color{black} The EHT observations of the BH shadows of M87* and Sagittarius A* offer an unprecedented opportunity to test gravitational theories and fundamental physics in the strong-field regime \cite{EHTL1,EHTL4,EHTL6,EHTL12}. In this context, exploring how Lorentz violation and string cloud affects BH shadows enables us to use observational data to place constraints on the possible magnitude of such violations.

The BH shadow radius as measured by an observer located at position $r_{O}$ is given by \cite{Perlick2022,Zhang2020}:
\begin{equation}
R_s=\beta_c \sqrt{f(r_{O})}=r_s \sqrt{\frac{f(r_O)}{f(r_s)}}.\label{b14} 
\end{equation}

In our case, the shadow size is given by
\begin{equation}
R_s=
\begin{cases}
\displaystyle r_s\sqrt{\frac{\frac{1-\alpha}{1-\ell} - \frac{2\,M}{r_O}- \frac{\Lambda}{3\,(1-\ell)}\, r^2_O}{\frac{1-\alpha}{1-\ell} - \frac{2\,M}{r_s}- \frac{\Lambda}{3\,(1-\ell)}\, r^2_s}}, & \text{for } Q=0,\\
\hfill
\displaystyle r_s \sqrt{\frac{\frac{1-\alpha}{1-\ell} - \frac{2\,M}{r_O}+\frac{Q^2}{(1-\ell)^2\,r^2_O} - \frac{\Lambda}{3\,(1-\ell)}\, r^2_O}{\frac{1-\alpha}{1-\ell} - \frac{2\,M}{r_s}+\frac{Q^2}{(1-\ell)^2\,r^2_s} - \frac{\Lambda}{3\,(1-\ell)}\, r^2_s}}, & \text{for } Q \neq 0.
\end{cases}\label{b15}
\end{equation}

By inserting the photon-sphere radius $r_s$ from Eq. (\ref{b13}) into Eq. \eqref{b15}, one obtains an explicit expression for the shadow radius of the chosen BH.} From this equation, it becomes evident that the shadow size $R_s$ is influenced by several geometric and physical parameters. These include LSB parameter \(\ell\), the string cloud parameter $\alpha$, the cosmological constant $\Lambda$, and the BH mass $M$. Moreover, for charged BH case, this shadow radius is affect by the electric charge $Q$.  {\color{black} In Figure \ref{fig:photon-sphere}, we illustrate the combined effects of KR-field parameter $\ell$ and string cloud parameter $\alpha$ on the photon sphere radius ($r_s$) and shadow radius ($R_s$) for charged BHs for an observer located at $r_O=20 M$. In Figures \ref{fig:shadow-radius}, we plot the shadow rings for uncharged/charged BH by varying the KR-field parameter $\ell$ (left panel) and the string cloud parameter $\alpha$ (right panel) for an observer located at $r_O=20M$ with cosmological constant $\Lambda=-0.001/M^2$. We observe that the presence of electric charge in BH solution reduces the shadow rings.

\begin{figure}[ht!]
    \centering
    \includegraphics[width=0.45\linewidth]{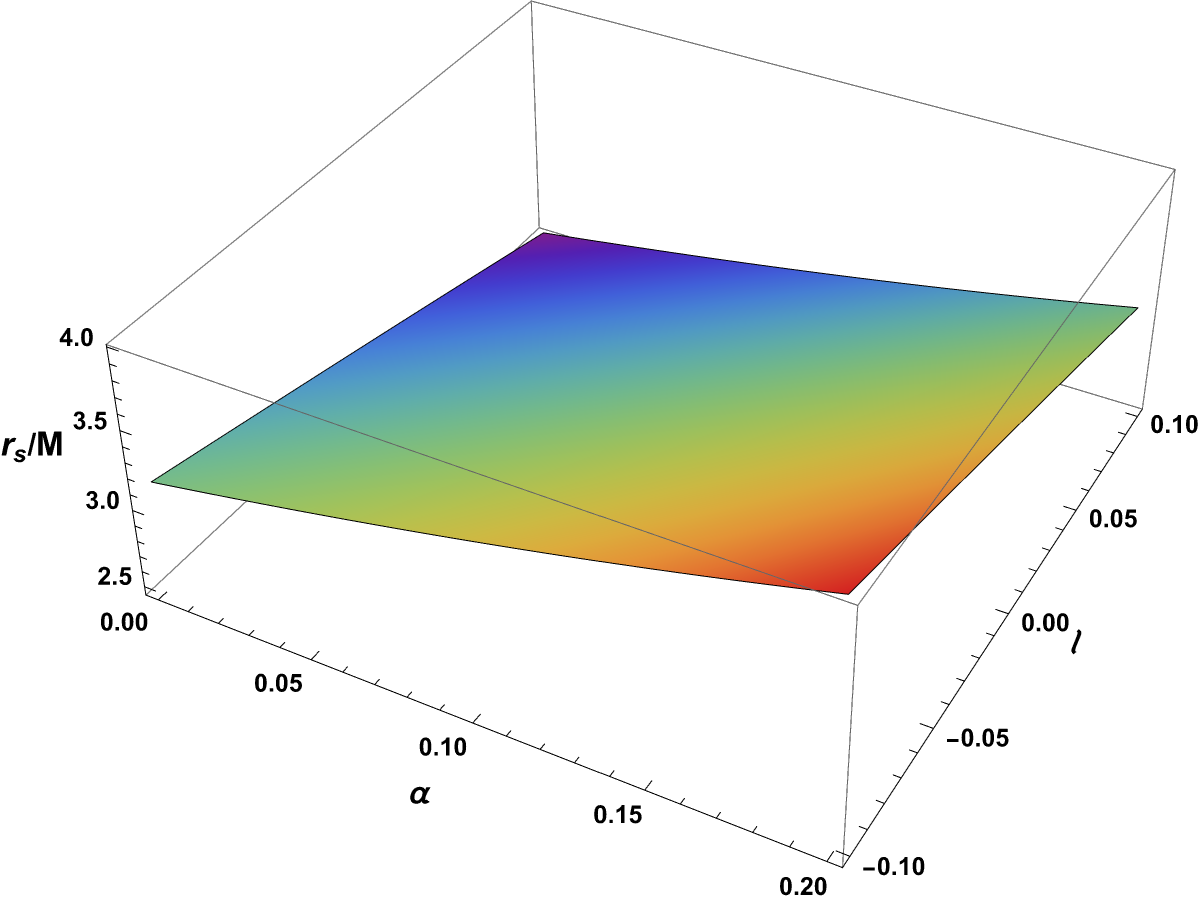}\qquad
    \includegraphics[width=0.45\linewidth]{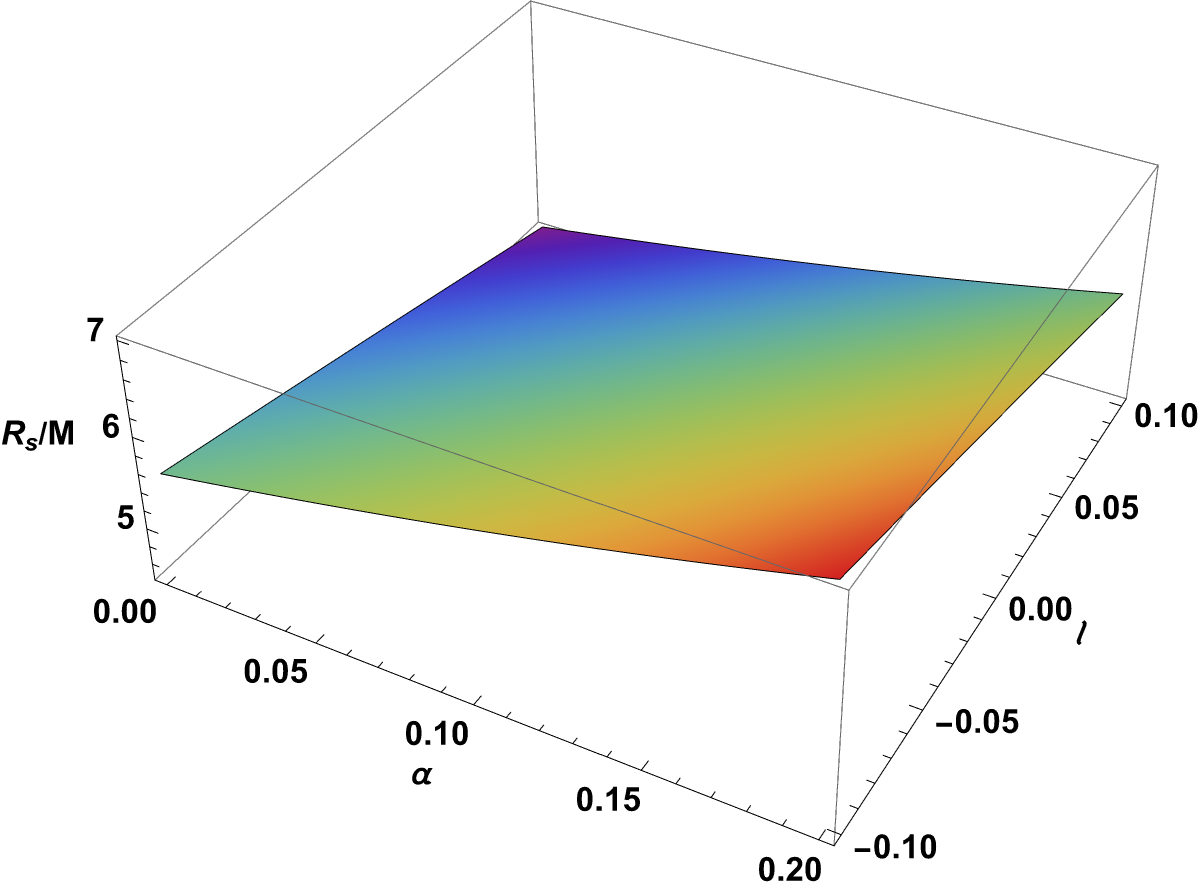}
    \caption{Photon sphere radius $r_s$ and shadow radius $R_s$ by varying KR field parameter $\ell$ and string cloud parameter $\alpha$. Here, $Q/M=0.5,\,\,r_O/M=20,\,\,\Lambda=-0.001/M^2$.}
    \label{fig:photon-sphere}
\end{figure}

\begin{figure}[ht!]
    \centering
    \includegraphics[width=0.45\linewidth]{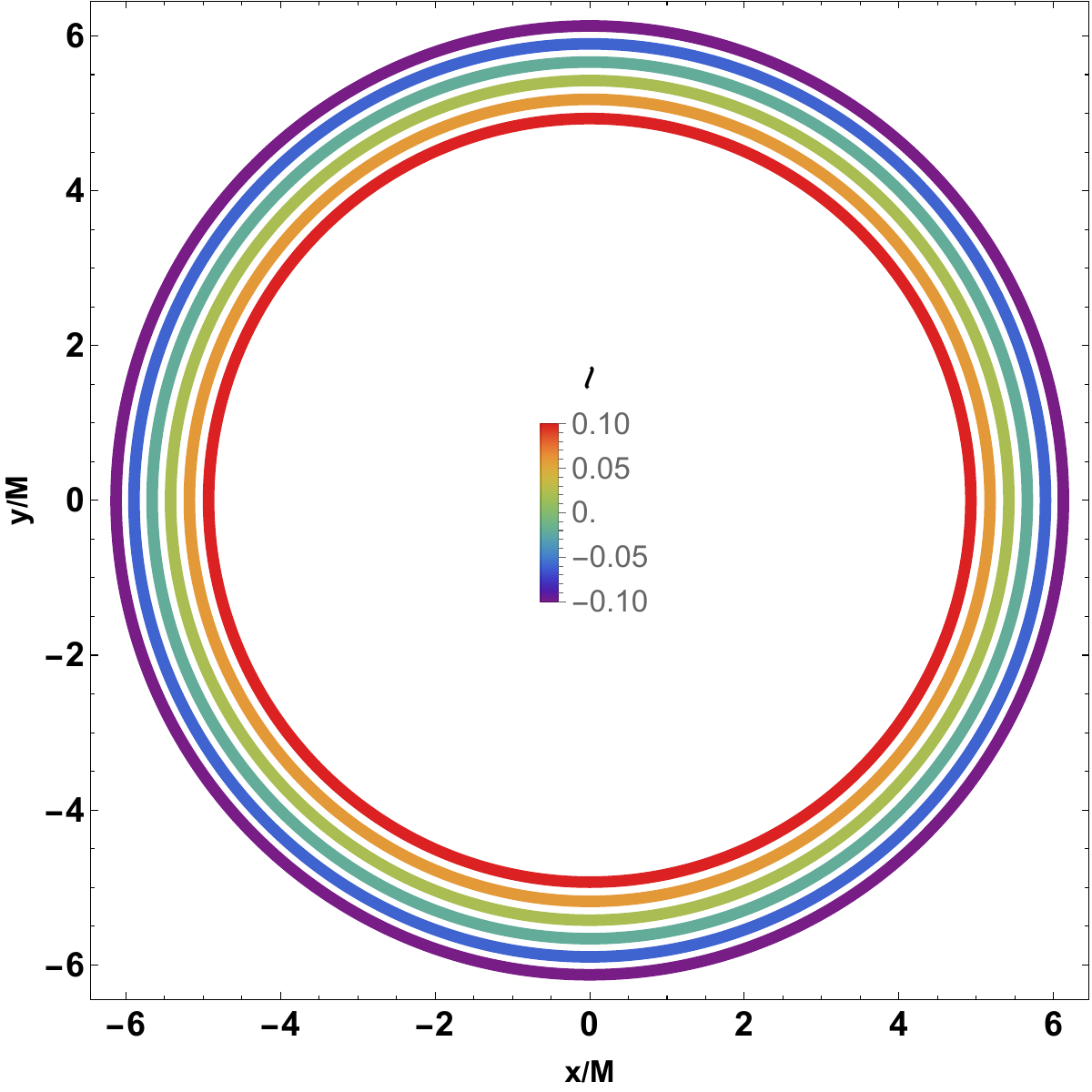}\qquad
    \includegraphics[width=0.45\linewidth]{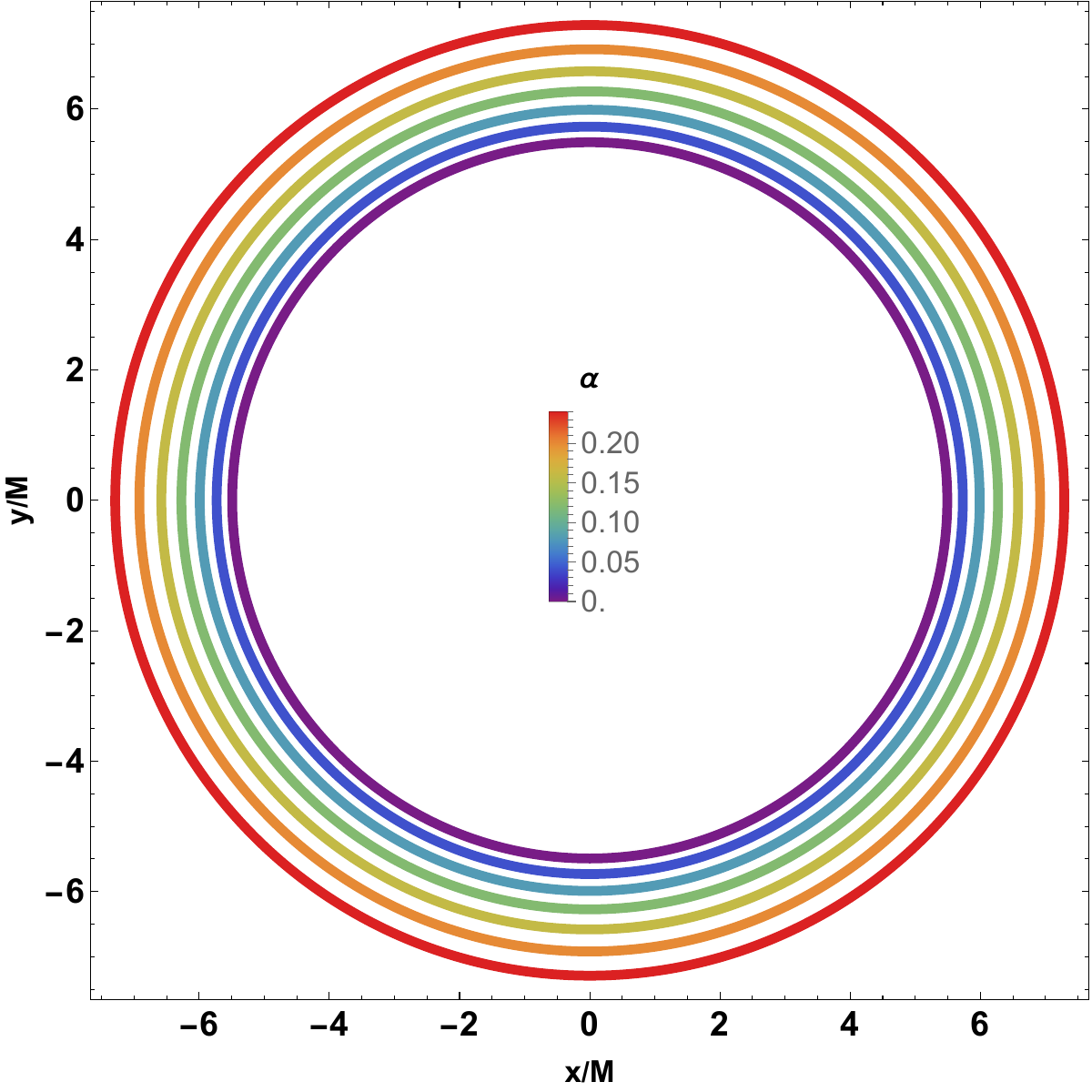}\\
    (i) $\alpha=0.1$ \hspace{6cm} (ii) $\ell=-0.1$
    \caption{Shadows $R_s\left(=\sqrt{x^2+y^2}\right)$ for charged BH by varying the KR field parameter $\ell$ and string cloud parameter $\alpha$. Here, $Q/M=0.5,\,\,r_0/M=20,\,\,\Lambda=-0.001/M^2$.}
    \label{fig:shadow-radius}
\end{figure}

For the RN-like BH without a cosmological constant, the function \( f(r_{O}) \) approaches \( \frac{1-\alpha}{1-\ell} \) for an observer located at infinity (\( r_{0} \to \infty \)). In this limit, the expression in Eq.~(\ref{b15}) simplifies to ($R_{\rm s\, SBH}=3\sqrt{3} M$)
\begin{equation}
R_s =
\begin{cases}
\displaystyle 
\frac{1-\ell}{1-\alpha}\,R_{\rm s\, SBH},
& \text{for } Q = 0, \\[1.2em]
\displaystyle 
\frac{1-\ell}{1-\alpha}\,
\frac{\left(1+\sqrt{1-\frac{8Q^2(1-\alpha)}{9M^2(1-\ell)^3}}\right)^{\!2}}
{2\sqrt{2}\sqrt{\,1+\sqrt{1-\frac{8Q^2(1-\alpha)}{9M^2(1-\ell)^3}}
-\frac{2Q^2(1-\alpha)}{3M^2(1-\ell)^3}}}\,R_{\rm s\, SBH}, 
& \text{for } Q \neq 0,
\end{cases}
\label{b15a}
\end{equation}

\noindent For $\alpha \to 0$, corresponding to the absence of string cloud in the charged BH solution, the shadow radius reduces to the result  obtained in Ref. \cite{CC10}.

\noindent For $\ell=0$ corresponding to the absence of KR-field effects, one can find
\begin{equation}
R_s =
\begin{cases}
\displaystyle 
\frac{R_{\rm s\, SBH}}{1-\alpha},
& \text{for } Q = 0, \\[1.2em]
\displaystyle 
\frac{R_{\rm s\, SBH}}{1-\alpha}\,
\frac{\left(1+\sqrt{1-\frac{8Q^2(1-\alpha)}{9M^2}}\right)^{\!2}}
{2\sqrt{2}\sqrt{\,1+\sqrt{1-\frac{8Q^2(1-\alpha)}{9M^2}}
-\frac{2Q^2(1-\alpha)}{3M^2}}}, 
& \text{for } Q \neq 0,
\end{cases}
\label{b15b}
\end{equation}

\noindent When $\ell=\alpha>0$, that is equal contribution from the KR-field and string cloud, the shadow radius simplifies as,
\begin{equation}
R_s =
\begin{cases}
\displaystyle 
R_{\rm s\, SBH},
& \text{for } Q = 0, \\[1.2em]
\displaystyle 
\frac{\left(1+\sqrt{1-\frac{8Q^2(1-\alpha)}{9M^2(1-\ell)^3}}\right)^{\!2}}
{2\sqrt{2}\sqrt{\,1+\sqrt{1-\frac{8Q^2(1-\alpha)}{9M^2(1-\ell)^3}}
-\frac{2Q^2(1-\alpha)}{3M^2(1-\ell)^3}}}\,R_{\rm s\, SBH}, 
& \text{for } Q \neq 0,
\end{cases}
\label{b15c}
\end{equation}

\noindent From the above equation (\ref{b15c}), we observe that when the contributions from the KR field and the string cloud are equal, the shadow radius of the uncharged BH coincides with the corresponding result for the Schwarzschild case.

\begin{table}[ht!]
\centering
\caption{Values of \(R_s/M\) for selected \(\alpha\) and \(\ell\)}
\begin{tabular}{|c|cccccc|}
\hline
$\alpha \backslash \ell$ & -0.10 & -0.05 & 0.05 & 0.10 & 0.15 & 0.20 \\
\hline
0.05 & 6.0166 & 5.74312 & {\bf 5.19615} & 4.92267 & 4.64919 & 4.37571 \\
0.10 & 6.35085 & 6.06218 & 5.48483 & {\bf 5.19615} & 4.90748 & 4.61880 \\
0.15 & 6.72443 & 6.41878 & 5.80746 & 5.50181 & {\bf 5.19615} & 4.89050 \\
0.20 & 7.14471 & 6.81995 & 6.17043 & 5.84567 & 5.52091 & {\bf 5.19615} \\
\hline
\end{tabular}
\label{tab:numerical}
\end{table}

For a more quantitative analysis, we aim to theoretically explore the upper limits of the parameters \(\ell\) and \(\alpha\). To do this, we constrain these parameters using observational data provided by the EHT collaboration for M87* and Sgr A* based on their respective shadow measurements. For M87*, it is well known \cite{EHTL1} that the angular diameter of the shadow, the distance from Earth, and the BH mass have been reported as \(\theta_{\mathrm{M87}^*} = (42 \pm 3)\, \mu\mathrm{as}\), \(D = 16.8\, \mathrm{Mpc}\), and \(M_{\mathrm{M87}^*} = (6.5 \pm 0.9) \times 10^{9} M_{\odot}\), respectively. For Sgr A*, the corresponding parameters are \(\theta_{\mathrm{Sgr}\, A^*} = (48.7 \pm 7)\, \mu\mathrm{as}\), \(D = (8.28 \pm 0.033)\, \mathrm{kpc}\), and \(M_{\mathrm{Sgr}\, A^*} = (4.3 \pm 0.013) \times 10^{6} M_{\odot}\) \cite{EHTL12, FF10}. Based on these data, we define the radius of the observed shadow for Sgr A* and M87* as follows:
\begin{equation}
R_{\mathrm{s}\,\mathrm{Sgr\, A}^*} = \frac{D_{\mathrm{Sgr\, A}^*} \, \Theta_{\mathrm{Sgr\, A}^*}}{2 M_{\mathrm{Sgr\, A}^*}}\quad,\quad 
R_{\mathrm{s}\,\mathrm{M87}^*} = \frac{D_{\mathrm{M87}^*} \, \Theta_{\mathrm{M87}^*}}{2 M_{\mathrm{M87}^*}}.\label{cond}
\end{equation}
The theoretical shadow diameter can be expressed as \(d_{\mathrm{s}}^{\mathrm{Theo}} = 2 R_{\mathrm{s}}\). From observational data, these diameters are \(d_{\mathrm{s}}^{\mathrm{M87}^*} = (11 \pm 1.5) M_{\odot}\,\mbox{and}\, d_{\mathrm{s}}^{\mathrm{Sgr\, A}^*} = (9.5 \pm 1.4) M_{\odot},\) respectively, where \(M_{\odot}\) is the BH mass.

Following the data results, we find the upper values of $\ell$ and $\alpha$ for the BHs in the galaxy M87* and Sgr A* and show their upper values in the Table \ref{tab:numerical}. Using the theoretical formula for the shadow radius with \(\ell = 0\) and \(\alpha = 0\), we obtain
\begin{equation}
R_{\mathrm{s}}^{\mathrm{theo}} = 3 \sqrt{3} M \approx 5.196\, M. \label{cond1}
\end{equation}
If we take \(\ell = 0.05\) and \(\alpha = 0.10\), the shadow radius becomes
\begin{equation}
R_{\mathrm{s}} = 5.485\, M_{\odot}, \label{cond2}
\end{equation}
which represents the upper limit of the shadow radius. Alternatively, setting \(\ell = 0.20\) and \(\alpha = 0.10\), we find
\begin{equation}
R_{\mathrm{s}} = 4.619\, M_{\odot}, \label{cond3}
\end{equation}
corresponding to the lower limit of the shadow radius.

In general ($\alpha \neq \ell$), to be consistent with the observational data of the shadow radius, for example, Sgr A*, the parameters \(\ell\) and \(\alpha\) must satisfy the following constraint:
\begin{equation}
0.889 \lesssim \frac{1 - \ell}{1 - \alpha} \lesssim 1.056. \label{cond4}
\end{equation}
This relation imposes upper and lower bounds on the parameters \(\ell\) and \(\alpha\).
}

\section{Topological Features of Photon Sphere}

In recent years, topological methods have become increasingly valuable for investigating BH  solutions. A significant advancement in this field is Duan’s $\phi$-mapping theory, which establishes a connection between topological defects and critical points in BH systems. These topological defects occur at points where a vector field vanishes, signifying key phase transitions. Such points give rise to a conserved topological current, which is intricately linked to the field's geometric properties. The topological current leads to a topological invariant, a crucial quantity that captures the system’s global phase structure. This invariant is computed using geometric quantities that describe how the vector field twists and turns in spacetime, providing insights into the system's underlying dynamics. For a comprehensive discussion and investigation across various BHs, readers can see \cite{pp1,pp2,pp3,pp4,pp5,pp6,pp7,pp8,pp9}.

\begin{figure}[ht!]
    \centering
    \includegraphics[width=0.4\linewidth]{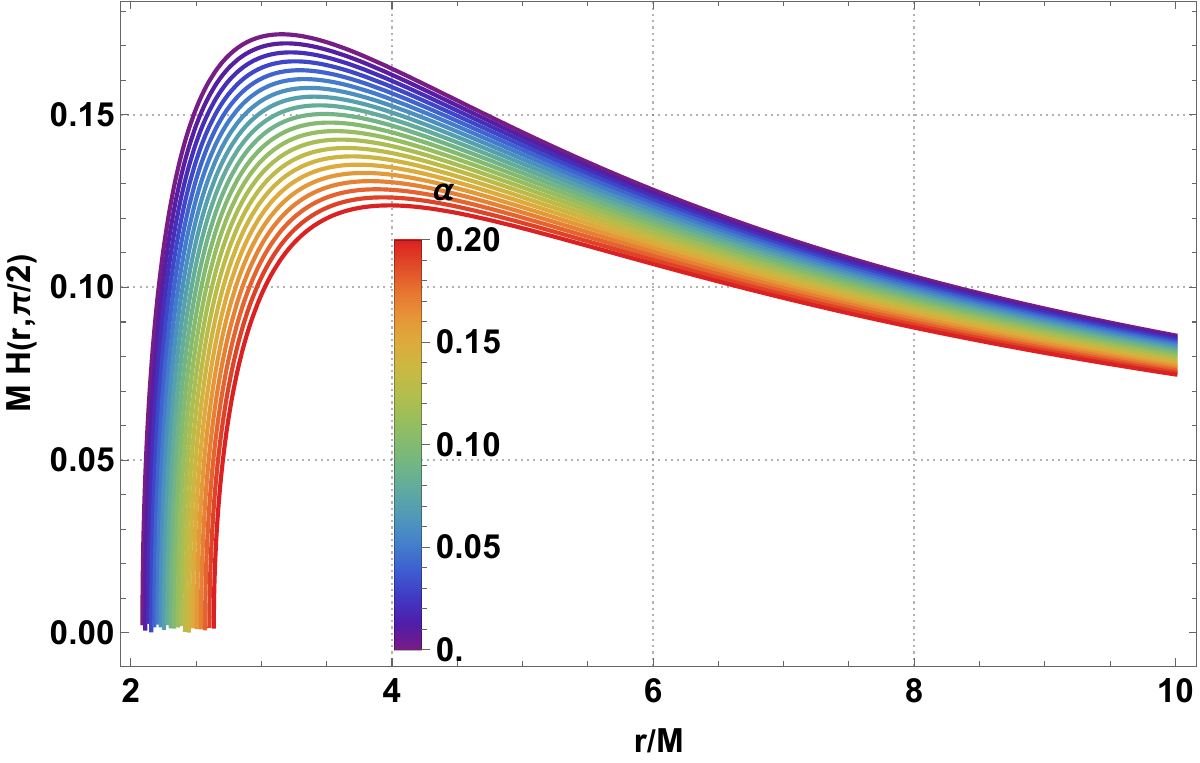}\quad\quad
    \includegraphics[width=0.4\linewidth]{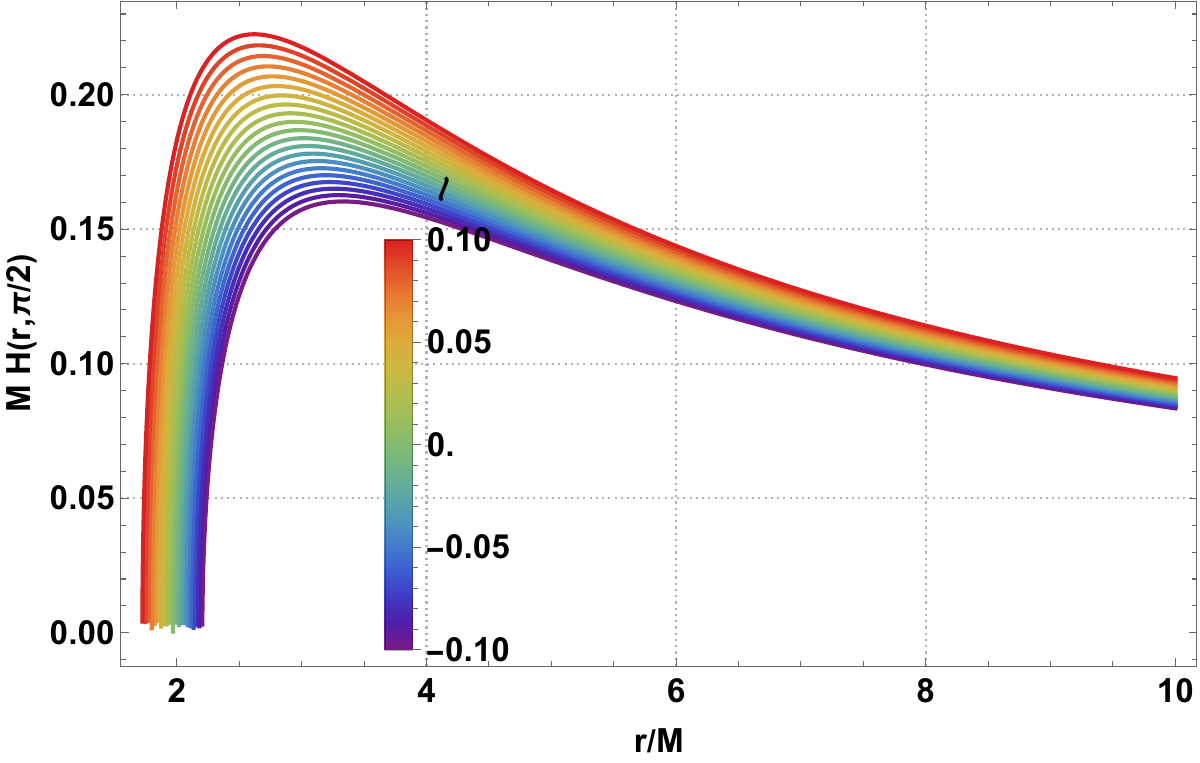}\\
    (i) $\ell=-0.1$ \hspace{6cm} (ii) $\alpha=0.05$
    \caption{ Behavior of the potential function $M\,H(r, \theta)$ as a function of $r/M$ by varying the KR-field parameter $\ell$ and string parameter $\alpha$. Here $Q/M=0.5,\,\,\Lambda=-0.001/M^2$.}
    \label{fig:potential-function}
\end{figure}

\begin{figure}[ht!]
    \centering
    \includegraphics[width=0.4\linewidth]{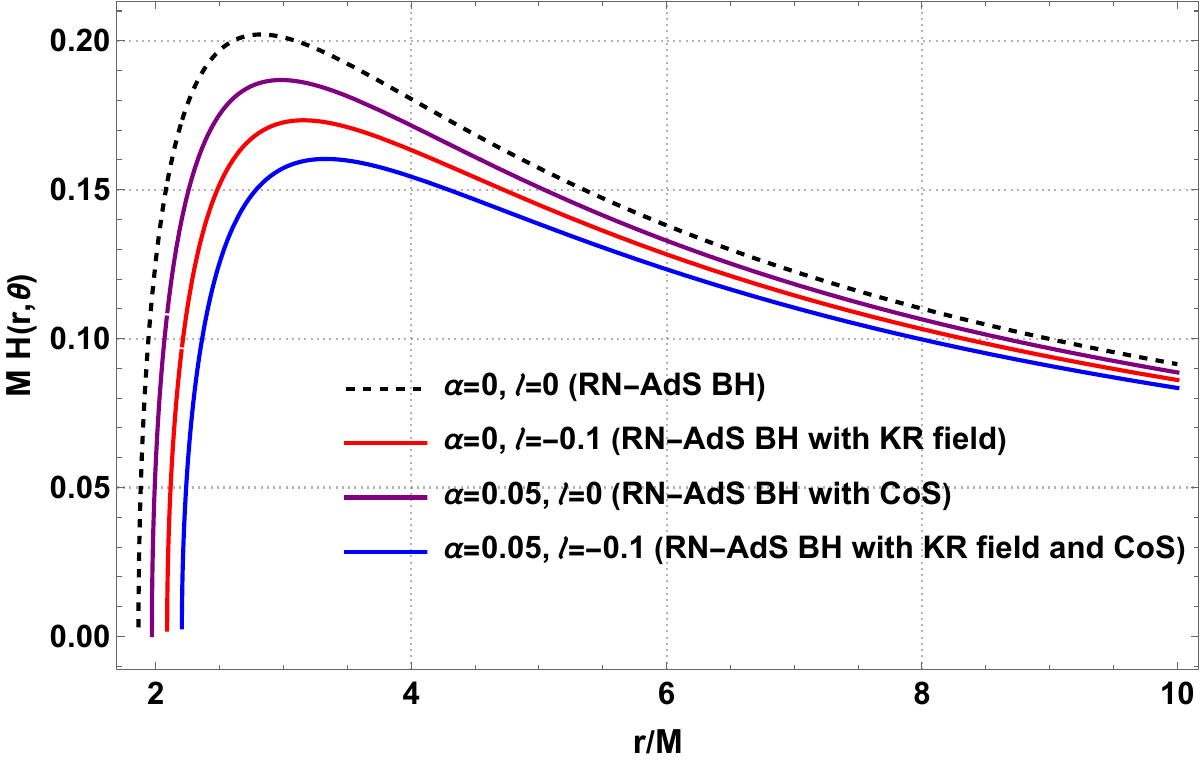}
    \caption{A comparison of the potential function $M\,H(r, \theta)$ as a function of $r/M$ for charged BHs in different configurations. Here $Q/M=0.5,\,\,\Lambda=-0.001/M^2$.}
    \label{fig:potential-function-comp}
\end{figure}

To investigate the characteristics of photon spheres within quantum-corrected BH geometries, we build upon the metric presented in Eq.\eqref{metric} and employ analytical techniques developed in previous studies \cite{pp1,pp2,pp3,pp4,pp5,pp6,pp7,pp8,pp9}. In order to study the topological property of the photon rings, one can first introduce a potential function as,
\begin{equation}
H(r,\theta) = \sqrt{\frac{-g_{tt}}{g_{\phi\phi}}} =\frac{\sqrt{f(r)}}{r\,\sin \theta}=
\begin{cases}
    \displaystyle \frac{1}{r\,\sin \theta}\,\sqrt{\frac{1 - \alpha}{1 - \ell} - \frac{2\,M}{r}- \frac{\Lambda}{3\,(1-\ell)} \, r^2}, & \text{for } Q=0,\\[1.2em]
    \displaystyle \frac{1}{r\,\sin \theta}\,\sqrt{\frac{1 - \alpha}{1 - \ell} - \frac{2\,M}{r} + \frac{Q^2}{r^2\,(1-\ell)^2} - \frac{\Lambda}{3\,(1-\ell)} \, r^2}, & \text{for } Q \neq 0
\end{cases}
\label{null18}
\end{equation}
Obviously, the radius of the PS locates at the root of $\partial_r\,H(r,\theta)=0$. 

In Figure \ref{fig:potential-function}, we illustrate the behavior of the potential function $M\,H(r, \theta)$ by varying the string parameter $\alpha$ and the KR field parameter $\ell$. We observed that as $\alpha$ rises, the topological potential function reduces. In contrast, the potential function rises as the value of $\ell$ increases. {\color{black} Figure \ref{fig:potential-function-comp} shows a comparison of the potential function $H(r, \theta)$ for charged BHs in different configurations: with and without KR field and string clouds.}

Following the definition in \cite{pp2,pp3}, using the topological vector potential given in Eq.(\ref{null18}), we introduce a vector ${\bf v}=(v_r,\,v_{\theta})$ field whose components are defined as,
\begin{align}
    v_r=\frac{\partial_r H(r,\theta)}{\sqrt{g_{rr}}}=\sqrt{f(r)}\,\partial_r H(r,\theta),\qquad  v_{\theta}=\frac{\partial_{\theta} H}{\sqrt{g_{\theta\theta}}}=\frac{1}{r}\,\partial_{\theta} H(r,\theta).\label{null19}
\end{align}
It follows that $\partial^{\mu} H\,\partial_{\mu} H=v_r^2+v^2_{\theta}=v^2$. 

Simplification of the above relations result
\begin{equation}
    v_r=
    \begin{cases}
        \displaystyle -\frac{1}{r^2\,\sin  \theta}\,\left(\frac{1-\alpha}{1-\ell}-\frac{3\,M}{r}\right), & \text {for } Q=0,\\[1.2em]
        \displaystyle -\frac{1}{r^2\,\sin  \theta}\,\left(\frac{1-\alpha}{1-\ell}-\frac{3\,M}{r}+\frac{2\,Q^2}{r^2\,(1-\ell)^2}\right), & \text{ for } Q \neq 0,
    \end{cases}
    \label{null25}
\end{equation}
and
\begin{equation}
    v_{\theta}=
    \begin{cases}
        \displaystyle -\frac{\cot \theta}{r^2\,\sin \theta}\,\sqrt{\frac{1 - \alpha}{1 - \ell} - \frac{2\,M}{r}- \frac{\Lambda}{3\,(1-\ell)} \, r^2}, & \text{for } Q=0,\\[1.2em]
        \displaystyle -\frac{\cot \theta}{r^2\,\sin \theta}\,\sqrt{\frac{1 - \alpha}{1 - \ell} - \frac{2\,M}{r} + \frac{Q^2}{r^2\,(1-\ell)^2} - \frac{\Lambda}{3\,(1-\ell)} \, r^2}, & \text{for } Q=\neq 0.
    \end{cases}
    \label{null26}
\end{equation}
Noted that at $r=r_\text{ph}$ and $\theta=\pi/2$ both the components vanish and hence, one will find null vector field.

The magnitude of the vector field is given by {\small
\begin{equation}
    v(r,\theta)=
    \begin{cases}
        \displaystyle \frac{1}{r^2\,\sin\theta}\,\sqrt{\cot^2 \theta\,\left(\frac{1 - \alpha}{1 - \ell} - \frac{2\,M}{r}- \frac{\Lambda}{3\,(1-\ell)} \, r^2\right)+\left(\frac{1-\alpha}{1-\ell}-\frac{3\,M}{r}\right)^2}, & \text{for } Q=0,\\[1.5em]
        \displaystyle \frac{1}{r^2\,\sin\theta}\,\sqrt{\cot^2 \theta\,\left(\frac{1 - \alpha}{1 - \ell} - \frac{2\,M}{r} + \frac{Q^2}{r^2\,(1-\ell)^2} - \frac{\Lambda}{3\,(1-\ell)} \, r^2\right)+\left(\frac{1-\alpha}{1-\ell}-\frac{3\,M}{r}+\frac{2\,Q^2}{r^2\,(1-\ell)^2}\right)^2}, & \text{for } Q \neq 0.
    \end{cases}
    \label{null27}
\end{equation}
}
\begin{figure}[ht!]
    \centering
    \includegraphics[width=0.3\linewidth]{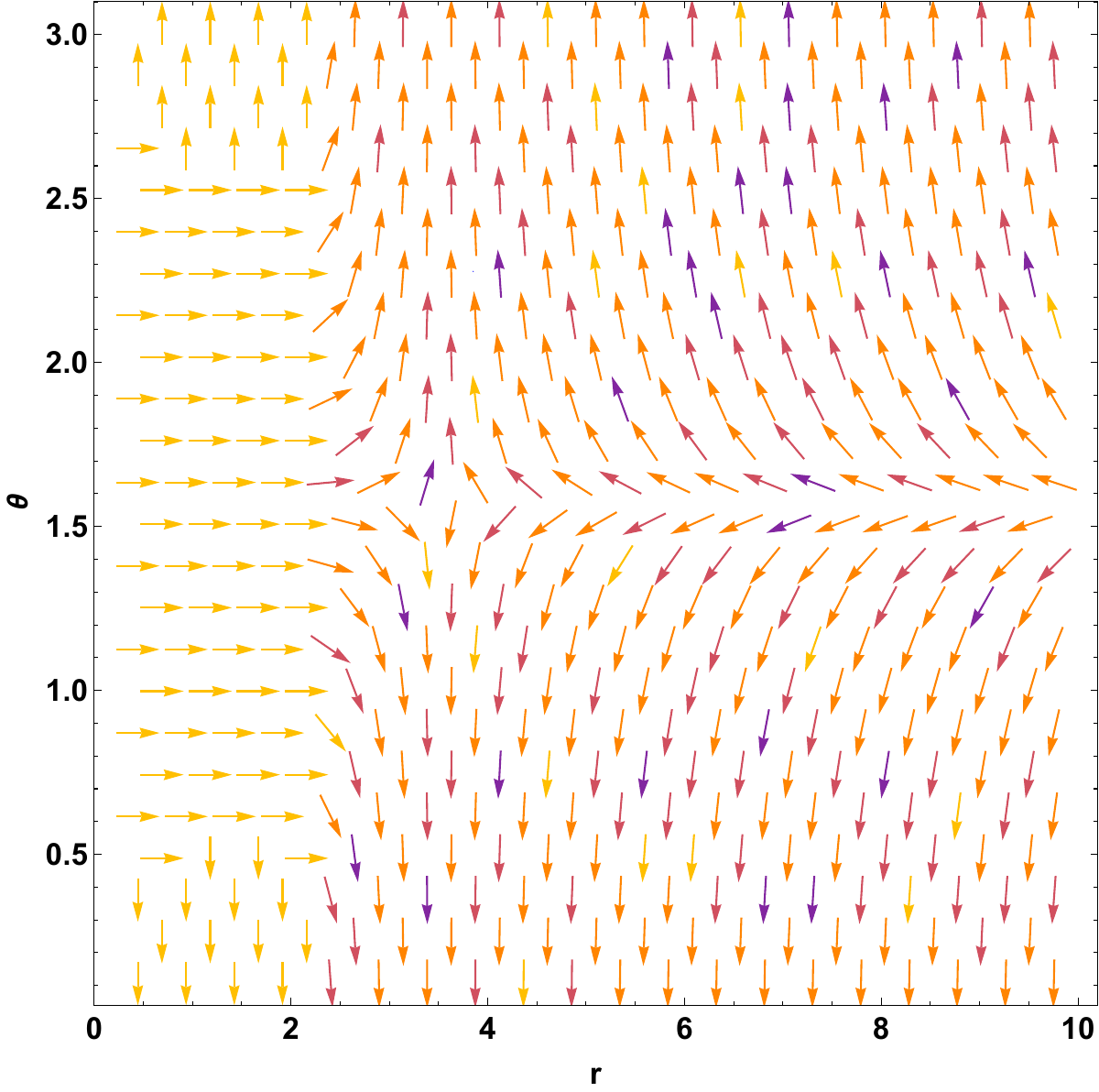}\qquad
    \includegraphics[width=0.3\linewidth]{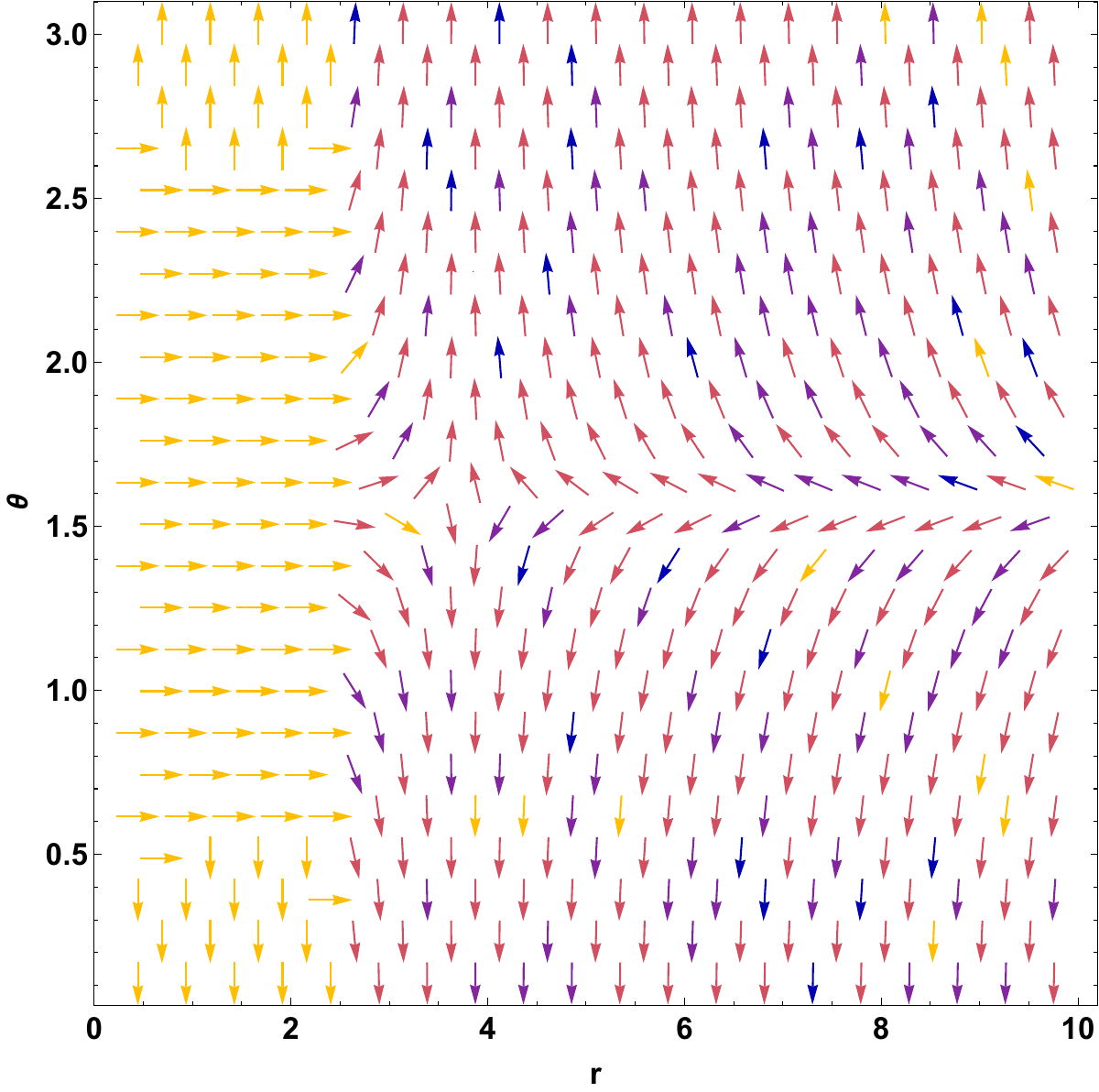}\\
    (i) $\alpha=0.10$ \hspace{4cm} (ii) $\alpha=0.15$ 
    \caption{The arrow represents the normalized vector field \({\bf n}\) on a portion of the $r-\theta$ plane for two values of the string cloud $\alpha$. Here $Q/M=0.5,\,\,\Lambda=-0.001/M^2,\,\,\ell=-0.1$.}
    \label{fig:vector-field-1}
    \hfill\\
    \includegraphics[width=0.3\linewidth]{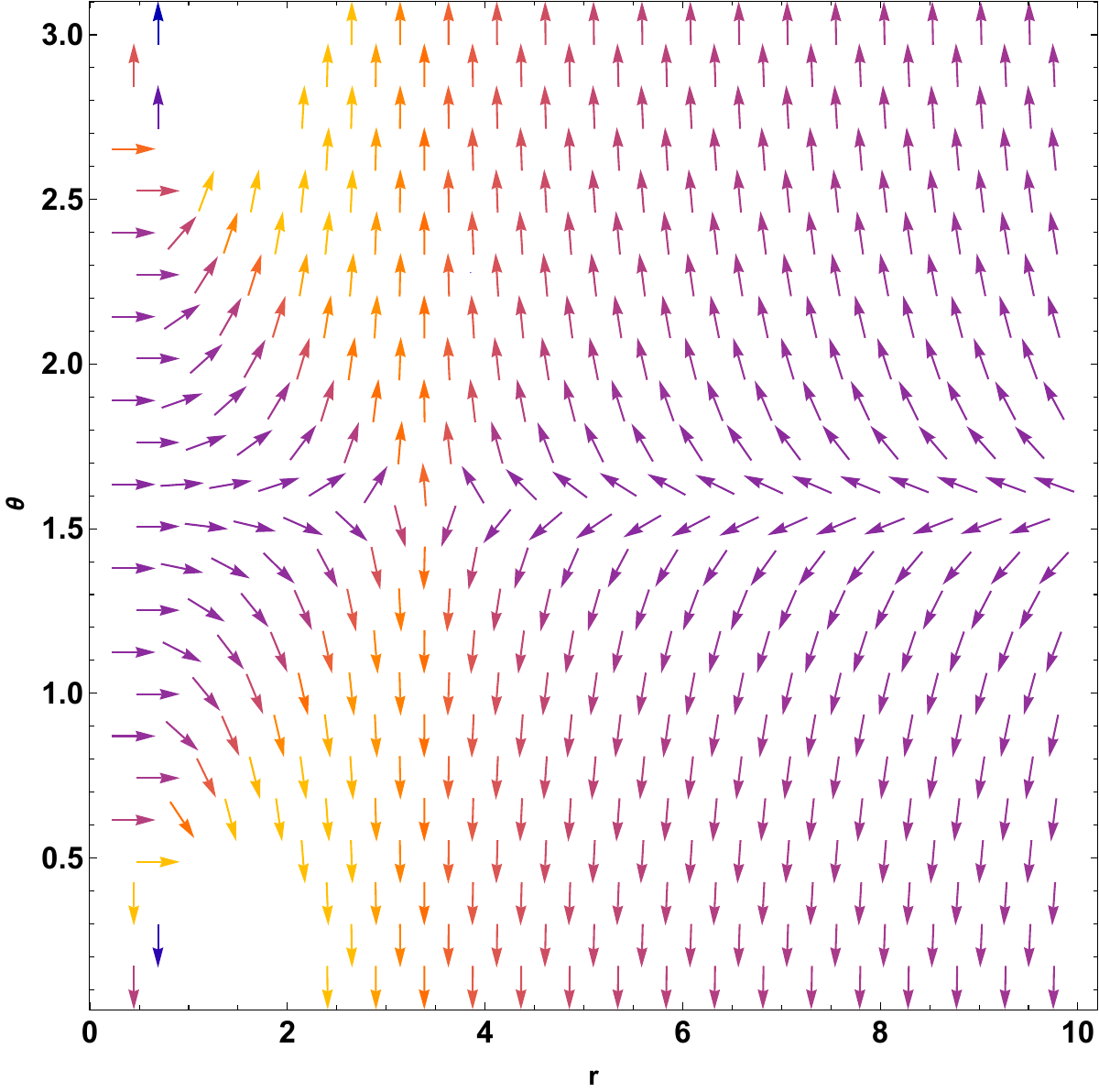}\qquad
    \includegraphics[width=0.3\linewidth]{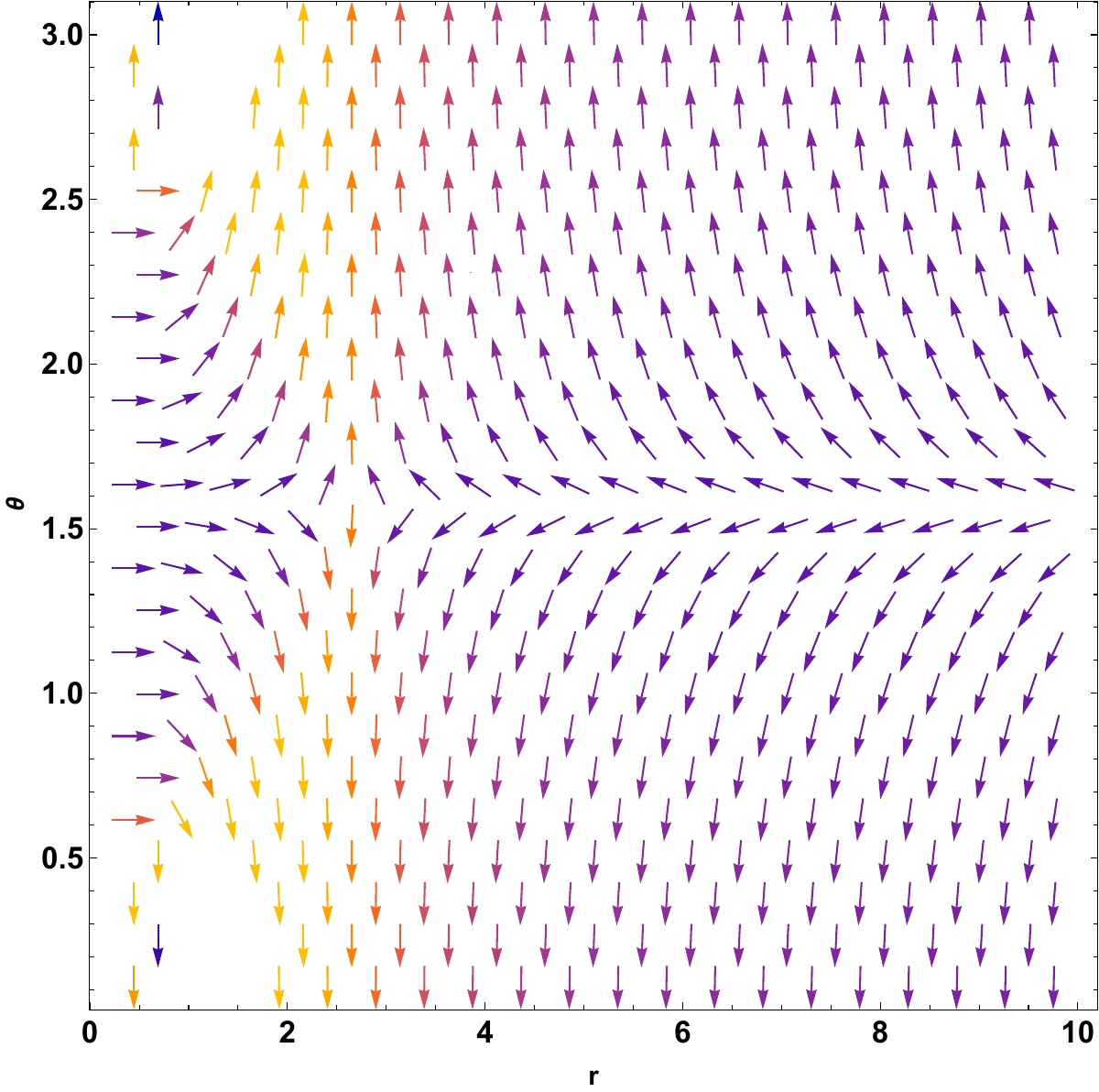}\\
    (i) $\ell=-0.1$ \hspace{4cm} (ii) $\ell=0.1$ 
    \caption{ The arrow represents the normalized vector field \({\bf n}\) on a portion of the $r-\theta$ plane for two values of the KR field parameter \(\ell\). Here $Q/M=0.5,\,\,\Lambda=-0.001/M^2,\,\,\alpha=0.05$.}
    \label{fig:vector-field-2}
\end{figure}

Although the circular photon orbit for a spherically symmetric BH is a photon sphere (PS), which is independent of the coordinate $\theta$, here we aim to investigate the topological property of the circular photon orbit, so we preserve $\theta$ in our discussions. Note that the vector field can also be reformulated as,
\begin{equation}
    {\bf v}=v\,e^{i\,\Omega}\quad,\quad\mbox{where} \quad v_r=v\,\cos \Omega\quad,\quad v_{\theta}=v\,\sin \Omega.\label{null20}
\end{equation}
The normalized vectors are defined as
\begin{equation}
    {\bf n}=(n_r\,,\,n_{\theta})=\frac{{\bf v}}{v}=\left(\frac{v_r}{\sqrt{v^2_r+v^2_{\theta}}}\,,\,\frac{v_{\theta}}{\sqrt{v^2_r+v^2_{\theta}}}\right).\label{null21}
\end{equation}

Thereby, the components of the normalized vector field are as follows:
\begin{align}
    n_r=
    \begin{cases}
        \displaystyle -\frac{\left(\frac{1-\alpha}{1-\ell}-\frac{3\,M}{r}\right)}{\sqrt{\cot^2 \theta\,\left(\frac{1 - \alpha}{1 - \ell} - \frac{2\,M}{r}- \frac{\Lambda}{3\,(1-\ell)} \, r^2\right)+\left(\frac{1-\alpha}{1-\ell}-\frac{3\,M}{r}\right)^2}}, & \text{for } Q=0,\\[2.5em]
        \displaystyle -\frac{\left(\frac{1-\alpha}{1-\ell}-\frac{3\,M}{r}+\frac{2\,Q^2}{r^2\,(1-\ell)^2}\right)}{\sqrt{\cot^2 \theta\,\left(\frac{1 - \alpha}{1 - \ell} - \frac{2\,M}{r} + \frac{Q^2}{r^2\,(1-\ell)^2} - \frac{\Lambda}{3\,(1-\ell)} \, r^2\right)+\left(\frac{1-\alpha}{1-\ell}-\frac{3\,M}{r}+\frac{2\,Q^2}{r^2\,(1-\ell)^2}\right)^2}}, & \text{for } Q =\neq 0.
    \end{cases}
    \label{null28}
\end{align}
And
\begin{align}
    n_{\theta}=
    \begin{cases}
        \displaystyle -\cot \theta\,\sqrt{\frac{\frac{1 - \alpha}{1 - \ell} - \frac{2\,M}{r}- \frac{\Lambda}{3\,(1-\ell)}\,r^2}{\cot^2 \theta\,\left(\frac{1 - \alpha}{1 - \ell} - \frac{2\,M}{r}- \frac{\Lambda}{3\,(1-\ell)} \, r^2\right)+\left(\frac{1-\alpha}{1-\ell}-\frac{3\,M}{r}\right)^2}}, & \text{for } Q=0,\\[2.5em]
        \displaystyle -\cot \theta\,\sqrt{\frac{\frac{1 - \alpha}{1 - \ell} - \frac{2\,M}{r} + \frac{Q^2}{r^2\,(1-\ell)^2} - \frac{\Lambda}{3\,(1-\ell)}\,r^2}{\cot^2 \theta\,\left(\frac{1 - \alpha}{1 - \ell} - \frac{2\,M}{r} + \frac{Q^2}{r^2\,(1-\ell)^2} - \frac{\Lambda}{3\,(1-\ell)} \, r^2\right)+\left(\frac{1-\alpha}{1-\ell}-\frac{3\,M}{r}+\frac{2\,Q^2}{r^2\,(1-\ell)^2}\right)^2}}, & \text{for } Q \neq 0.
    \end{cases}
    \label{null29}
\end{align}

From the above expressions, we see that the unit vector field ${\bf n}$ is influenced by the Rastall parameter \((\lambda)\), the CoS parameter \((\alpha)\), the QF parameters \((\mathcal{N}, \omega)\). Additionally, the BH’s mass \(M\) and electric charge \(Q\) alter this unit vector field.

{\color{black}
Now, we discuss a special case corresponds to $\alpha=\ell>0$, that is equal contribution from the KR-field and string cloud effects, the components of the normalized vector field (${\bf n}$) simplifies to:
\begin{align}
    n_r=
    \begin{cases}
        \displaystyle -\frac{\left(1-\frac{3\,M}{r}\right)}{\sqrt{\cot^2 \theta\,\left(1- \frac{2\,M}{r}- \frac{\Lambda}{3\,(1-\ell)} \, r^2\right)+\left(1-\frac{3\,M}{r}\right)^2}}, & \text{for } Q=0,\\[2.5em]
        \displaystyle -\frac{\left(1-\frac{3\,M}{r}+\frac{2\,Q^2}{r^2\,(1-\ell)^2}\right)}{\sqrt{\cot^2 \theta\,\left(1- \frac{2\,M}{r} + \frac{Q^2}{r^2\,(1-\ell)^2} - \frac{\Lambda}{3\,(1-\ell)} \, r^2\right)+\left(1-\frac{3\,M}{r}+\frac{2\,Q^2}{r^2\,(1-\ell)^2}\right)^2}}, & \text{for } Q =\neq 0.
    \end{cases}
    \label{null28a}
\end{align}
And
\begin{align}
    n_{\theta}=
    \begin{cases}
        \displaystyle -\cot \theta\,\sqrt{\frac{1- \frac{2 M}{r}- \frac{\Lambda}{3(1-\ell)}\,r^2}{\cot^2 \theta\,\left(1- \frac{2 M}{r}- \frac{\Lambda}{3\,(1-\ell)} \, r^2\right)+\left(1-\frac{3\,M}{r}\right)^2}}, & \text{for } Q=0,\\[2.5em]
        \displaystyle -\cot \theta\,\sqrt{\frac{1- \frac{2\,M}{r} + \frac{Q^2}{r^2\,(1-\ell)^2} - \frac{\Lambda}{3\,(1-\ell)}\,r^2}{\cot^2 \theta\,\left(1- \frac{2\,M}{r} + \frac{Q^2}{r^2\,(1-\ell)^2} - \frac{\Lambda}{3\,(1-\ell)} \, r^2\right)+\left(1-\frac{3\,M}{r}+\frac{2\,Q^2}{r^2\,(1-\ell)^2}\right)^2}}, & \text{for } Q \neq 0.
    \end{cases}
    \label{null29a}
\end{align}

}

One can plot the normalized or unit vector field \(\mathbf{n}\) on a portion of the $r-\theta$ plane by varying values of various parameters. In Figures \ref{fig:vector-field-1}-\ref{fig:vector-field-2}, we depict the normalized vector field for two different values of CoS parameter $\alpha$ and the KR-field parameter $\ell$. {\color{black} For these plots, one can see the variation of the normalized vector field for different values of parameters.}

\section{Dynamics of Neutral Test Particles}

The dynamics of neutral test particles around a BH in the background of KR gravity coupled with a CoS is significantly influenced by the combined effects of the KR field and the string cloud. The KR field modifies the space-time geometry, while the CoS introduces additional stress-energy contributions, altering the gravitational potential. As a result, the ISCO and particle trajectories differ from those in standard General Relativity, exhibiting shifts in orbital radii and stability conditions. These modifications affect accretion processes and the motion of matter near the BH, providing potential observational signatures of KR gravity and string cloud effects.

The motion of a neutral test particle in the curved spherically symmetric space-time can be analyzed using the Hamiltonian approach, as discussed in Refs. \cite{AB3,AB4,AB5,AB6,AB7,AB12} given by
\begin{equation}
    \mathrm{H}=\frac{1}{2}\,g^{\mu\nu}\,p_{\mu}\,p_{\nu}+\frac{1}{2}\,m^2,\label{ss1}
\end{equation}
where $m$ is the mass of neutral particles, $p^{\mu}=m\,u^{\mu}$ is the four-momentum, $u^{\mu}=dx^{\mu}/d\tau$ is the four-velocity equation, and $\tau$ is the appropriate time of the neutral particle. Also, the Hamilton equations of motion are given: 
\begin{equation}
    \frac{dx^{\mu}}{d\lambda}\equiv m\,u^{\mu}=\frac{dH}{dp_{\mu}},\label{ss2}
\end{equation}
and
\begin{equation}
    \frac{dp_{\mu}}{d\lambda}=-\frac{\partial H}{\partial x^{\mu}},\label{ss3}
\end{equation}
where the affine parameter is given by $\lambda=\tau/m$.

As stated earlier, the chosen BH solution is static and spherically symmetric. Consequently, the metric tensor $g_{\mu\nu}$ is independent of the cyclic coordinates, namely the temporal coordinate $t$ and the azimuthal coordinate $\phi$. Due to this symmetry, the corresponding components of the particle’s four-momentum, specifically $p_t$ and $p_{\phi}$, are conserved along its geodesics. These conserved quantities are given by
\begin{eqnarray}
    &&\frac{p_t}{m}=-f(r)\,\dot{t}=-\mathcal{E},\label{ss4}\\
    &&\frac{p_{\phi}}{m}=r^2\,\sin^2 \theta\,\dot{\phi}=\mathcal{L}_0,\label{ss5} 
\end{eqnarray}
where $\mathcal{E}=\mathrm{E}/m$ and $\mathcal{L}_0=\mathrm{L}/m$, respectively are the specific energy and angular momentum per unit mass of the neutral test particles. Moreover, the conjugate momentum associated with $\theta$ coordinate is given by
\begin{equation}
    \frac{p_{\theta}}{m}=r^2\,\dot{\theta}.\label{ss6}
\end{equation}
The components of the four-velocity $u^{\mu}$ in the time \(u^t\), azimuthal \(u^{\phi} \) and angular \(u^{\theta}\) directions obey the governing equations given in the following forms:
\begin{align}
    \dot{t}&=\frac{\mathcal{E}}{f(r)},\label{ss6a}\\
    \dot{\phi}&=\frac{\mathcal{L}_0}{r^2\,\sin^2 \theta},\label{ss6bb}\\
    \dot{\theta}&=\frac{p_{\theta}}{m\,r^2}.\label{ss6cc}
\end{align}

Moreover, massive test particles follow the time-like trajectories and thereby using the normalization condition $g_{\mu\nu}\,u^{\mu}\,u^{\nu}=-1$ and metric (\ref{metric}), we find
\begin{equation}
    -f(r)\,\dot{t}^2+\frac{1}{f(r)}\,\dot{r}^2+r^2\,\sin^2\theta\,\dot{\phi}^2+r^2\,\dot{\theta}^2=-1.\label{ss7}
\end{equation}
Substituting $\dot{t}$, $\dot{\theta}$ and $\dot{\phi}$, we finally arrive
\begin{equation}
    -\frac{\mathcal{E}^2}{f(r)}+\frac{\dot{r}^2}{f(r)}+\frac{\mathcal{L}^2_0}{r^2\,\sin^2 \theta}+\frac{p^2_{\theta}}{m^2\,r^2}=-1.\label{ss8}
\end{equation}
The above equation can be re-written as,
\begin{equation}
    \dot{r}^2+\left(\frac{\mathcal{L}^2_0}{r^2\,\sin^2 \theta}+1\right)\,f(r)+\frac{p^2_{\theta}}{m^2\,r^2}\,h(r)=\mathcal{E}^2.\label{ss9}
\end{equation}

With these, the Hamiltonian Eq. (\ref{ss1}) can be expressed as follows:
\begin{equation}
    H=\frac{1}{2}\,f(r)\,p^2_r+\frac{1}{2}\,\frac{p^2_{\theta}}{r^2}+\frac{1}{2}\,\frac{m^2}{f(r)}\,\left[U_\text{eff}(r,\theta)-\mathcal{E}^2\right],\label{ss9aa}
\end{equation}
where \(U_\text{eff}(r,\theta)\) is the effective potential and takes the following form:
\begin{equation}
    U_\text{eff}(r,\theta)=\left(\frac{\mathcal{L}^2_0}{r^2\,\sin^2 \theta}+1\right)\,f(r).\label{ss10a}
\end{equation}

In the equatorial plane defined by $\theta=\pi/2$ and $p_{\theta}=0$, the effective potential of time-like geodesic reduces as
\begin{equation}
    U_\text{eff}(r)=\left(\frac{\mathcal{L}^2_0}{r^2}+1\right)\,f(r)=
    \begin{cases}
        \displaystyle \left(1+\frac{\mathcal{L}^2_0}{r^2}\right)\,\left[\frac{1-\alpha}{1-\ell}-\frac{2\,M}{r}-\frac{\Lambda}{3\,(1-\ell)}\,r^2\right], & \text{for } Q=0,\\[1.5em]
        \displaystyle \left(1+\frac{\mathcal{L}^2_0}{r^2}\right)\,\left[\frac{1-\alpha}{1-\ell}-\frac{2\,M}{r}+\frac{Q^2}{(1-\ell)^2\,r^2}-\frac{\Lambda}{3\,(1-\ell)}\,r^2\right], & \text{for } Q \neq 0
    \end{cases}
    \label{ss10}
\end{equation}

\begin{figure}[ht!]
    \centering
    \includegraphics[width=0.4\linewidth]{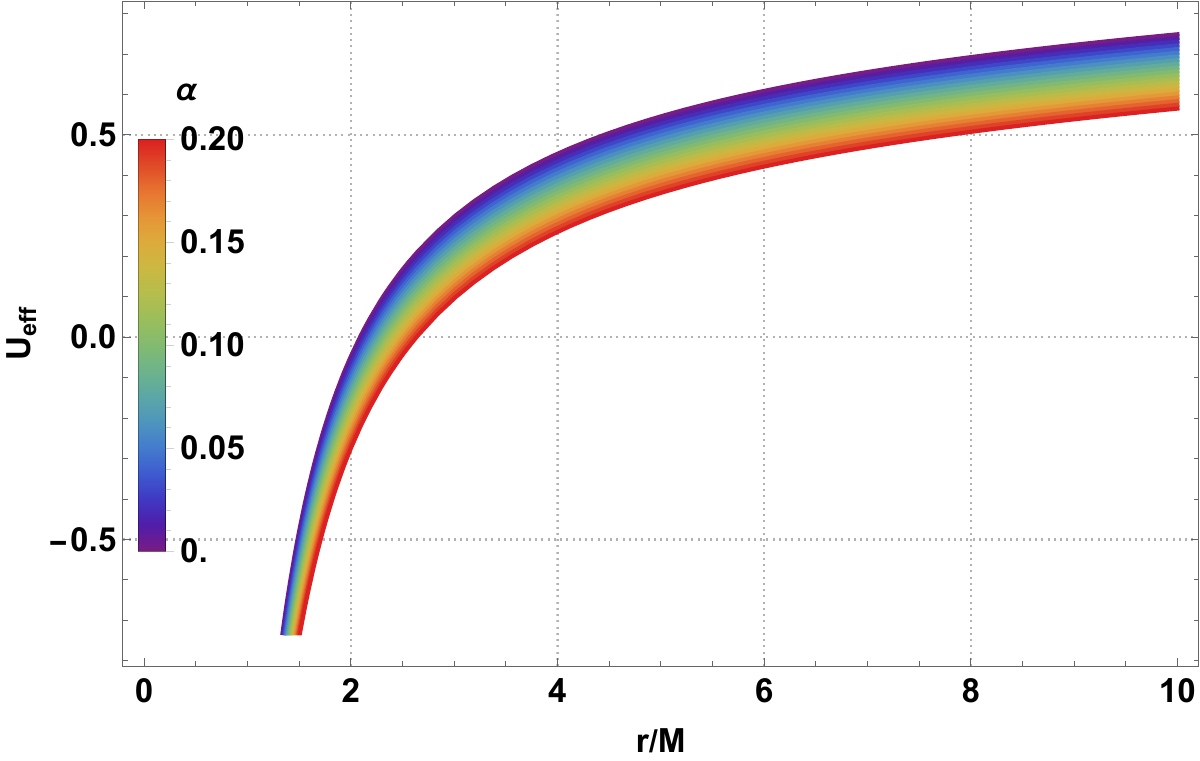}\qquad
    \includegraphics[width=0.4\linewidth]{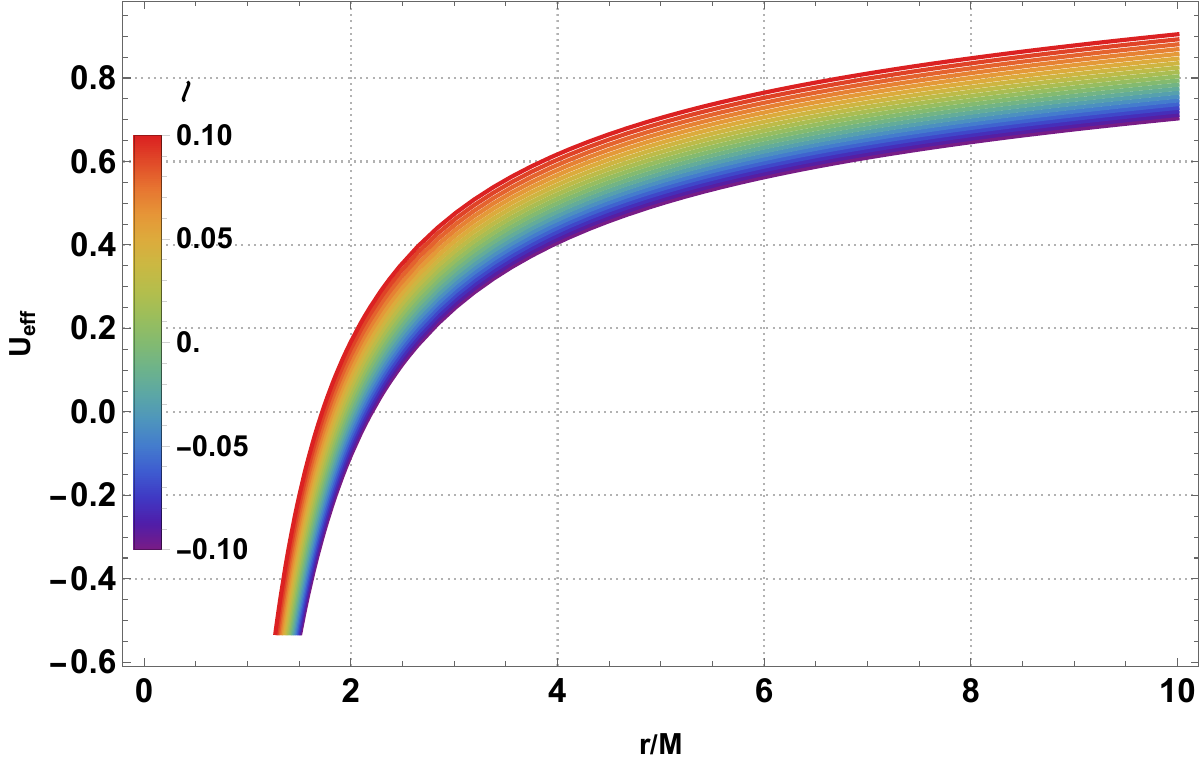}\\
    (i) $\ell=-0.1$ \hspace{6cm} (ii) $\alpha=0.05$
    \caption{ Behavior of the effective potential $U_\text{eff}(r)$ as a function of $r/M$ for different values of KR field parameter $\ell$ and CoS parameter $\alpha$. Here $Q/M=0.05,\,\,\mathcal{L}_0/M=2$ and $\Lambda=-0.001/M^2$.}
    \label{fig:potential-neutral}
\end{figure}

\begin{figure}[ht!]
    \centering
    \includegraphics[width=0.4\linewidth]{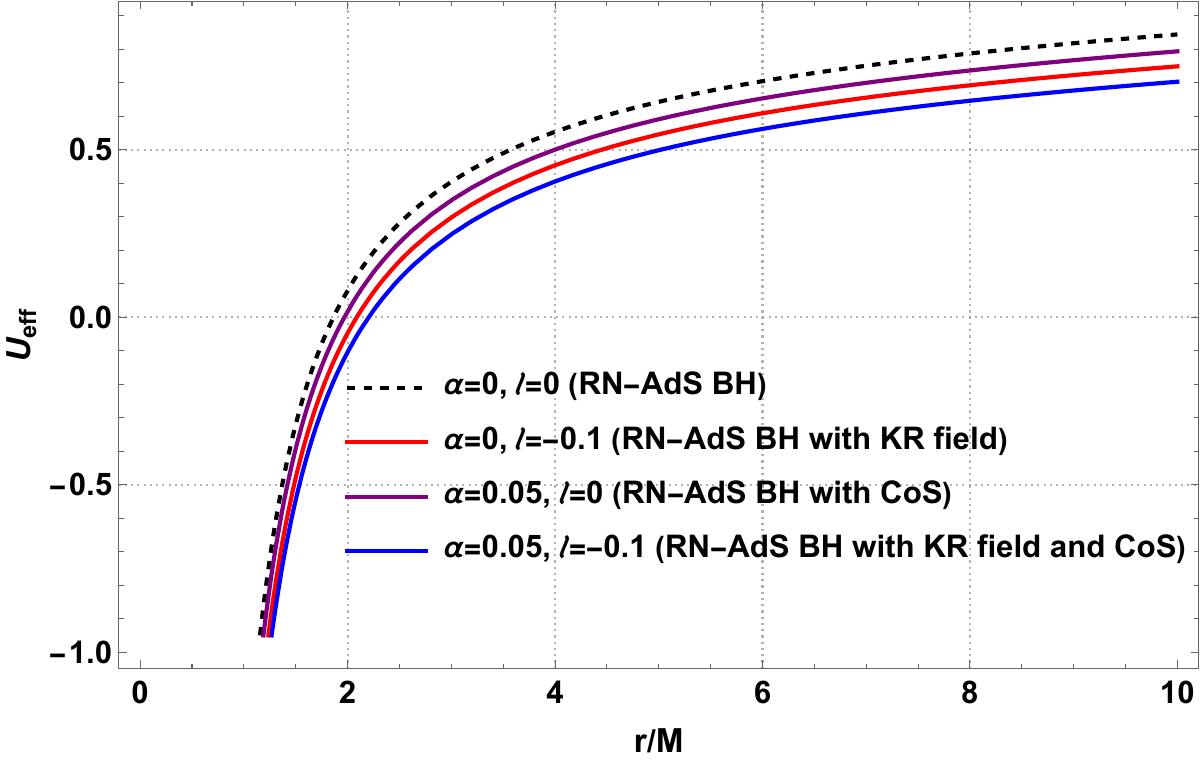}
    \caption{ A comparison of the effective potential $U_\text{eff}(r)$ as a function of $r/M$ for charged BHs in different configurations. Here $Q/M=0.05,\,\,\mathcal{L}_0/M=2$ and $\Lambda=-0.001/M^2$.}
    \label{fig:potential-neutral-comp}
\end{figure}

In Figure \ref{fig:potential-neutral}, we depict the effective potential governs the dynamics of neutral particles by varying the string parameter $\alpha$ and the KR field parameter $\ell$ for two values of the electric charge $Q=0.5$ and $Q=0.75$. In panels (i)-(ii), we observed that as the value of \(\alpha\) rises, the effective potential decreases. In contrast, in panels (iii)-(iv) this potential rises with increasing the value of $\ell$.  {\color{black} Figure \ref{fig:potential-neutral-comp} shows a comparison of the effective potential for test particles in charged BHs in different configurations: with and without KR field and string clouds.}

For the ISCO, the following conditions must be satisfied:
\begin{equation}
    \mathcal{E}^2=U_\text{eff}(r),\quad\quad U'_\text{eff}(r)=0,\quad\quad U''_\text{eff}(r) \geq 0,\label{cc4}
\end{equation}
where prime denotes partial derivative w. r. to $r$.

Using the first two conditions results the following:
\begin{equation}
    \mathrm{L}_\text{sp}=\sqrt{\frac{r^3\,f'}{2\,f-r\,f'}}=
    \begin{cases}
        \displaystyle r\,\sqrt{\frac{ \dfrac{M}{r}- \dfrac{\Lambda}{3\,(1-\ell)} r^2 }{ \dfrac{1 - \alpha}{1 - \ell} - \dfrac{3\,M}{r}}}, & \text{ for } Q=0\\[2.5em]
        \displaystyle r\,\sqrt{\frac{ \dfrac{M}{r} - \dfrac{Q^2}{(1-\ell)^2\,r^2} - \dfrac{\Lambda}{3\,(1-\ell)} r^2 }{ \dfrac{1 - \alpha}{1 - \ell} - \dfrac{3\,M}{r} + \dfrac{2 Q^2}{(1-\ell)^2\,r^2} }}. & \text{ for } Q \neq 0 \label{cc5}
    \end{cases}
\end{equation}
And
\begin{equation}
    \mathrm{E}_\text{sp}=\sqrt{\frac{2\,f^2}{2\,f-r\,f'}}=
    \begin{cases}
        \displaystyle \frac{\left(\frac{1-\alpha}{1-\ell}-\frac{2\,M}{r}-\frac{\Lambda}{3\,(1-\ell)}\,r^2\right)}{\sqrt{ \dfrac{1 - \alpha}{1 - \ell} - \dfrac{3\,M}{r}}}, & \text {for } Q=0\\[2.5em]
        \displaystyle \frac{\left(\frac{1-\alpha}{1-\ell}-\frac{2\,M}{r}+\frac{Q^2}{(1-\ell)^2\,r^2}-\frac{\Lambda}{3\,(1-\ell)}\,r^2\right)}{\sqrt{ \dfrac{1 - \alpha}{1 - \ell} - \dfrac{3\,M}{r} + \dfrac{2\,Q^2}{(1-\ell)^2\,r^2} }}, & \text {for } Q \neq 0
    \end{cases}
    .\label{cc6}
\end{equation}

{\color{black}
Here, we discuss a special case corresponds to $\alpha=\ell>0$, that is equal contribution from the KR-field and string cloud effects, the specific energy and the specific angular momentum simplifies to:
\begin{equation}
    \mathrm{L}_\text{sp}=
    \begin{cases}
        \displaystyle r\,\sqrt{\frac{ \dfrac{M}{r}- \dfrac{\Lambda}{3\,(1-\ell)} r^2 }{1 - \dfrac{3\,M}{r}}}, & \text{ for } Q=0\\[2.5em]
        \displaystyle r\,\sqrt{\frac{ \dfrac{M}{r} - \dfrac{Q^2}{(1-\ell)^2\,r^2} - \dfrac{\Lambda}{3\,(1-\ell)} r^2 }{1- \dfrac{3\,M}{r} + \dfrac{2 Q^2}{(1-\ell)^2\,r^2} }}. & \text{ for } Q \neq 0 \label{cc5a}
    \end{cases}
\end{equation}
And
\begin{equation}
    \mathrm{E}_\text{sp}=
    \begin{cases}
        \displaystyle \frac{\left(1-\frac{2\,M}{r}-\frac{\Lambda}{3\,(1-\ell)}\,r^2\right)}{\sqrt{ 1 - \dfrac{3\,M}{r}}}, & \text {for } Q=0\\[2.5em]
        \displaystyle \frac{\left(1-\frac{2\,M}{r}+\frac{Q^2}{(1-\ell)^2\,r^2}-\frac{\Lambda}{3\,(1-\ell)}\,r^2\right)}{\sqrt{ 1 - \dfrac{3\,M}{r} + \dfrac{2\,Q^2}{(1-\ell)^2\,r^2} }}, & \text {for } Q \neq 0
    \end{cases}
    .\label{cc6a}
\end{equation}
}

\begin{figure}[ht!]
    \centering
    \includegraphics[width=0.4\linewidth]{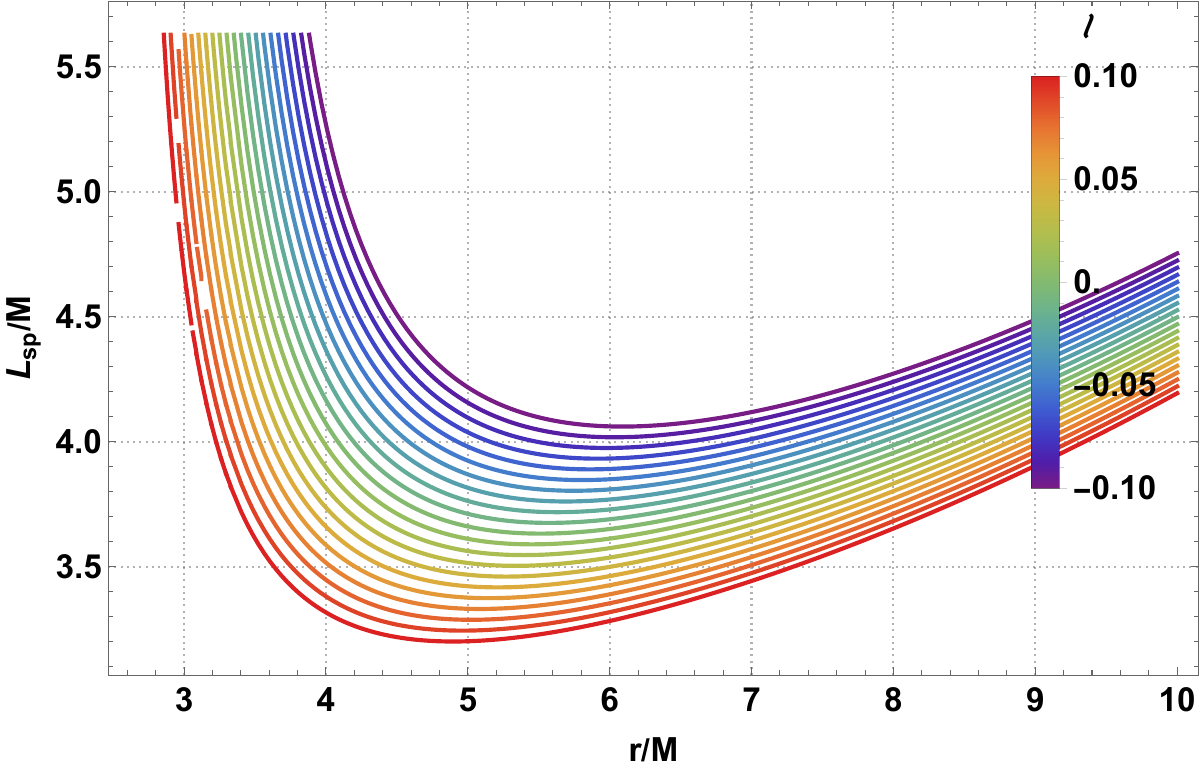}\qquad
    \includegraphics[width=0.4\linewidth]{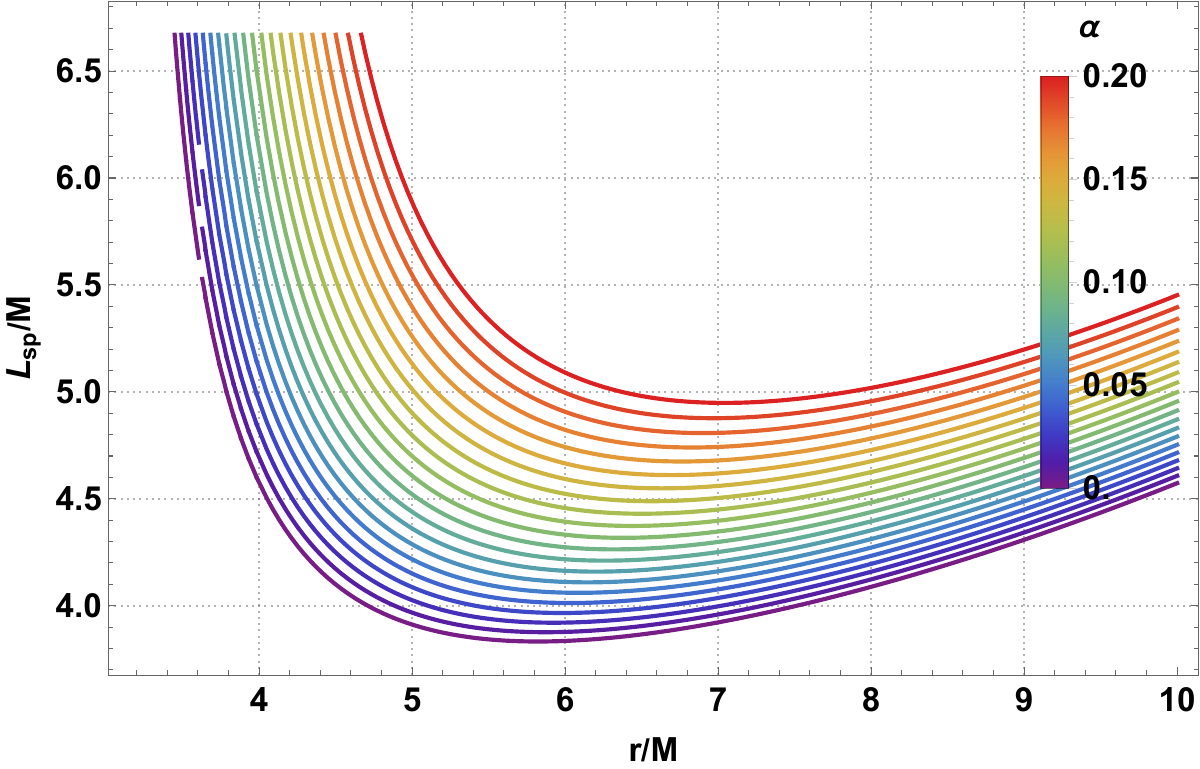}\\
    (i) $\alpha=0.05$ \hspace{6cm} (ii) $\ell=-0.1$
    \caption{ Behavior of the specific angular momentum for different values of CoS parameter $\alpha$. Here $Q/M=0.5,\,\,\Lambda=-0.001/M^2$.}
    \label{fig:momentum}
\end{figure}

\begin{figure}[ht!]
    \centering
    \includegraphics[width=0.4\linewidth]{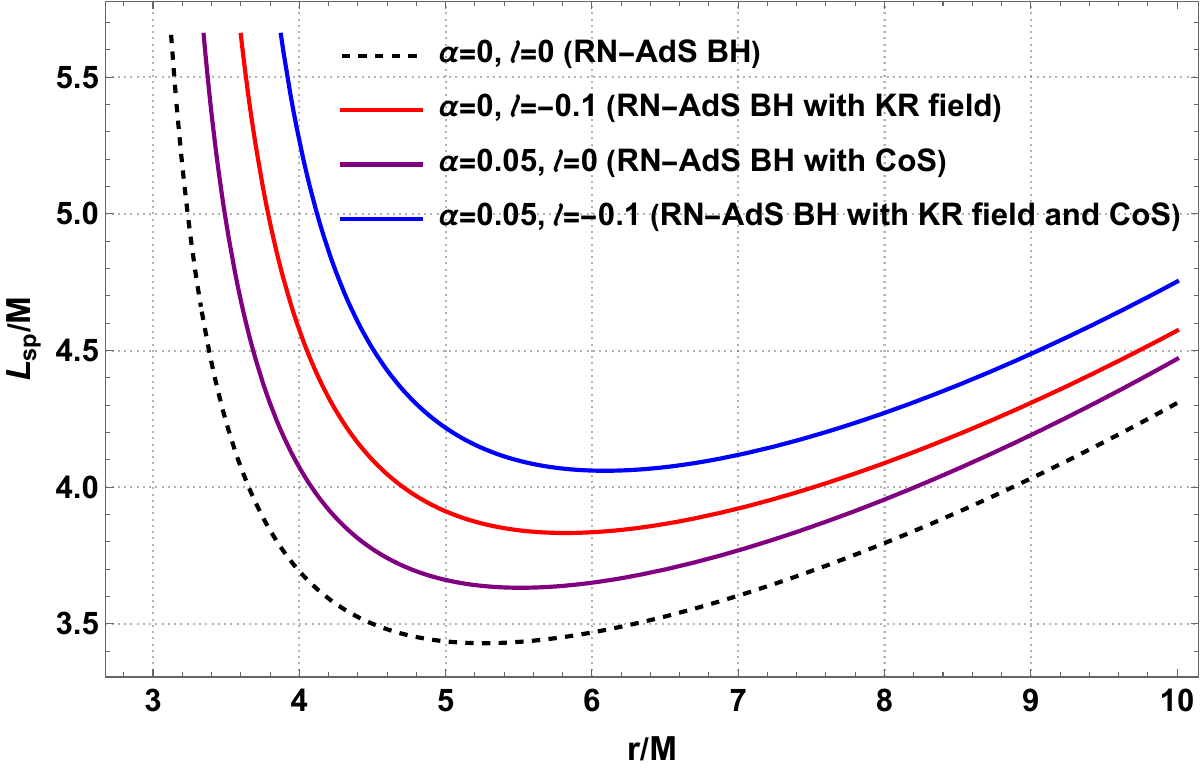}
    \caption{ A comparison of the specific angular momentum in charged BHs under different configurations. Here $Q/M=0.5,\,\,\Lambda=-0.001/M^2$.}
    \label{fig:momentum-comp}
\end{figure}

\begin{figure}[ht!]
    \centering
    \includegraphics[width=0.4\linewidth]{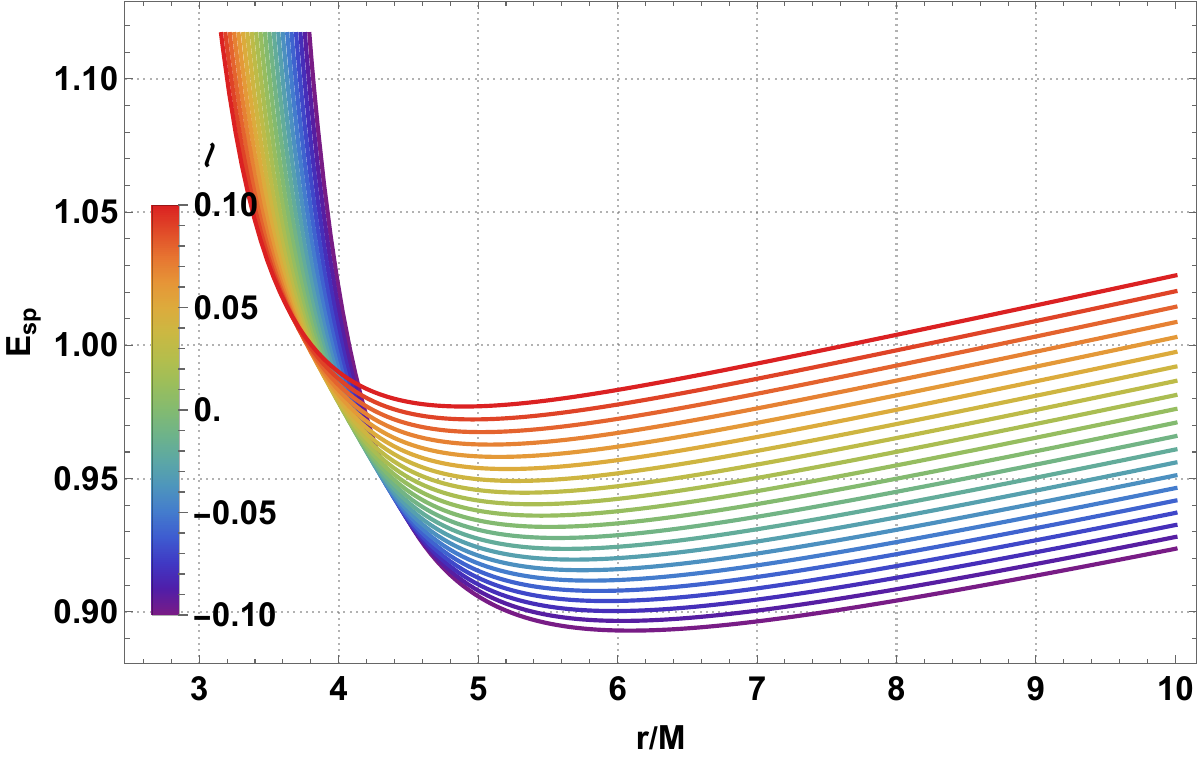}\qquad
    \includegraphics[width=0.4\linewidth]{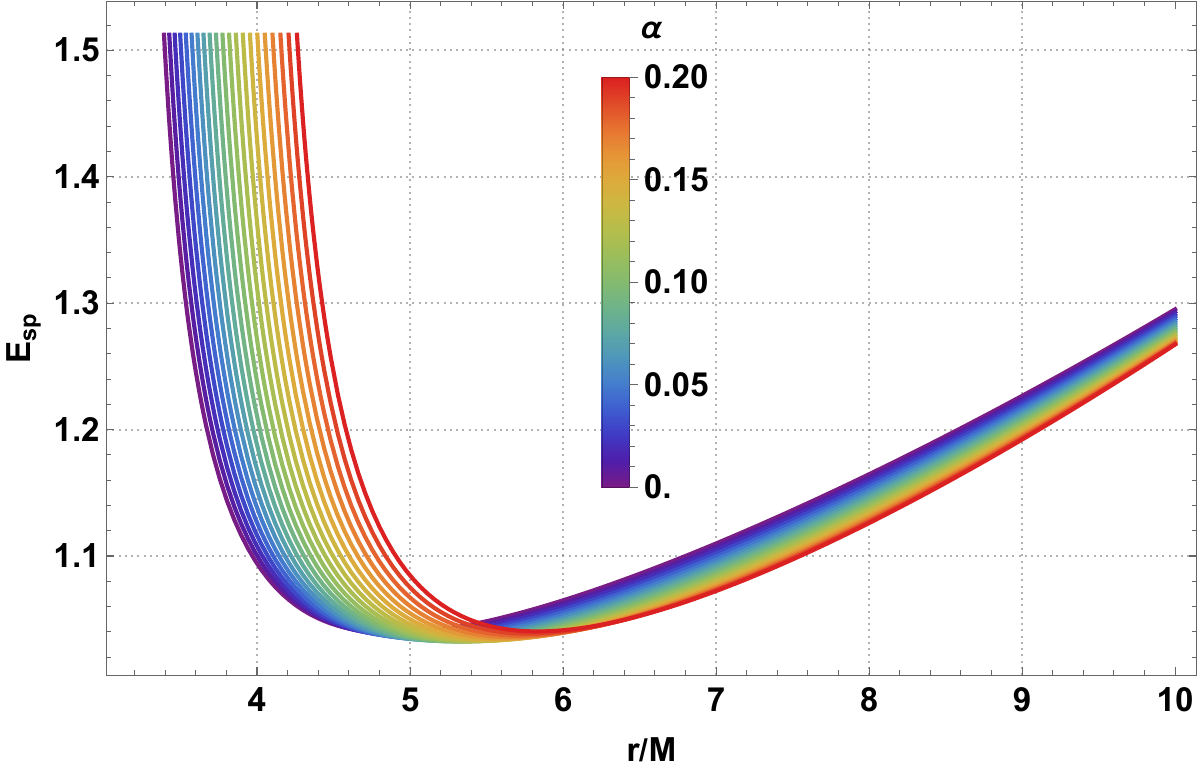}\\
    (i) $\alpha=0.05$ \hspace{6cm} (ii) $\ell=-0.1$
    \caption{ Behavior of the specific energy for different values of LSB parameter $\ell$. Here $Q/M=0.5,\,\,\Lambda=-0.001/M^2$.}
    \label{fig:energy}
\end{figure}

\begin{figure}[ht!]
    \centering
    \includegraphics[width=0.4\linewidth]{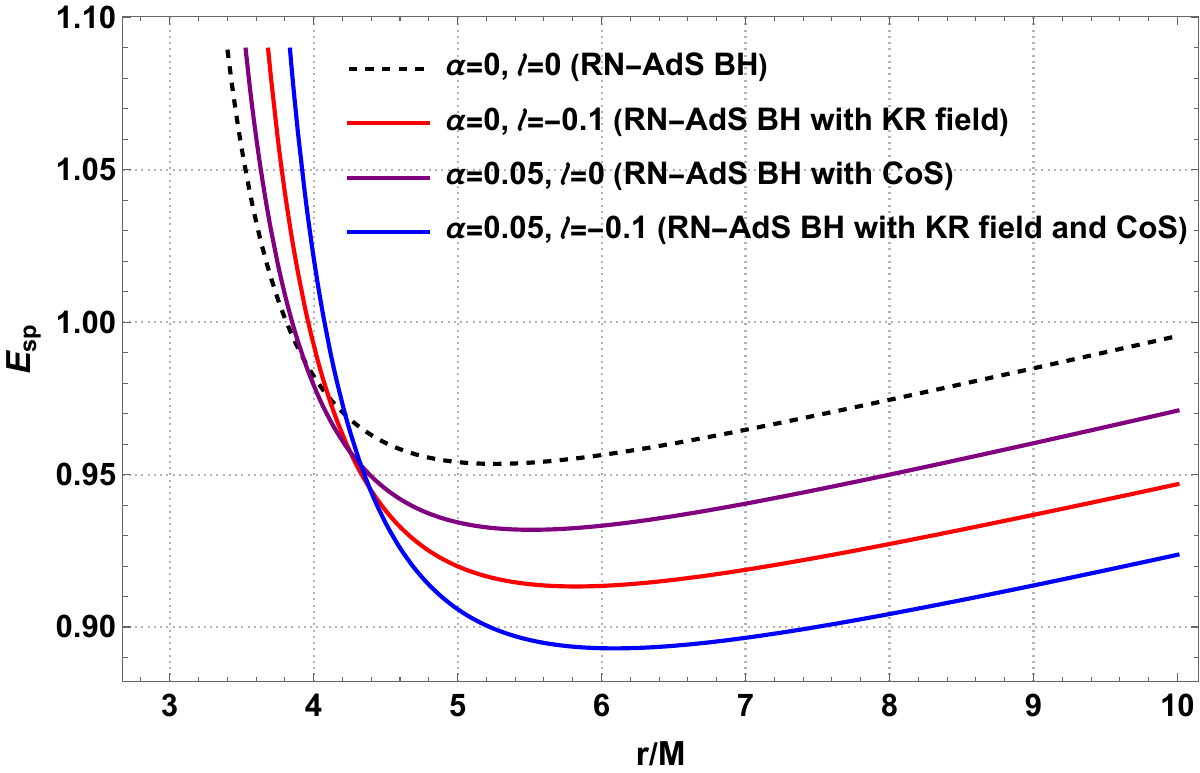}
    \caption{ A comparison of the specific energy in charged BHs under different configurations. Here $Q/M=0.5,\,\,\Lambda=-0.001/M^2$.}
    \label{fig:energy-comp}
\end{figure}

Here, \(\mathrm{E}_\text{sp}\) and \(\mathrm{L}_\text{0\,sp}\) represents the specific energy and angular momentum of massive particles moving in circular orbits around the BH. These quantities are fundamental in characterizing the dynamical properties of test particles in curved space-time. Our analysis shows that both \(\mathrm{E}_\text{sp}\) and \(\mathrm{L}_\text{sp}\) are significantly influenced by the KR field parameter \(\ell\), as well as by the string cloud \((\alpha\).

In Figures \ref{fig:momentum}, we illustrate the behavior of the specific angular momentum \(\mathrm{L}_{\rm sp}\) by varying the string parameter $\alpha$ and the KR field parameter $\ell$ for two different values of the electric charge $Q$ of the BH. In the left panels, we observed that \(\mathrm{L}_{\rm sp}\) decreases with increasing $\ell$. In contrast, in the right panels \(\mathrm{L}_{\rm sp}\) increases with increasing $\alpha$. {\color{black} Figure \ref{fig:momentum-comp} shows a comparison of the specific angular momentum of test particles orbiting in circular paths around charged BHs under different configurations: with and without KR field and string clouds. We observed that the test particles required more specific angular momentum orbiting charged BH under consideration in comparison to the BHs in other configurations.}
 
In Figures \ref{fig:energy}, we illustrate the behavior of the specific energy \(\mathrm{E}_{\rm sp}\) by varying the string parameter $\alpha$ and the KR field parameter $\ell$ for two different values of the electric charge $Q$ of the BH. In both panels, we observed that \(\mathrm{E}_{\rm sp}\) increases with increasing $\alpha$ and $\ell$. {\color{black} Figure \ref{fig:energy-comp} shows a comparison of the specific energy of test particles orbiting in circular paths around charged BHs under different configurations: with and without KR field and string clouds.}

The last condition in Eq.(\ref{cc4}) will give the innermost stable circular orbits satisfying the following equation:
\begin{equation}\label{cc7}
    3\,f(r)\,f'(r)+r\,f(r)\,f''(r)-2\,r\,(f'(r))^2=0
\end{equation}
{\color{black} Substituting the metric function $f(r)$ given in Eq. (\ref{function}) in the Eq. (\ref{cc7}) yields
\begin{align}
    &\left(M r-\frac{Q^2}{(1 - \ell)^2}-\frac{\Lambda}{3\,(1-\ell)}\,r^4\right)\left(\frac{3 (1-\alpha)}{1-\ell} r^2
- 10M r
+ \frac{7 Q^{2}}{(1-\ell)^{2}}
+ \frac{\Lambda}{3(1-\ell)}\, r^{4}\right)\nonumber\\
&+\left(\frac{1-\alpha}{1 - \ell} r^2 - 2 M r +\frac{Q^2}{(1 - \ell)^2}-\frac{\Lambda}{3\,(1-\ell)}\,r^4\right)\left(-2\,M r+3 \frac{Q^2}{(1 - \ell)^2}-\frac{\Lambda}{3\,(1-\ell)}\,r^4\right)=0.\label{special-1}
\end{align}
Several limiting cases of the parameter dependencies are as follows:
\begin{itemize}
    \item For $Q=0$ (uncharged case): 
    \begin{align}
    -4 (1-\alpha)\Lambda r^{4}
+ 15 M \Lambda (1-\ell) r^{3}
+ 3 M (1-\alpha)(1-\ell) r
- 18 M^{2} (1-\ell)^2
=0.\,\, \mbox{(AdS-BH in KR-gravity with CoS)}\label{special-2}
\end{align}

\item For $\ell=0$ (no LV effects): 
\begin{align}
    -4\Lambda (1-\alpha) r^{6}
+ 15\Lambda M r^{5}
- 12\Lambda Q^{2} r^{4}
+ 3M (1-\alpha) r^{3}
- 18 M^{2} r^{2}
+ 27 M Q^{2} r
- 12 Q^{4}
= 0 .\,\, \mbox{(RN-AdS BH with CoS)}\label{special-4}
\end{align}

\item For $\Lambda=0$ (without the cosmological constant):
\begin{align}
    \frac{M (1-\alpha)}{1-\ell}\, r^{3}
- 6 M^{2} r^{2}
+ \frac{9 M Q^{2}}{(1-\ell)^{2}}\, r
- \frac{4 Q^{4}}{(1-\ell)^{4}}
=0.\quad \mbox{(Charged-BH in KR background with CoS)}\label{special-5}
\end{align}

\item For $\alpha=\ell>0$, that is the equal contribution form the KR-field and string cloud effects, the above polynomial relation (\ref{special-1}) simplifies to
\begin{align}
    &\left(M r-\frac{Q^2}{(1 - \ell)^2}-\frac{\Lambda}{3\,(1-\ell)}\,r^4\right)\left(3 r^2
- 10M r + \frac{7 Q^{2}}{(1-\ell)^{2}}
+ \frac{\Lambda}{3(1-\ell)}\, r^{4}\right)\nonumber\\
&+\left(r^2 - 2 M r +\frac{Q^2}{(1 - \ell)^2}-\frac{\Lambda}{3\,(1-\ell)}\,r^4\right)\left(-2\,M r+3 \frac{Q^2}{(1 - \ell)^2}-\frac{\Lambda}{3\,(1-\ell)}\,r^4\right)=0.\label{special-6}
\end{align}
\end{itemize}

}

\begin{figure}[ht!]
    \centering
    \includegraphics[width=0.45\linewidth]{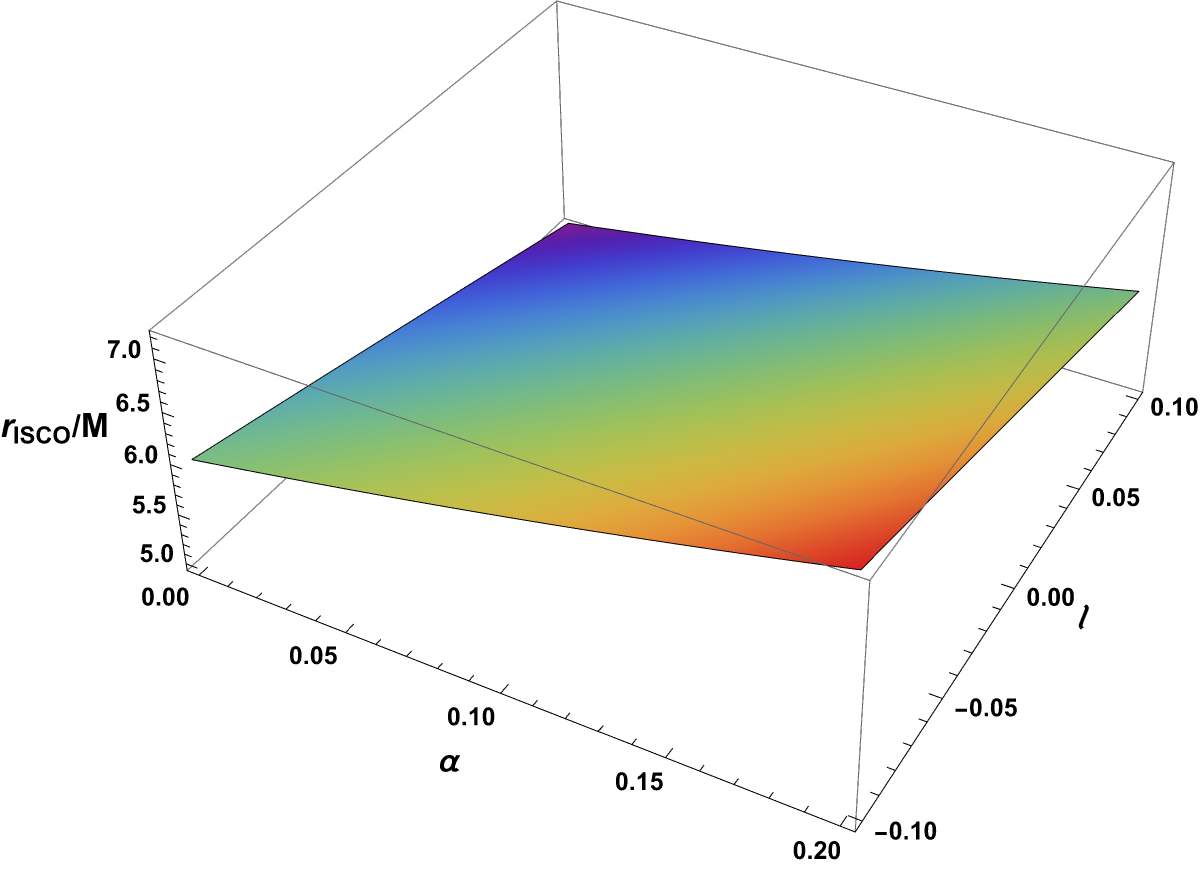}\qquad
    \includegraphics[width=0.45\linewidth]{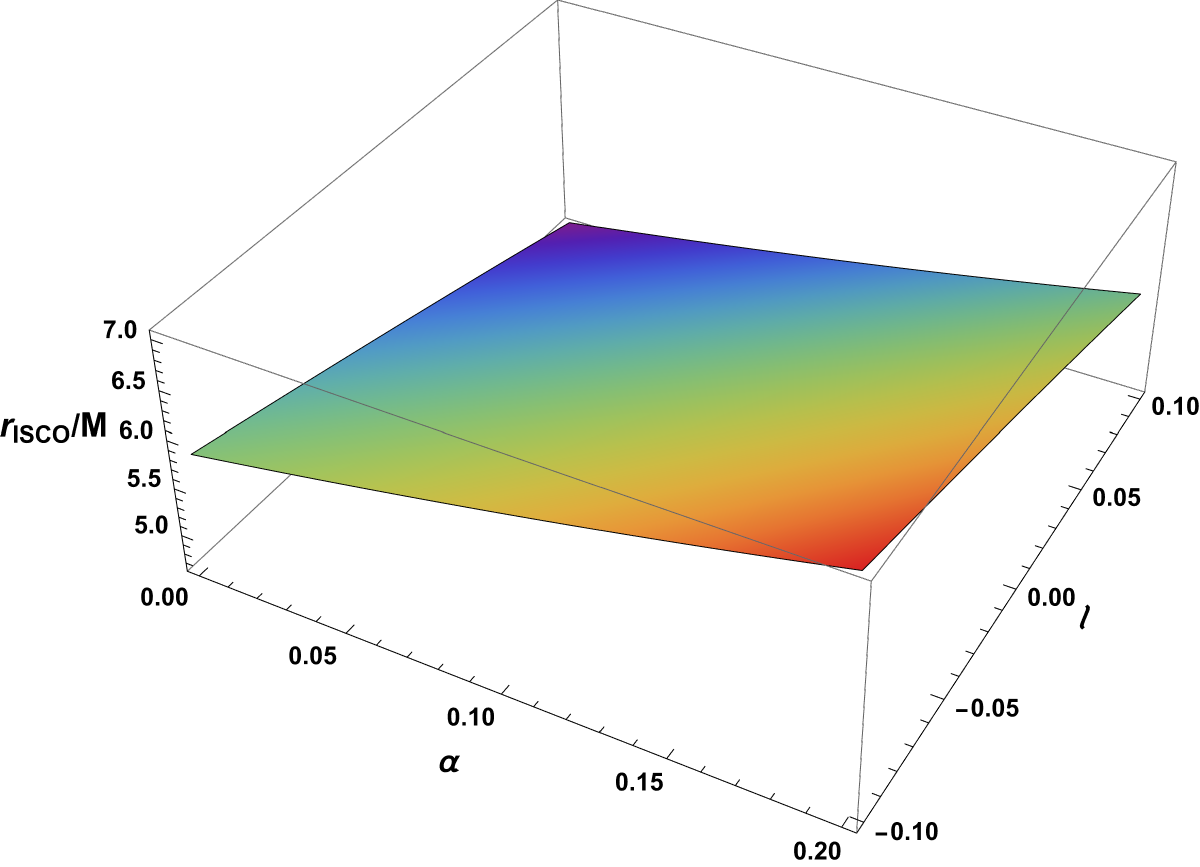}\\
    (i) $Q/M=0.25$ \hspace{6cm} (ii) $Q/M=0.50$
    \caption{ISCO radius $r_{\rm ISCO}$ by varying KR field parameter $\ell$ and string cloud parameter $\alpha$ for two values of the electric charge $Q/M$. Here, $\Lambda=-0.001/M^2$.}
    \label{fig:ISCO}
\end{figure}

It is important to note that obtaining an analytical solution to the resulting polynomial equation for \( r = r_{\text{ISCO}} \) is highly nontrivial due to its complexity. Nevertheless, by selecting appropriate parameter values, one can numerically evaluate the ISCO radius while varying the KR field parameter \(\ell\) and the CoS parameter \(\alpha\). Table~\ref{tab:2} present the numerical results for the ISCO radius per unit mass ($r_{\rm ISCO}/M$) corresponding to \( Q/M=0.5 \) an $\Lambda=-0.001/M^2$.

\begin{table}[ht!]
\centering
\caption{Numerical solutions of ISCO radius \(r_\text{ISCO}/M\) for varying \(\alpha\) and \(\ell\). Here $Q/M=0.5,\,\,\Lambda=-0.001/M^2$.}
\label{tab:2}
\begin{tabular}{|c|cccc|}
\hline
$\alpha (\downarrow) \backslash \ell (\rightarrow)$ 
& $-0.10$ & $-0.05$ & $0.05$ & $0.10$ \\
\hline
0.05 
& 6.0014 & 5.7599 & 5.2605 & 5.0020 \\
0.10 
& 6.2895 & 6.0411 & 5.5269 & 5.2605 \\
0.15 
& 6.6030 & 6.3471 & 5.8172 & 5.5424 \\
0.20 
& 6.9458 & 6.6818 & 6.1349 & 5.8511 \\
\hline
\end{tabular}
\end{table}

From this Table, it becomes evident that both the KR field parameter \(\ell\) and the CoS parameter \(\alpha\) have a significant impact on the ISCO radius. For a fixed value of the CoS parameter, such as \(\alpha = 0.05\), the ISCO radius tends to decrease as the KR field parameter \(\ell\) increases. This indicates that stronger KR field effects lead to a more compact ISCO. A similar reduction behavior is observed when the electric charge per unit mass \(Q/M\) is increased, holding \(\alpha\) and \(\ell\) fixed (see Fig. \ref{fig:ISCO}). Conversely, when fixing the KR field parameter, for example, at \(\ell = -0.1\), the ISCO radius increases with increasing values of the CoS parameter \(\alpha\). This suggests that the CoS parameter tends to enlarge the ISCO radii. These trends highlight the intricate interplay between the KR field, CoS effects, and electric charge in shaping the orbital dynamics around the compact object.

\section{Fundamental Frequencies}

The motion of test particles in the space-time around BHs can be decomposed into radial, azimuthal, and vertical oscillations. The frequencies associated with these oscillations, known as fundamental frequencies, are essential to understanding the dynamics of accretion disks and the origin of QPOs. In the presence of QED gravity, CoS and PFDM, the fundamental frequencies are modified due to changes in the spacetime geometry. The three fundamental frequencies are:

\begin{center}
    {\bf I.\, Keplerian Frequency}
\end{center}

The Keplerian frequency, $\nu_{K,i}$, is a fundamental frequency associated with the orbital motion of a test particle around a central massive object. It describes the azimuthal angular velocity of the particle along a stable circular orbit and is crucial to understanding the dynamics of the accretion disk and the QPO phenomenon. The Keplerian frequency is given by::
\begin{equation}
    \nu _{K,r,\theta} = \frac{1}{2\,\pi }\frac{c^3}{G\,M}\, \Omega _{r,\theta, \phi }\ , [{\textrm{Hz}}]\ . \label{pp1}
\end{equation}

Azimuthal frequency \(\Omega_{\phi}\), which is the frequency of Keplerian orbits of test particles in the azimuthal direction, is given by:
\begin{equation}
    \nu_K=\Omega_{\phi}=\frac{d\phi}{dt}=\frac{\dot{\phi}}{\dot{t}}=\frac{\omega_{\phi}}{\dot{t}}=\sqrt{\frac{f'(r)}{2\,r}}\Bigg{|}_{r=r_c}=
    \begin{cases}
        \displaystyle \frac{1}{r}\,\left(\frac{M}{r}-\frac{\Lambda}{3\,(1-\ell)}\,r^2\right)^{1/2}, & \text{for } Q=0,\\[1.2em]
        \displaystyle \frac{1}{r}\,\left(\frac{M}{r}-\frac{Q^2}{(1-\ell)^2\,r^2}-\frac{\Lambda}{3\,(1-\ell)}\,r^2\right)^{1/2}, & \text{for } Q \neq 0.
    \end{cases}
    \label{pp2}
\end{equation}

\begin{center}
    {\bf II.\, Radial and Vertical Frequencies}
\end{center}

The radial and vertical angular frequencies \(\nu_r=\Omega_r\) and \(\nu_{\theta}=\Omega_{\theta}\) are the frequency of oscillations of the neutral test particles in the radial direction along the stable orbits, which can be determined from the second derivatives of the effective potential by $r$ and $\theta$ coordinates, respectively are:
\begin{align} 
\Omega_r^2=-\frac{1}{2\, g_{rr}\,(u^t)^2}\, \frac{\partial^2 U_{\text {eff}}}{\partial r^2} \bigg{|}_{r = r_c}.\label{pp3} 
\end{align}
And
\begin{align} 
\Omega^2_{\theta}=- \frac{1}{2\, g_{\theta \theta }\, (u^t)^2} \frac{\partial ^2 U_{\text {eff}}}{\partial \theta ^2} \bigg{|}_{r = r_c}\ . \label{pp4}
\end{align}

Noted that these are the frequencies of neutral test particles as measured by a static distant observer. According to a local observer, the harmonic oscillatory motion frequencies are provided by: 
\begin{align}
    \omega_r^2&=-\frac{1}{2\, g_{rr}}\, \frac{\partial^2 U_{\text {eff}}}{\partial r^2} \bigg{|}_{r = r_c},\label{pp5} \\
    \omega^2_{\theta}&=-\frac{1}{2\, g_{\theta \theta }} \frac{\partial ^2 U_{\text {eff}}}{\partial \theta ^2} \bigg{|}_{r = r_c}\ . \label{pp6}
\end{align}
Here $\Omega^2_i=\frac{\omega^2_i}{(u^t)^2}$ with $u^t=\dot{t}=\sqrt{\frac{2}{2\,f(r)-r\,f'(r)}}$ is the relation between these two types of frequencies as measured by a local and a static distant observer. 

\begin{figure}[ht!]
    \centering
    \includegraphics[width=0.45\linewidth]{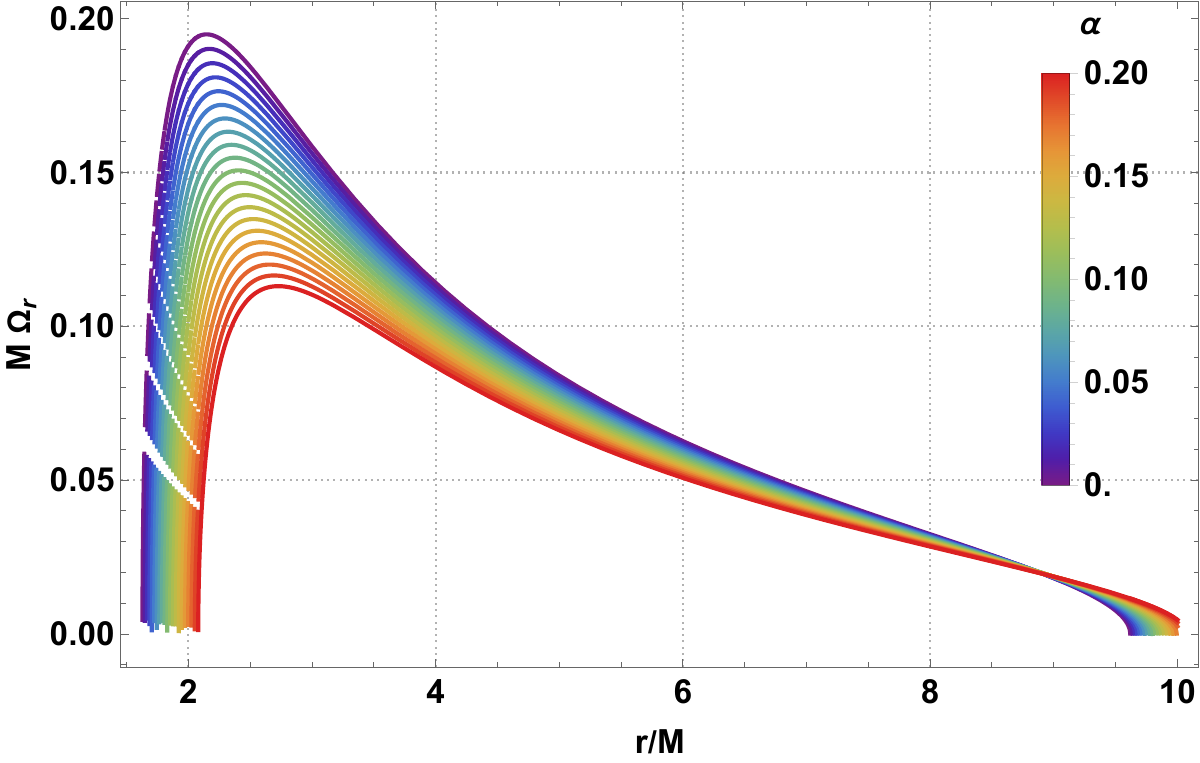}\qquad
    \includegraphics[width=0.45\linewidth]{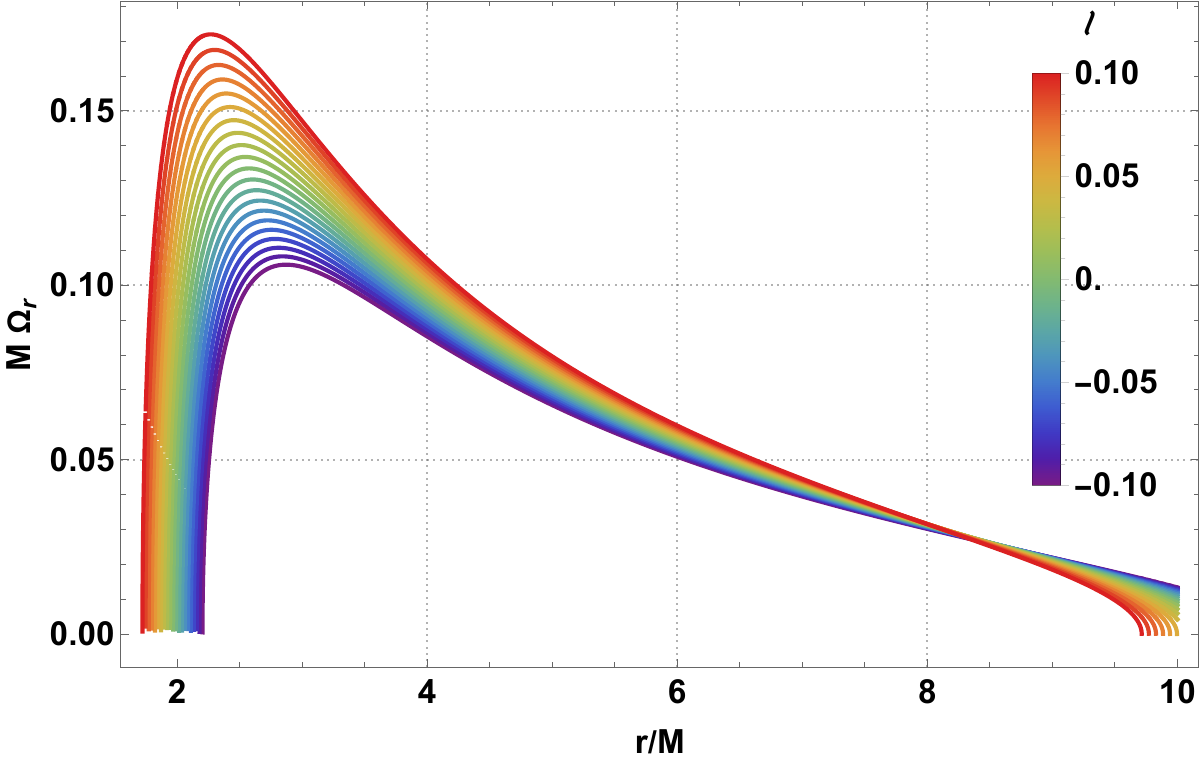}\\
    (i) $\ell=0.1$ \hspace{6cm} (ii) $\alpha=0.05$
    \caption{ Behavior of the fundamental frequency $\Omega_r$ for different values of CoS parameter $\alpha$ and LSB parameter $\ell$. Here $Q/M=0.5,\,\,\Lambda=-0.001/M^2$.}
    \label{fig:frequency-1}
    \hfill\\
    \includegraphics[width=0.45\linewidth]{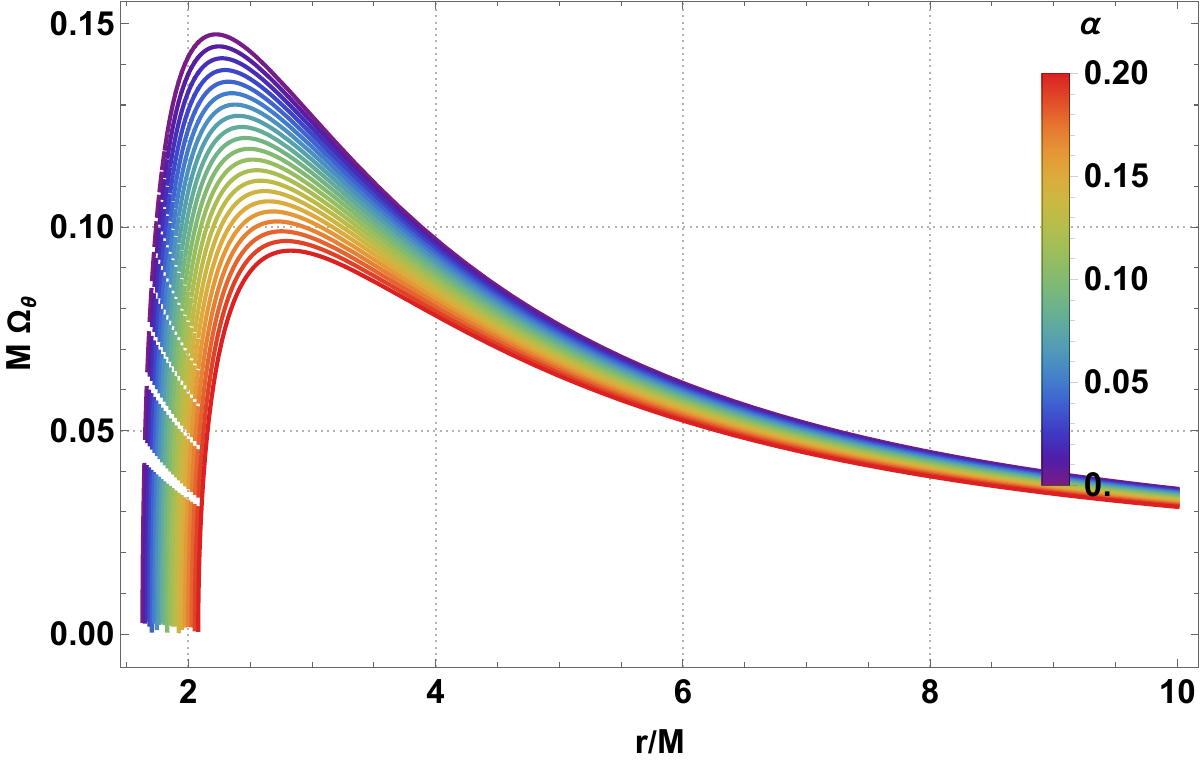}\qquad
    \includegraphics[width=0.45\linewidth]{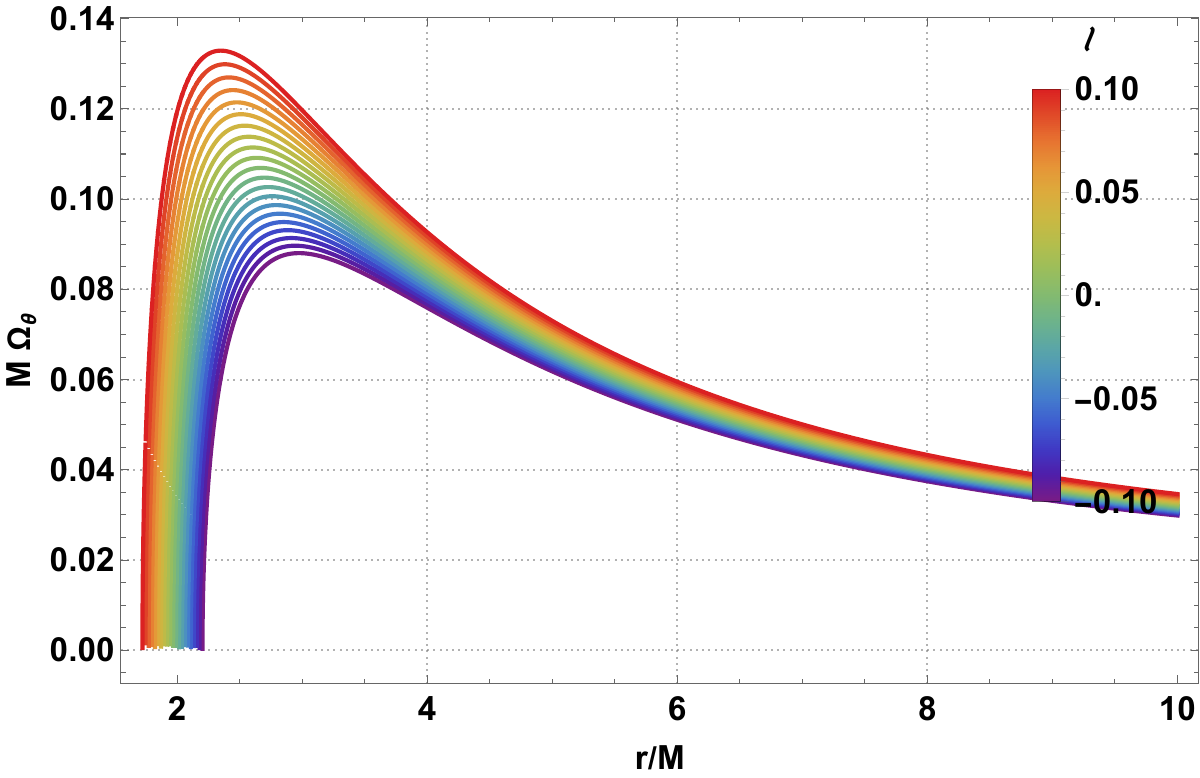}\\
    (i) $\ell=0.1$ \hspace{6cm} (ii) $\alpha=0.05$
    \caption{ Behavior of the fundamental frequency $\Omega_{\theta}$ for different values of CoS parameter $\alpha$ and LSB parameter $\ell$. Here $Q/M=0.5,\,\,\Lambda=-0.001/M^2$.}
    \label{fig:frequency-2}
\end{figure}

In our case, we find these frequencies for charged BHs ($ Q \neq 0$)  as follows: {\footnotesize
\begin{align}
    \Omega_r^2&=-\frac{1}{2}\,\left(\frac{1-\alpha}{1-\ell}-\frac{2\,M}{r}+\frac{Q^2}{(1-\ell)^2\,r^2}-\frac{\Lambda}{3\,(1-\ell)}\,r^2\right)
    \Bigg[\left(\frac{1-\alpha}{1-\ell}-\frac{2\,M}{r}+\frac{Q^2}{(1-\ell)^2\,r^2}-\frac{\Lambda}{3\,(1-\ell)}\,r^2\right)\left(-\frac{4\,M}{r^3}+\frac{6\,Q^2}{(1-\ell)^2\,r^4}-\frac{2\,\Lambda}{3\,(1-\ell)}\right)\nonumber\\
    &+\frac{2}{r^2}\,\left(\frac{M}{r} - \frac{Q^2}{(1-\ell)^2\,r^2}-\frac{\Lambda}{(1-\ell)\,r}\right)\left(\frac{3\,(1-\alpha)}{1-\ell}-\frac{10\,M}{r}+\frac{7\,Q^2}{(1-\ell)^2\,r^2}+\frac{\Lambda}{3\,(1-\ell)}\,r^2\right)\Bigg],\label{pp7}\\[1.5em]
    \Omega^2_{\theta}&=-\frac{1}{r^2}\,\left(\frac{M}{r}-\frac{Q^2}{(1-\ell)^2\,r^2}-\frac{\Lambda}{3\,(1-\ell)}\,r^2\right)\,\left(\frac{1-\alpha}{1-\ell}-\frac{2\,M}{r}+\frac{Q^2}{(1-\ell)^2\,r^2}-\frac{\Lambda}{3\,(1-\ell)}\,r^2\right),\label{pp8}
\end{align}
} And {\footnotesize
\begin{align}
    \omega_r^2&=-\frac{1}{2}\,\frac{\left(\frac{1-\alpha}{1-\ell}-\frac{2\,M}{r}+\frac{Q^2}{(1-\ell)^2\,r^2}-\frac{\Lambda}{3\,(1-\ell)}\,r^2\right)}{\left(\frac{1-\alpha}{1-\ell}-\frac{3\,M}{r}+\frac{2\,Q^2}{(1-\ell)^2\,r^2}\right)}
    \Bigg[\left(\frac{1-\alpha}{1-\ell}-\frac{2\,M}{r}+\frac{Q^2}{(1-\ell)^2\,r^2}-\frac{\Lambda}{3\,(1-\ell)}\,r^2\right)\,\left(-\frac{4\,M}{r^3}+\frac{6\,Q^2}{(1-\ell)^2\,r^4}-\frac{2\,\Lambda}{3\,(1-\ell)}\right)\nonumber\\
    &+\frac{2}{r^2}\,\left(\frac{M}{r} - \frac{Q^2}{(1-\ell)^2\,r^2}-\frac{\Lambda}{(1-\ell)\,r}\right)\,\left(\frac{3\,(1-\alpha)}{1-\ell}-\frac{10\,M}{r}+\frac{7\,Q^2}{(1-\ell)^2\,r^2}+\frac{\Lambda}{3\,(1-\ell)}\,r^2\right)\Bigg],\label{pp7a}\\[1.5em]
    \omega^2_{\theta}&=-\frac{1}{r^2}\,\left(\frac{M}{r}-\frac{Q^2}{(1-\ell)^2\,r^2}-\frac{\Lambda}{3\,(1-\ell)}\,r^2\right)\,\left(\frac{1-\alpha}{1-\ell}-\frac{2\,M}{r}+\frac{Q^2}{(1-\ell)^2\,r^2}-\frac{\Lambda}{3\,(1-\ell)}\,r^2\right)\left(\frac{1-\alpha}{1-\ell}-\frac{3\,M}{r}+\frac{2\,Q^2}{(1-\ell)^2\,r^2}\right)^{-1},\label{pp8a}
\end{align}
}
{\color{black} Here one may also consider several special cases, such as $\ell = 0$ (absence of KR-field effects), 
$\alpha = 0$ (absence of string cloud effects), 
and $\alpha = \ell > 0$ (equal contribution from the KR field and the string cloud). For each of these cases, one can compute the corresponding pairs of frequency expressions $(\Omega_r\,,\, \Omega_{\theta})$ and $(\omega_r\,,\, \omega_{\theta})$.} In Figures \ref{fig:frequency-1} to \ref{fig:frequency-2}, we illustrate the behavior of the fundamental frequencies $\Omega_r$, and $\Omega_{\theta}$ as measured by a static distant observer by varying the string parameter $\alpha$ and the KR field parameter $\ell$, while keeping other parameters are held constant.

\begin{figure}[ht!]
    \centering
    \includegraphics[width=0.45\linewidth]{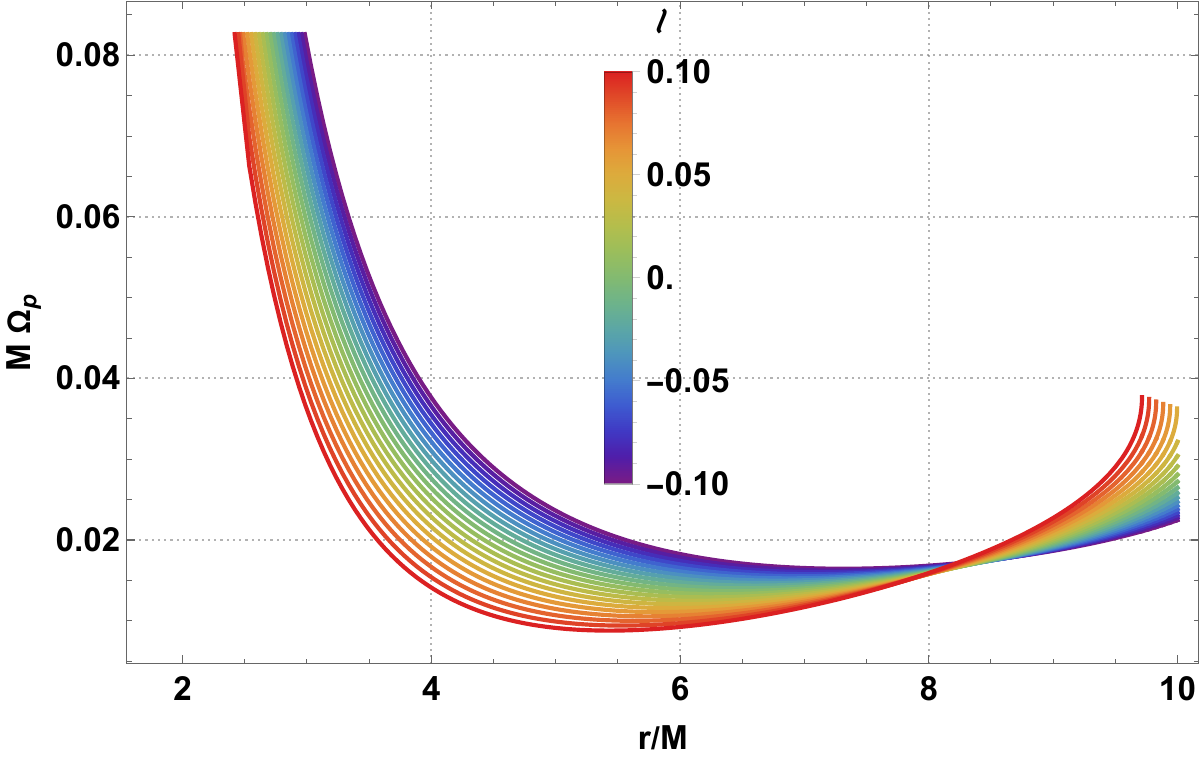}\qquad
    \includegraphics[width=0.45\linewidth]{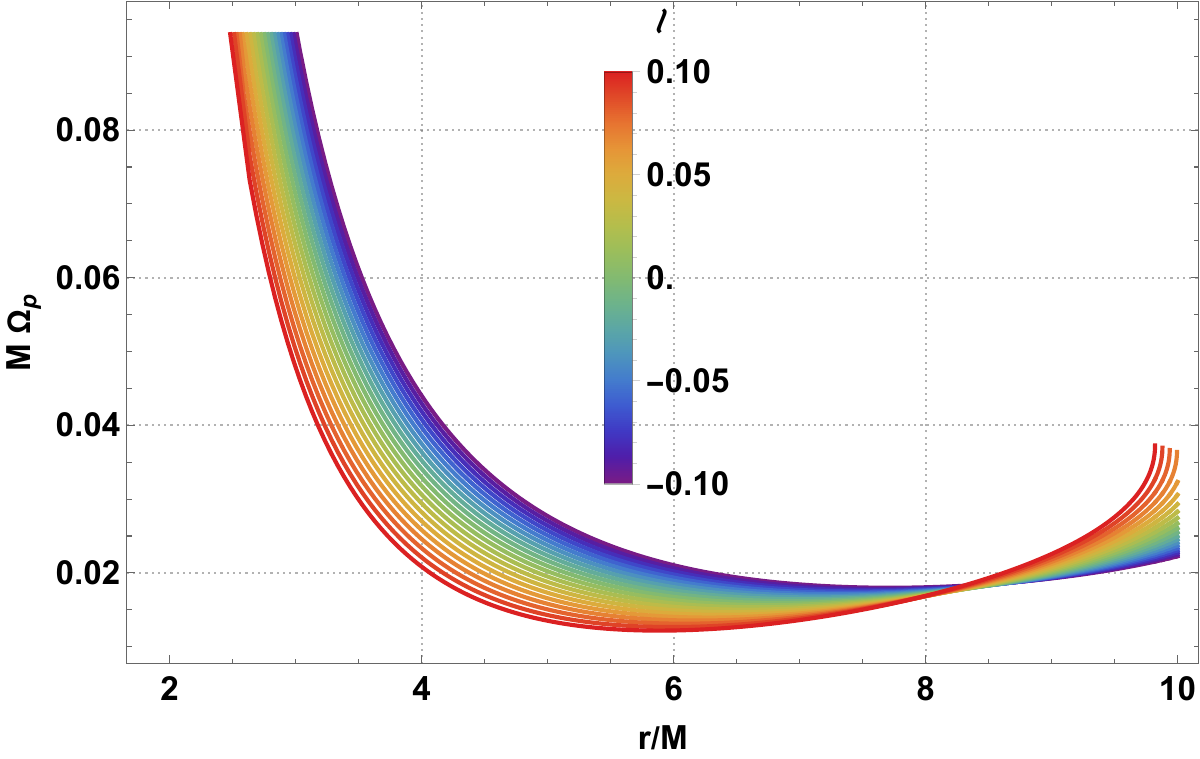}\\
    (i) $\alpha=0.05$ \hspace{6cm} (ii) $\alpha=0.10$
    \caption{ Behavior of the periastron frequency for different values of LSB parameter $\ell$. Here $Q/M=0.5,\,\,\Lambda=-0.001/M^2$.}
    \label{fig:frequency-4}
\end{figure}

\begin{center}
    {\bf III.\, Periastron Frequency }
\end{center}

Finally, we determine the periastron frequency of a neutral test particle orbiting a non-rotating charged EB AdS BH, with particular attention to small perturbations near the equatorial plane at 
\( \theta = \frac{\pi}{2} \). We perturb the particle slightly from its stable circular orbit to analyze the resulting periastron precession. This perturbation induces oscillations about the equilibrium position, which are characterized by a radial frequency \( \Omega_r \). According to the  following relationship, the periastron frequency, denoted by \( \Omega_p \), is defined as the difference between the orbital (azimuthal) frequency \( \Omega_\phi \) and the radial frequency \( \Omega_r \), as given by the following relation~:
\begin{equation}
   \Omega_p = \Omega_{\phi} - \Omega_r.\label{pp10} 
\end{equation}

In Figure \ref{fig:frequency-4}, we illustrate the behavior of the periastron frequency $\Omega_p$ by varying the KR field parameter $\ell$ for two values of the electric charge $Q=0.25$ and $Q=0.5$, while keeping the string parameter $\alpha$ as well as the others held fixed.

\section{Summary and conclusions}\label{sec4}

In this study, we have analyzed the optical properties and dynamical behavior of test particles in the vicinity of a charged AdS BH surrounded by a CoS within the framework of KR gravity. Our investigation focused on exploring the influence of key parameters, including the KR field parameter $\ell$, the CoS parameter $\alpha$, the BH charge $Q$, and the cosmological constant $\Lambda$, on various physical and observational quantities. Specifically, we examined their effects on the spacetime curvature, the effective potentials governing both photon and test particle dynamics, and the topological characteristics of photon rings. In addition, we analyzed the specific angular momentum, energy profiles, and the fundamental frequencies associated with QPOs of neutral test particles.

Our findings show that the KR field, in conjunction with the CoS and cosmological constant, induces several significant modifications to the BH spacetime. Changes in the spacetime geometry are evidenced by modifications in the Ricci scalar, the quadratic Ricci tensor, and the Kretschmann scalar, reflecting the combined influence of the KR field and CoS on the gravitational field near the BH. The effective potential for photons is also modified, which directly affects the PS radius and observable features such as the BH shadow. These changes can lead to measurable differences in the optical appearance of BHs, potentially detectable by EHT or future observational facilities. Using observational data from M87* and Sgr A*, we have constrained the parameters $\ell$ and $\alpha$ and established bounds on their allowed values.

The effective potential for neutral test particles is similarly perturbed, resulting in deviations in the specific energy $\mathrm{E}_{\rm sp}$ and angular momentum $\mathrm{L}_{\rm sp}$ required for circular orbits. Consequently, the ISCO location is shifted, which has direct implications for accretion disk dynamics and associated radiation spectra. Furthermore, the fundamental frequencies of QPOs, including the Keplerian frequency $\Omega_\phi$, radial frequency $\Omega_r$, vertical frequency $\Omega_\theta$, and periastron frequency $\Omega_p$, are all affected by the presence of the KR field and CoS. This suggests that these effects may imprint observable signatures on the timing properties of X-ray emissions from accreting systems.

Taken together, these results demonstrate that KR gravity leads to measurable and potentially observable deviations from GR. The effects on key astrophysical observables-such as the BH shadow, photon trajectories, and QPO frequencies-highlight the potential of KR gravity as a viable framework for probing new physics in strong gravity regimes. Our work underscores the importance of using high-precision astrophysical observations to constrain alternative theories of gravity, particularly through imaging and timing techniques.

Future research could extend the present analysis by incorporating more realistic models of accretion disks, including magnetohydrodynamic effects and radiative transfer, to better simulate astrophysical environments. Additionally, further investigation into other observable signatures-such as the polarization of emitted radiation, gravitational waveforms from perturbed BHs, and echoes in gravitational wave signals-may deepen our understanding of the role of the KR field and CoS in modified gravity. We also plan to study perturbations of various spin fields and analyze the associated QNMs, thereby offering a more complete picture of the physical and observational implications of KR gravity in the presence of a CoS and cosmological constant. Finally, we aim to investigate the greybody factors and analyze the combined effects of the geometric parameters.

{\color{black}
\section*{Appendix} \label{app:A}

In this appendix, we present the theoretical framework of the modified gravity model in which a KR antisymmetric tensor field $B_{\mu\nu}$ couples nonminimally to the Ricci tensor, allowing for spontaneous LSB. This formulation follows the approach developed in Refs.~\cite{Yang:2023wtu,ref3}.

\subsection{Action and Field Equations}

The action for this theory is given by \cite{Yang:2023wtu,ref3}:
\begin{align}
S &= \int d^4x \, \sqrt{-g} \left[ \frac{1}{2\kappa} \left( R - 2\Lambda + \varepsilon\, B^{\mu\lambda} B^{\nu}{}_{\lambda}\, R_{\mu\nu} \right) 
- \frac{1}{12} H_{\lambda\mu\nu} H^{\lambda\mu\nu} 
- V\left(B_{\mu\nu} B^{\mu\nu} \pm b^2\right) \right] \nonumber \\
&\quad + \int d^4x\,\sqrt{-g}\,\mathcal{L}_{\rm CoS},
\label{action}
\end{align}
where $\kappa = 8\pi G_N$, with $G_N$ being the Newtonian gravitational constant, $\Lambda$ is the cosmological constant, and $\varepsilon$ represents the coupling constant between the Ricci tensor and the KR field $B_{\mu\nu}$. The mass dimension of this coupling constant in natural units is $[\varepsilon] = M^{-2}$. The term $\mathcal{L}_{\rm CoS}$ denotes the Lagrangian density associated with the CoS.

The field strength of $B_{\mu\nu}$ is defined as:
\begin{equation}
H_{\lambda\mu\nu} = \partial_{\lambda} B_{\mu\nu} + \partial_{\mu} B_{\nu\lambda} + \partial_{\nu} B_{\lambda\mu}.
\label{field_strength}
\end{equation}

It should be noted that, as must be for a bumblebee model, the action lacks gauge symmetry due to the nonminimal coupling and the potential $V(B_{\mu\nu}B^{\mu\nu} \pm b^2)$, which is a smooth function that triggers the spontaneous LSB, leading to a nonvanishing VEV for the KR field.

Varying the action~\eqref{action} with respect to the metric yields the modified Einstein equations:
\begin{equation}
R_{\mu\nu} + \Lambda\, g_{\mu\nu} - \frac{1}{2} g_{\mu\nu}\, R 
= \kappa\, T^{B}_{\mu\nu} + T^{\varepsilon}_{\mu\nu} + T^{\rm CoS}_{\mu\nu},
\label{einstein_eq}
\end{equation}
where $T^{\rm CoS}_{\mu\nu}$ is the energy-momentum tensor for the CoS \cite{PSL}, $T^{B}_{\mu\nu}$ is for the KR field, and $T^{\varepsilon}_{\mu\nu}$ arises from the nonminimal coupling.

\subsection{Energy-Momentum Tensors}

The energy-momentum tensor concerning the pure bumblebee sector is given by:
\begin{equation}
T_{\mu\nu}^B = \frac{1}{2} H_{\alpha\beta\mu} H_{\nu}^{\alpha\beta} 
- \frac{1}{12} g_{\mu\nu} H_{\lambda\alpha\beta} H^{\lambda\alpha\beta} 
- g_{\mu\nu} V + 4 B_{\alpha\mu} B^{\alpha}_{\nu} V',
\label{T_bumblebee}
\end{equation}
while the nonminimal coupling contributes with the energy-momentum tensor:
\begin{align}
T_{\mu\nu}^{\varepsilon} &= \frac{\varepsilon}{\kappa} \bigg( 
\frac{1}{2} g_{\mu\nu} B^{\alpha\lambda} B^{\beta}_{\lambda} R_{\alpha\beta} 
- B_{\alpha\mu} B_{\beta\nu} R^{\alpha\beta} 
- B_{\alpha\beta} B^{\alpha}_{\mu} R^{\beta}_{\nu} 
- B_{\alpha\beta} B^{\alpha}_{\nu} R^{\beta}_{\mu} \nonumber \\
&\quad + \frac{1}{2} \nabla^{\alpha} \nabla_{\mu} B^{\beta}_{\nu} R_{\alpha\beta} 
+ \frac{1}{2} \nabla^{\alpha} \nabla_{\nu} B^{\beta}_{\mu} R_{\alpha\beta} 
- \frac{1}{2} \Box B^{\alpha}_{\mu} B_{\alpha\nu} 
- \frac{1}{2} g_{\mu\nu} \nabla_{\alpha} \nabla_{\beta} B^{\alpha\lambda} B^{\beta}_{\lambda} \bigg).
\label{T_epsilon}
\end{align}

\subsection{KR Field Equation}

Variation of the action~\eqref{action} with respect to $B_{\mu\nu}$ gives the field equation:
\begin{equation}
\nabla_{\lambda} H^{\lambda\mu\nu} 
= 4 V'(X)\, B^{\mu\nu} + \frac{\varepsilon}{\kappa} 
\left( B^{\mu\lambda} R^{\nu}{}_{\lambda} - B^{\nu\lambda} R^{\mu}{}_{\lambda} \right),
\label{KR_field_eq}
\end{equation}
where $X = B_{\mu\nu} B^{\mu\nu} \pm b^2$, and the prime denotes the derivative with respect to the argument of the potential. We note that we are not considering any coupling between the matter fields and the background KR field. At the same time, other geometrical couplings could be considered, such as $\varepsilon B^{\mu\nu} B_{\mu\nu} R$ and $\varepsilon B^{\mu\nu} B^{\alpha\beta} R_{\mu\nu\alpha\beta}$. In this work, we are only interested in the coupling between the KR field and the Ricci tensor.

\subsection{Spontaneous LSB}

To induce spontaneous LSB, the potential $V$ is chosen to give a nonzero VEV for $B_{\mu\nu}$:
\begin{equation}
\langle B_{\mu\nu} \rangle = b_{\mu\nu}, \quad \text{with} \quad b_{\mu\nu}\, b^{\mu\nu} = -|b|^2.
\label{VEV}
\end{equation}

Assuming a purely pseudo-electric KR field configuration:
\begin{equation}
b_{01} = -b_{10} = \frac{|b|}{\sqrt{2}}, \quad \text{others zero},
\label{KR_config}
\end{equation}
leads to a vanishing field strength $H_{\lambda\mu\nu} = 0$, which simplifies the analysis considerably. With this configuration, the equations of motion for $B_{\mu\nu}$ are automatically satisfied.

The strength of the Lorentz violation is encoded in the dimensionless parameter:
\begin{equation}
\ell = \frac{\varepsilon\, |b|^2}{2}.
\label{ell_def}
\end{equation}

\subsection{Effective Field Equations}

The Einstein field equations can be rewritten in a convenient form with an effective energy-momentum tensor:
\begin{equation}
R_{\mu\nu} - \Lambda\, g_{\mu\nu} = T_{\mu\nu} - \frac{1}{2} g_{\mu\nu}\, T,
\label{field}
\end{equation}
where the total energy-momentum tensor is:
\begin{equation}
T_{\mu\nu} = \kappa\, T^{B}_{\mu\nu} + T^{\varepsilon}_{\mu\nu} + T^{\rm CoS}_{\mu\nu},
\label{total_Tmunu}
\end{equation}
and $T = g^{\mu\nu} T_{\mu\nu}$ denotes its trace. It is noteworthy that this total energy-momentum tensor is conserved due to the Bianchi identities.

\subsection{Electromagnetic Field Coupling}

For charged solutions, one must add the electromagnetic field contribution to the action~\eqref{action}:
\begin{equation}
S \to S + \int d^4x\,\sqrt{-g}\,\mathcal{L}_{\rm M}.
\label{action_charged}
\end{equation}

The matter Lagrangian $\mathcal{L}_{\rm M}$ is considered to be the electromagnetic field with a coupling to the KR field:
\begin{equation}
\mathcal{L}_{\rm M} = -\frac{1}{2} F^{\mu\nu} F_{\mu\nu} - \varepsilon\, B^{\alpha\beta} B^{\gamma\rho} F_{\alpha\beta} F_{\gamma\rho},
\label{EM_lagrangian}
\end{equation}
where $F_{\mu\nu} = \partial_{\mu} A_{\nu} - \partial_{\nu} A_{\mu}$ represents the field strength of the electromagnetic field, and $\varepsilon$ is a coupling constant. When the KR field acquires a nonzero VEV, the second term in Eq.~\eqref{EM_lagrangian} induces LSB of the electromagnetic field and allows for the existence of electrically charged BH solutions.

The modified field equation for the charged case is given by:
\begin{equation}
R_{\mu\nu} - \frac{1}{2} g_{\mu\nu}\, R + \Lambda\, g_{\mu\nu} 
= \kappa \left( T^{\rm EM}_{\mu\nu} + T^{B}_{\mu\nu} \right) + T^{\varepsilon}_{\mu\nu} + T^{\rm CoS}_{\mu\nu},
\label{charged_field_eq}
\end{equation}
where $T^{\rm EM}_{\mu\nu}$ is the energy-momentum tensor of the electromagnetic field.

\subsection{Comparison with Global Monopole Case}

It is instructive to compare the CoS matter content with the global monopole case studied in Ref.~\cite{ref3}. For the global monopole (GM), the stress-energy tensor components are:
\begin{equation}
T^{t}_{t}(\text{GM}) = T^{r}_{r}(\text{GM}) = \frac{\kappa\eta^2}{r^2}, \quad 
T^{\theta}_{\theta}(\text{GM}) = T^{\phi}_{\phi}(\text{GM}) = 0,
\label{GM_stress}
\end{equation}
where $\eta$ is the symmetry-breaking energy scale of the global monopole. For the CoS [cf. Eq.~(9) in the main text]:
\begin{equation}
T^{t}_{t}(\text{CoS}) = T^{r}_{r}(\text{CoS}) = \frac{\alpha}{r^2}, \quad 
T^{\theta}_{\theta}(\text{CoS}) = T^{\phi}_{\phi}(\text{CoS}) = 0,
\label{CoS_stress}
\end{equation}
where $\alpha$ is the string cloud parameter. Comparing Eqs.~\eqref{GM_stress} and \eqref{CoS_stress}, we observe the parameter correspondence:
\begin{equation}
\kappa\eta^2 \quad \longleftrightarrow \quad \alpha.
\label{correspondence}
\end{equation}

This correspondence is reflected in the metric solutions. For the global monopole in KR gravity \cite{ref3}:
\begin{equation}
f(r)\big|_{\text{GM}} = \frac{1 - \kappa\eta^2}{1 - \ell} - \frac{2M}{r} - \frac{\Lambda}{3(1-\ell)}\, r^2,
\label{metric_GM}
\end{equation}
while for the CoS:
\begin{equation}
f(r)\big|_{\text{CoS}} = \frac{1 - \alpha}{1 - \ell} - \frac{2M}{r} - \frac{\Lambda}{3(1-\ell)}\, r^2,
\label{metric_CoS}
\end{equation}
confirming the correspondence $\kappa\eta^2 \leftrightarrow \alpha$.

}

\section*{Acknowledgments}

F.A. acknowledges the Inter University Centre for Astronomy and Astrophysics (IUCAA), Pune, India for granting visiting associateship. \.{I}.~S. expresses gratitude to T\"{U}B\.{I}TAK, ANKOS, and SCOAP3 for their academic support. He also acknowledges COST Actions CA22113, CA21106, CA21136, CA23130, and CA23115 for their contributions to networking.

\section*{Data Availability Statement}

No data is associated with this manuscript [Author's comment: No new data were generated or created in this study].

\section*{Code/Software}

No new code/software were developed in this manuscript [Author's comment: No code/software were developed or created in this study].

\end{document}